\documentclass[11pt,aps,amsmath,amssymb,nofootinbib,notitlepage,longbibliography]{revtex4-1}
\usepackage{amsmath,amssymb}              
\usepackage{slashed}
\usepackage{graphicx}
\usepackage{bm}
\usepackage[top=1.0in,bottom=1.0in,left=1.0in,right=1.0in]{geometry}
\usepackage[colorlinks,linkcolor=blue,citecolor=blue]{hyperref}

\newcommand{\be}{\begin{equation}}
\newcommand{\ee}{\end{equation}}
\newcommand{\bea}{\begin{eqnarray}}
\newcommand{\eea}{\end{eqnarray}}
\newcommand{\ba}{\begin{eqnarray}}
\newcommand{\ea}{\end{eqnarray}}

\newcommand{\Dslash}{D\hspace{-1.6ex}/\hspace{0.6ex} }

\newcommand{\pslash}{p\hspace{-1.ex}/\hspace{0.6ex} }

\begin{document}

\title{
Nonperturbative quark-antiquark interactions\\ in mesonic  form  factors  }
\author{ 
 Edward Shuryak and Ismail Zahed }
 
\affiliation{ Department of Physics and Astronomy, \\ Stony Brook University,\\
Stony Brook, NY 11794, USA}

\begin{abstract} 
The existing theory of hard exclusive  QCD processes is based on two assumptions: (i) $factorization$ into a  {\em hard block} times   light front distribution amplitudes (DA's); (ii) use of perturbative gluon exchanges within  the hard block. However, unlike DIS and jet physics, the characteristic  momentum transfer $Q$ involved in the factorized block 
is  not large enough for this theory  to be phenomenologically successful. 
In  this work, we revisit the latter assumption (ii), by  explicitly calculating the {\em instanton-induced} contributions to the hard block, and show that they contribute 
substantially to the vector, scalar and gravitational form factors of the pseudoscalar, scalar and vector mesons, over a wide range of momentum transfer.
 \end{abstract}

\maketitle

\section{Introduction}
\subsection{The main goals and plan of the paper}
The field of hadronic physics going back to the  pioneering theoretical and experimental works of  the 1960's,
continues to be a field of active development till  today.  Remarkably,  it remains still deeply divided along two conceptually
different approaches.
 
 One approach is focused on the nontrivial vacuum properties, with more specifically the central aspects of
 {\em chiral symmetry breaking and confinement}.
  The discovery of instantons and the development of numerical lattice gauge theory have put the  Euclidean formulation of QCD 
  at the center stage. The theory and phenomenology of multiple Euclidean correlation functions, 
 became the primary source of information about quark-quark interactions. 
 The inter-relation of perturbative and nonperturbative contributions in various channels, as a function of the distance between
the operators, were clarified already in 1990's (see e.g. a review \cite{Shuryak:1993kg}). Models, with ``constituent quark"
masses, confining and ``residual" 4-fermion forces, provided a good description of most aspects of hadronic spectroscopy.
More recently, the discussion has shifted to the properties of  operators made of  4-, 5- and 6-quarks and their mixture with gluons.
 
Another approach is focused  on  {\em partonic physics}, with more specifically inclusive and exclusive reactions.
The reader hardly needs to be reminded of the importance of deep inelastic scattering (DIS) and jet physics,
where perturbatively calculated  hard cross sections are assumed to factor out from the structure and fragmentation functions,
which are empirically  fitted to large sample of data. These functions, defined on a light front,  
are not readily amenable to an  Euclidean formulation.  The  light front distribution amplitudes and functions of the lightest hadrons have been discussed in the 
context of the QCD sum rules~\cite{Radyushkin:1994xv}, bottom-up holographic models~\cite{Brodsky:2011yv}, bound state resummations~\cite{Chang:2013pq}, basis light front quantization~\cite{Jia:2018ary,Shuryak:2019zhv},   and covariant  constituent quark models~\cite{Broniowski:2017wbr,Petrov:1997ve,Anikin:2000bn}. 
Recently, an Euclidean formulation was put forth to extract  the light front distributions from equal-time quasi-distributions~\cite{Ji:2013dva,arXiv:1506.00248}. Its implementation on the 
lattice~\cite{Zhang:2017bzy},  and in the random instanton vacuum model~\cite{Kock:2020frx} have been reasonably successful, providing a first principle approach.

The theory of {\em exclusive QCD reactions} (the subject  of this work)  follows a similar reasoning, see the early works
\cite{Brodsky:1973kr,Chernyak:1977as,Radyushkin:1977gp}.
It is also based on two assumptions:\\

\noindent (i) the {\em separation of scales},  based on the assumption that the momentum transfer 
$Q$ (the scale in the
``hard block"), is large compared to the typical quark mass and transverse momenta inside hadrons;\\
(ii) the
``hard block" can be calculated {\em perturbatively using  gluon exchanges}.\\

However, the theory based on these two assumptions is insofar not  successful, as there remains a wide gap 
between the semi-hard  domain of $Q$  in which experimental/lattice  results are available and the ``asymptotic" theory at $Q\rightarrow \infty$. 
(The latest lattice results with very fine lattices that we will discuss at the end of the paper are starting  to fill this gap).  The empirical values of 
 the mesonic 
form factors times $Q^2$  are well above the one-gluon exchange predictions, even with (what we consider maximally possible) flat distribution amplitudes and
higher twist contributions  included. 

This is not  surprising, as there is an important difference 
between the scales in DIS and jet physics on one hand, and exclusive processes on the other.
 The former operates  in the range $Q^2=10^2-10^4 \, {\rm GeV}^2$, while the
 exclusive processes operates in a different range with  $Q^2=2-10\, {\rm GeV}^2$ (sometimes referred to as a  $semi-hard$ regime).
 
 We accept the assumption (i) mentioned above: the $Q^2$ scale 
is  indeed large compared to the typical squared transverse momentum $\langle p_\perp^2 \rangle \sim 0.1\, {\rm GeV}^2$ within a hadron,
or the constituent quark mass $M_Q^2\sim 0.1-0.15\, {\rm GeV}^2$. In the Breit frame description of the form factors, conventional ``collinear"
kinematics  should still hold.  So we still have a notion of a ``hard block operator", sandwiched between two wave functions.

Yet we do not accept the second assumption (ii), showing that at such momentum transfer $Q$,  the nonperturbative 
quark interactions   are $not$ at all negligible in comparison  to gluon exchanges. Therefore a purely perturbative treatment of the 
``hard block"  needs to be supplemented by calculations of {\em leading nonperturbative contributions} from first principles, 
and  this paper makes the first steps in this direction.

\begin{figure}[h!]
\begin{center}
\includegraphics[width=12cm]{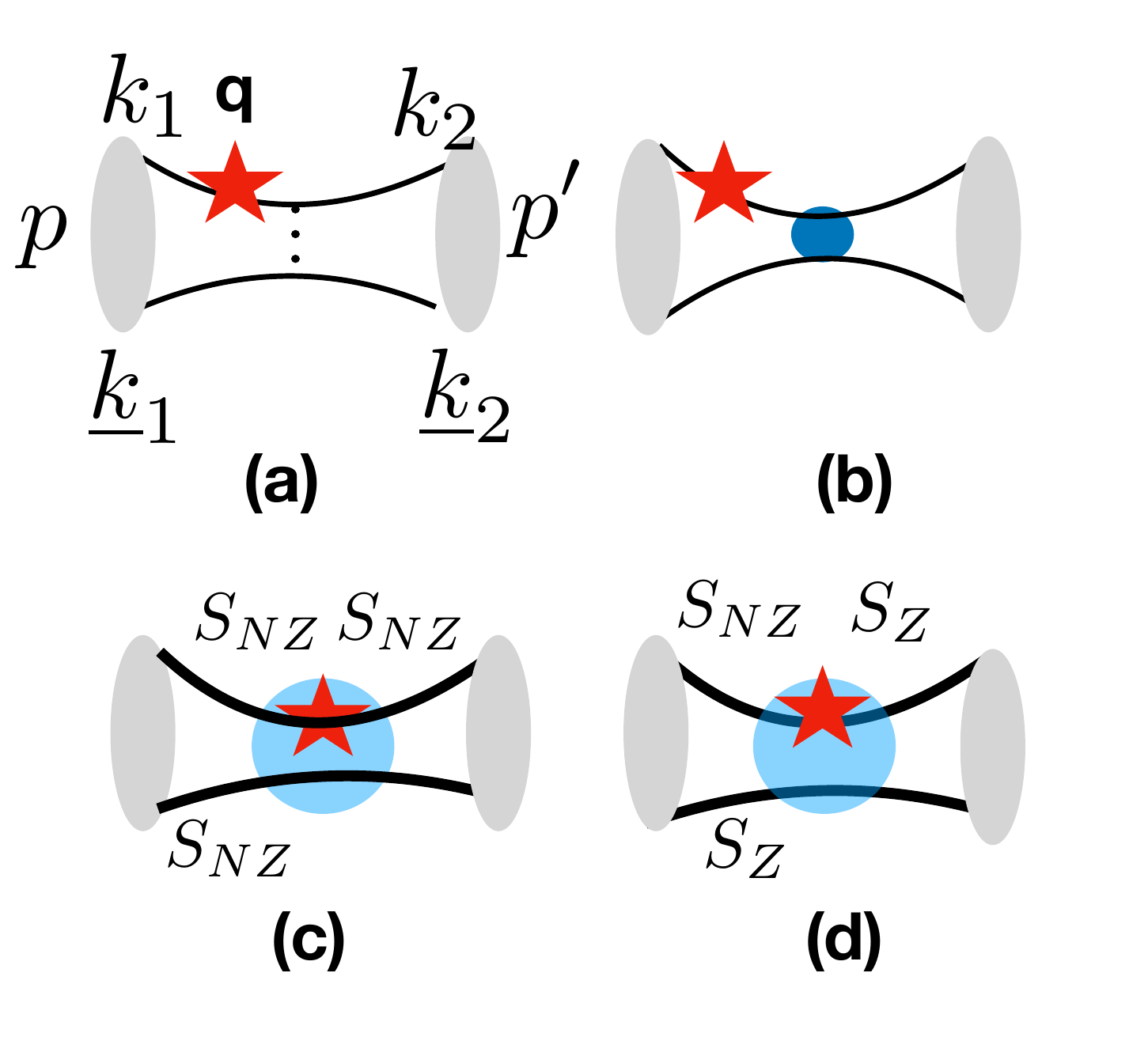}
\caption{The perturbative one-gluon exchange  diagrams: (a) explains our notations for momenta of quarks and mesons. The thin solid lines are free quark propagators, the red star indicates the virtual photon (or scalar) vertex, bringing in  large momentum $q^\mu$, and 
the	shaded ovals represent the (light-front) mesonic density matrices (distributions).
	The diagram (b) indicates a ``Born-style contribution",
	in which the gluon propagator is substituted by
	the Fourier transform of the instanton field.
	The  diagrams (c,d) contain three propagators
	in the instanton background (thick lines). In (c) all of them are $S_{NZ}$, made of nonzero Dirac modes, while in (d) two of them are $S_{Z}$ made of quark zero modes. This last contribution will be referred to  as  the ``'t Hooft-style term".) 
}
\label{fig_diags}
\end{center}
\end{figure}

Among various exclusive reactions, the  recent literature is focused mainly on decays of heavy quark
mesons  such as D- and B-mesons, much studied at electron colliders. However, in this work we restrict  our analysis to only elastic form factors 
of $light$ mesons. We will
 consider two types
of  ``hard blocks",  induced either by virtual photons or scalars. (Of course, scattering of Higgs bosons cannot
  be experimentally achieved, but it has been studied on the lattice, and it is rather interesting.)

Before we outline the content of this work, let us comment on other approaches  aimed at
the mesonic form factors. Instead of discussing on-shell light-cone
distribution amplitudes, one may start with two- or three-point
correlation functions, containing the electromagnetic current and two local currents with meson quantum numbers. 
One of them or both may carry large virtuality, in which case 
the three points probed are all close to
each other. After evaluating the correlation functions
in the deeply virtual regimes, one relates them to on-shell 
form factors via  dispersion
relations or QCD sum rules.

For the pion form factor
these approaches can be split into two categories, based on the different pion currents used. The first one (e.g.\cite{Braun:1999uj}) uses the $axial$ currents $(\bar d \gamma_\mu\gamma_5 u)$,  while the second one uses the pseudoscalar one  
$(\bar d\gamma_5 u)$. Note that in the former 
both quarks carry the same chirality (LL+RR), while in the latter they carry opposite chiralities (LR+RL). 

In the first category, the  correlators are analyzed using the QCD sum rule methods, previously developed for the two-point  axial
current correlators. They make use of  the operator product expansion (OPE) which assumes that the distances between the 
 currents are small in comparison to the typical vacuum fields,
represented by the gluon and quark condensates. The small-distance correlators are then connected to
 integrals over on-shell contributions using the pertinent  spectral densities, with the pion plus the $A_1$ meson plus a ``high energy continuum".     
In the second category, the distances between the currents are also assumed to be small, but the calculation is based on the so called single-instanton approximation (SIA), see \cite{Forkel:1994pf,Blotz:1996ad,Faccioli:2002jd,Braguta:2003hd}. 

Instead of comparing the specific results of these works, we 
make a more general comment on the same ideas previously applied to the $two-point$ functions. 
Since  the  axial spectral density is  known
experimentally from $\tau$-lepton decays,  these correlation functions  are   phenomenologically known (see e.g. analysis in \cite{Schafer:2000rv} and many others). The OPE expressions may only be used  at rather small distances $x <  0.4 \, {\rm fm}$, while the pion contribution becomes visible only at much larger distances 
 $x>1\, {\rm fm}$. In between,  the  contribution of the $A_1$ meson dominates. Therefore we are very sceptical of the approach in the first category.
 (Note also that  the  ``instanton liquid model" works at all distances, see Fig.2 in~\cite{Schafer:2000rv}.)  

The pseudoscalar two-point function is  also known, and was 
calculated on the lattice in multiple works (see e.g. \cite{Negele:1997na}). Unlike  the axial case, here the pion contribution
is  large and dominant already at small distances $x \sim 0.3 \, {\rm fm}$. It is also well reproduced by the single instanton contribution.  
Therefore, one should perhaps trust the accuracy of the  approach in the second category (with the pseudoscalar currents) more. 
(The relevance of these comments to our work  will be evident below,
 in the relative contributions of the mesonic distribution amplitudes (DA) with different chiral structure).

The outline of our paper is as follows: the next introductory section compares the magnitude of one-gluon exchange with 
a generic  4-fermion interaction of the Nambu-Jona-Lasinio type, to get an  initial qualitative idea on  the relative strength of the perturbative and nonperturbative effects. Clearly, the nonperturbative contributions will wane out 
at larger momentum transfer. This section also includes a  subsection \ref{sec_instantons} with  a brief introduction to the
salient  instanton effects and their key parameters.

Since  the paper contains a lot of technical details, 
not so important for a first reading, we decided to collect all the results for the pion, the rho vector meson, the  scalar and gravitational form factors 
in section \ref{sec_results}. The actual  calculations start from the perturbative ones in  section \ref{sec_pert}. They include the twist-2 and twist-3 contributions, most of which have been discussed in some form in the litterature for the vector form factors, but
not in a fully quantitative manner.

As discussed in subsection \ref{sec_including_NJL},
these results can be generalized to a large set 
of effective 4-quark scattering operators, as a substitute for  one-gluon exchange.  A simple 
 warm-up calculation of this kind consists in taking the Fourier transform of the instanton field instead
of a gluon propagator,  as discussed in section \ref{sec_inst_Born}.  We  do not consider
such an approach internally consistent, and  for this reason we will not include
it in the "results" section.

The core  calculations of the instanton-induced effects are collected in section~\ref{sec_inst}.
We start by explaining an  LSZ algorithm (short of a Hamiltonian formulation), whereby 
 full multiple quark propagators in the instanton field are amputated from their  trivial free propagation,
 and leading to  {\it hard block} operators. We discuss separately the contributions to  the propagators
due the Dirac zero modes and the Dirac non-zero modes.

Section \ref{sec_wfs} contains a discussion of the mesonic light cone distribution amplitudes (DA's),
which are the wave functions integrated over transverse momenta. Following a  brief review of the  literature, 
we introduce the  pion, rho and scalar meson amplitudes with  different chiral structures, which we use consistently in the results of the calculations. 
The paper ends with a discussion section \ref{sec_summary}, in which the phenomenological and current lattice results  about the mesonic form factors are compared. A number of Appendices 
are added to include more technical details of the calculations.

\subsection{Comparing the one-gluon exchange with the 4-fermion interaction of the Nambu-Jona-Lasinio model} 
\label{sec_NJL}

Historically, the 1961 paper by Nambu and Jona-Lasinio  \cite{Nambu:1961tp}  was the first
breakthrough,  that established  the notion of chiral symmetry of the strong interactions, as well as  its spontaneous breaking.
Furthermore, it also suggested a particular mechanism for it to occur, by postulating the existence of a certain 4-fermion interaction with a given coupling $G_{NJL}$, 
strong enough to make a superconductor-like gap in the fermionic vacuum. The second important
parameter of the model  is the UV cutoff $\Lambda_{NJL} \sim 1\, {\rm GeV}$, below which their hypothetical
attractive  4-fermion interaction operates. Their magnitude  were determined from the
empirical quark condensate and pion properties~\cite{Bernard:1987im}, for a review see \cite{Klevansky:1992qe}.

 With time there were many application of the NJL model with different operators and parameters. For definiteness
we use the parameter set from  Ref.\cite{Hutauruk:2018zfk} (and other papers of the same authors) as
an example.  Those were consistently used  for the description of aspects of chiral symmetry breaking, such as the quark constituent masses, the pion and kaon masses, and those of other bound states like
nucleons (made of a constituent  quark and a diquark). The central part to all NJL applications is  the so called ``gap equation" for the effective quark mass

\begin{equation}\label{gap}M_Q=m+{3 G_{NJL} M_Q \over \pi^2} \int_{1/\Lambda_{UV}^2}^{1/\Lambda_{IR}^2} {d\tau \over \tau^2} \,e^{-\tau M^2}\end{equation}
where $m$ is the current quark mass, and $M_Q$ the constituent quark mass following from (\ref{gap}).  Note that when $m=0$, $M\neq 0$  cancels out in the l.h.s. and  the r.h.s,
 and remains only in
the (regulated) loop integral. For the input parameters used in these works
\begin{equation}G_{NJL}=19 \, \,{\rm GeV}^{-2},\,\,\,
\ \Lambda_{IR}=0.24\, {\rm GeV},\,\,\,  \Lambda_{UV}=0.645\,\, {\rm GeV }\end{equation}
the constituent mass is found to be $M_Q\approx 0.4\, {\rm GeV}$, close to half of the mass of  the ``usual"
$\rho$ meson mass or $1/3$ of the $\Delta$ baryon mass.

\begin{figure}[htbp]
\begin{center}
\includegraphics[width=7cm]{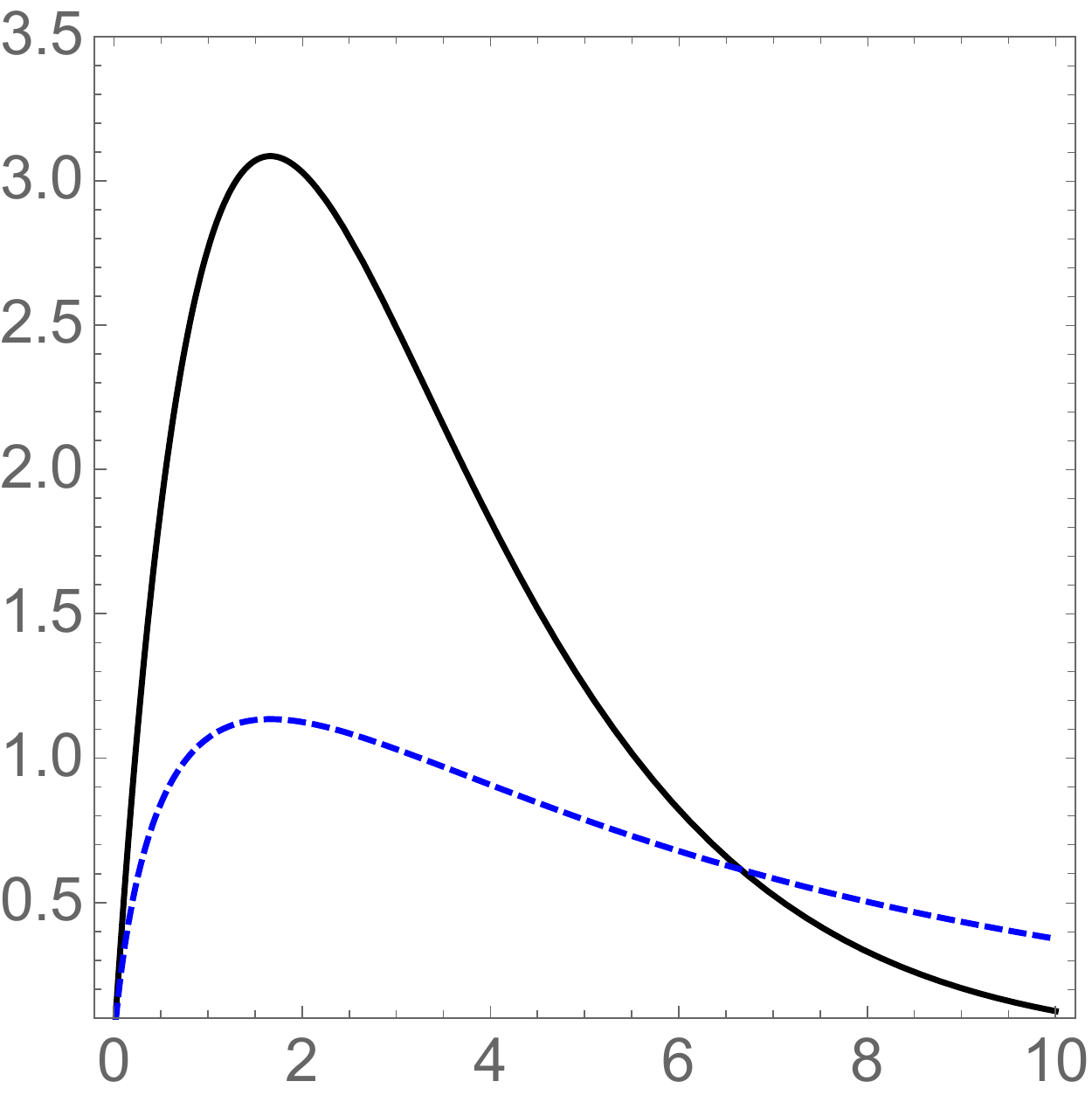}
\caption{The ratio of the nonperturbative-to-perturbative 4-fermion effective vertex (\ref{nonpert_to_pert}),  with a Gaussian form factor (solid) and exponential form factor  (dashed),
versus the momentum transfer squared $Q^2 \, ({\rm GeV}^2)$. 
}
\label{fig_nonpert_to_pert}
\end{center}
\end{figure}

For an estimate, it is useful  to use the  magnitude  of the NJL nonperturbative force, and compare it to 
the force from one-gluon  exchange or  $F_{\rm gluon}(k^2)=g^2/ k^2$.  For a typical exchange within a meson with 

\begin{equation}
k^2=x \overline x  Q^2\approx Q^2/4 
\end{equation} 
the ratio of the NJL to gluon exchange forces is

\begin{equation} 
{ G_{NJL} \over F_{\rm gluon} }\,{\rm exp}\bigg(-{k^2 \over \Lambda_{UV}^2}\bigg)
\label{nonpert_to_pert}
\end{equation} 
where we assumed a Gaussian or exponential form factor with $\Lambda_{UV}$. 
Fig.~\ref{fig_nonpert_to_pert} shows the dependence of (\ref{nonpert_to_pert}) on $Q^2$. 
While this ratio drops towards large momenta due to the 
 form factors, the ratio remains above one in a wide  range of momentum transfers.
 
 Thus, on a qualitative level one may think that the puzzling large value of the form factors
 at intermediate $Q$ can perhaps be understood, by adding to the perturbative
 diagram Fig.\ref{fig_diags}(a) the nonperturbative diagram (b), with the NJL effective quark scattering of appropriate 
  magnitude. 
  
  However, it is impossible to do it consistently. The electromagnetic, 
scalar or gravitational vertex (indicated by stars in this figure)  may occur well $inside$ the 
cutoff region (indicated by a blue circle, diagram (c)) with strong nonperturbative fields, and  the hypothetical nature of the local NJL interaction provides no obvious clues on
how to handle propagation in it.   For the minimal vector insertion, one may argue for gauge invariance to compensate
for the lack-thereof in the presence of a non-local 4- or 6-quark interaction~\cite{Plant:1997jr}, but this contraint does not ensure quantum 
U$_A$(1) explicit breaking (see below), and does not extend consistently to the scalar, gravitational  and more general  vertices. So, 
one needs a more microscopic approach, providing a consistent  description of quark propagation in   the
nonperturbative backgrounds.

\subsection{Brief introduction to instanton effects}
\label{sec_instantons}

So far we focused on the (historically first)
nonperturbative approach to
 physics of chiral symmetry breaking, namely the NJL model.
With the advent of QCD in 1970$^\prime$s, this hypothetical interaction
between quarks  obtained more fundamental explanation
which
 came from the understanding of gauge topology, Chern-Simons number
and topological tunneling events,  semiclassically described by  $instantons$ \cite{Belavin:1975fg}.  As discovered by t'Hooft \cite{tHooft:1976snw}, 
instantons indeed  generate 4- and  6-fermion effective interactions of quarks. Those qualitatively differ from the NJL operator in the fact that they
explicitly violate U$_A(1)$ chiral symmetry, see below. By the end of 1970's most ingredients of the instanton theory --
fermionic zero modes and  the propagators in the  instanton field we will be using -- were constructed~\cite{Gross:1980br}.

In 1980's the main question then was whether those instanton-induced inter-quark
forces are strong enough to generate chiral symmetry breaking. $Assuming$ it is so,
one of us \cite{Shuryak:1981ff} developed the so called  {\em instanton liquid model}  (ILM), using as inputs the values of the quark and gluon
condensates. It assumes that
 the instanton ensemble  have the following parameters

\begin{equation}
n_{I+\bar I}\approx 1\, {\rm fm}^{-4}, \,\, \,\,\, \rho\sim 1/3 \, {\rm fm} \sim 1/(0.6\, {\rm GeV})  \label{eqn_ILM} 
\end{equation}
for the instanton plus anti-instanton density and size, respectively.
Their combination known as the
{\em diluteness parameter} of the  instanton ensemble is defined by 
\begin{equation}\kappa\equiv \pi^2\rho^4 n_{I+\bar I} \label{eqn_diluteness} \end{equation} 
These two parameters of the  ILM correlates well with the parameters of the NJL model, in particular the size  $\rho$ corresponds to the 
inverse UV cutoff.   Years later,  these  parameters were confirmed, both by lattice studies and numerical simulation
of the Interacting Instanton Liquid Model of the ensemble, for a review see \cite{Schafer:1996wv}.

The prevailing picture of the nonperturbative fields populating the QCD vacuum in 1970's -- reflected in the wide use of the OPE 
in the QCD sum rule framework --
 was a near-homogeneous vacuum
fields, with characteristic momenta $$p\sim \Lambda_{QCD}\sim 1/fm $$ 
The ILM had drastically changed the picture, emphasizing
instead the role of the small-size instantons with relatively strong fields.
For  a qualitative estimate, let us mention the color summed field strength of 
these  fields inside the instantons 

\begin{equation}\big( G^a_{\mu\nu}(x) \big)^2 = {192 \rho^4 \over (x^2+\rho^2)^4} 
\end{equation} 
Its magnitude at the center of  a typical instanton 
with $\rho=1/3\, fm$ is  large
$$\sqrt{ \big( G^a_{\mu\nu}(0) \big)^2 }=\sqrt{192}/\rho^2\approx 5 \, {\rm GeV}^2$$ 
As we will show below, the averaging over the instanton size distribution of the induced
interactions in the hard block, causes a shift towards smaller instanton sizes  and stronger fields, 
with typically  $\rho=1/6\, fm$ and  $ G\sim 20 \, {\rm GeV}^2$.
 Such fields are by no means small 
  compared to the
the scale of the momentum transfer between quarks
in the semi-hard domain under consideration. 
(Note also, that it is larger than the charm quark mass  $m_c^2\approx 2\, {\rm GeV}^2$.
At the end of the paper we will speculate that our light-meson results can be extrapolated in quark mass, 
to the strange  and perhaps even charm sectors.)

Instanton fields were incorporated directly into many physical effects. The simplest are the 
heavy quark potentials \cite{Callan:1978ye} and high energy scattering \cite{Shuryak:2000df}, in which
quark trajectories can be described by straight lines. 
Many more applications follow from  t'Hooft  effective Lagrangian \cite{tHooft:1976snw}, following from
zero modes of the Dirac equation in the instanton background field, as briefly recalled in Appendix~\ref{app_zero}.
It is important to note  that the existence of zero modes is a consequence of topological 
theorems, and cannot be changed by any smooth deformation of the instanton field. 

The
multi-quark effective Lagrangian
for two quark flavors ($N_f=2$) consists of  certain 4-quark operators. Like the NJL interaction,
they preserve SU(N$_f)$ chiral symmetry, but $unlike$ the NJL interaction, they explicitly violate the U$_A(1)$
chiral symmetry. While in ``mesonic" notations, with $\sigma\equiv(\bar q q)$, $\vec \pi \equiv (\bar q i\gamma_5 \vec \tau q)$,
$\eta\equiv (\bar qi\gamma_5  q)$,  $\vec \delta\equiv (\bar q  \vec \tau q)$, the NJL Lagrangian has the structure

\begin{equation}
L_{\rm NJL}\sim  (\vec{\pi}^2+\sigma^2) 
\end{equation}
the instanton-induced one has the structure ($N_f=2$)

\begin{equation} \label{eqn_hooft_structure}
L_{\rm tHooft}\sim (\vec{\pi}^2+\sigma^2-\vec{ \delta}^2-\eta^2) 
\end{equation}
 It is the minus sign of
 the last two terms which indicates the explicit  breaking of  $U_A(1)$. Therefore in the $\eta$ channel
(called $\eta'$ for three flavors and in PDG meson tables) 
the interaction is not attractive but repulsive, making it heavy. 
In passing, we also note that  the 
light-front wave function of the $\eta'$ was recently calculated in~\cite{Shuryak:2019zhv}, see Fig.\ref{fig_pionwf}, and it is drastically different from that of the pions.)

With the original ILM parameters, the diluteness parameter
is  $\kappa\sim 1/10$, and multiple lattice studies using ``deep cooling" towards the action minima
have reproduced this value.  This conclusion however
was put in doubt by some more recent studies,
which studied the dependence on the cooling time
by  extrapolating  to  its zero value
time (that is,  to the quantum vacuum itself).
This dependence is related to  instanton-antiinstanton annihilation processes
during cooling. As a result, they suggested  a larger value
for  $\kappa$.

In particular,  lattice-based study  \cite{Athenodorou:2018jwu}  focused on the instanton contribution to  3- and 4-point 
Green functions in the full quantum vacuum and with cooling.
Their original motivation  was to extract the gluon coupling $\alpha_s(k)$,
so the observable on which this work was focused is the
the  ratio of the 3-point to 2-point Green function (in configurations transformed to Landau gauge)

\begin{equation}
\alpha_{MOM}(k)={k^6 \over 4\pi} {\langle  G^{(3)}(k^2)  \rangle^2 \over    \langle  G^{(2)}(k^2)    \rangle^3 }  
\label{ratioG3toG2}
\end{equation}
In the ``uncooled" quantum vacuum (with gluons)
 the effective coupling  starts running downward at large  $k>1\, {\rm GeV}$, as required
by asymptotic freedom. However  at
low $k\rightarrow 0$, one finds a persisting positive power of $k$, with a slope that  matches 
exactly the one following from an instanton ensemble~\cite{Boucaud:2002fx}

\begin{equation}
\alpha_{MOM}(k)\rightarrow \frac{k^4}{18\pi n}
\end{equation}
Furthermore, after cooling
for different cooling  time $\tau$,  it was observed that
the same power spreads to all momenta, even for $k> 1\, {\rm GeV}$.  This corresponds to the expectation that cooling eliminates perturbative 
gluons (the plain waves) but preserves 
(certain time-dependent fraction) of instantons.

Here, we will not cover 
 the details of this analysis, but rather mention their main conclusion: the 
total instanton density (extrapolated to $zero$ cooling time)  is $n\sim 10 \, {\rm fm}^{-4}$, an order of magnitude larger than
in the original ILM. In order words, 
this analysis suggests that the vacuum
instanton diluteness parameter (\ref{eqn_diluteness}) is actually not small, but rather large $\kappa\sim 1$. This conclusion  does not in fact
contradict our understanding of the underlying of  chiral symmetry breaking 
and the  parameters of the  ILM, since this large density  includes close $I\bar{I}$ pairs, with 
 zero topological charge. These molecules have a small impact on chiral symmetry breaking and related observables,
  and therefore  were not included in the  ILM. However,  their  internal gauge fields 
are very  strong, and should affect nonperturbative quark scattering of the type we  discuss in this paper.

\section{Results} \label{sec_results}

In this paper we consider a larger set of form factors
than it is usually done in the available literature. In particular, we discuss the  {\em pseudoscalar, vector} and {\em scalar} mesons, 
and calculate  the {\em vector, scalar, graviton} and {\em dilaton} form factors.

We also
 include in the distribution functions several possible Dirac/chiral structures allowed by parity.
 We calculate the contributions to the {\em hard block}
 corresponding to all four diagrams  of Fig.\ref{fig_diags}. Specifically, those are: (a) the perturbative one-gluon exchange; (b) 
the  Born-  style contribution of the instanton gauge field;  (c) the contribution of the nonzero mode quark propagators in the instanton background; (d) the contribution  of the instanton zero modes to the propagators,  or t'Hooft effective 4-fermion quark interaction. 
 
  There are many technical details about these contributions, the relative values 
 of the  various parameters, etc., all of which 
are relegated to  subsequent sections.  
  Therefore, we decided to present the final results first, with
 the step-by-step derivation to be given later in the paper. Here we do not discuss the 
 subleading contributions, the uncertainties
 of all parameters involved, etc. For that one has to read the paper in full.
 Also, the  discussion of the various (light-front) wave functions (distributions), will be discussed
 in section \ref{sec_wfs}.  
  In order to avoid
 too many plots, we selected a single ``reasonable" example  for the light-front distributions of the pion and  rho mesons. For the former is is just a  ``flat" distribution, $\phi_\pi(x)=1$, and for the latter
 we use  a simple parametrization 
 \begin{equation}\varphi_\rho(\xi)\sim {\rm exp}\bigg(-{0.7 \over 1-\xi^2}\bigg) \end{equation}
with $\xi=x-\bar x$,  the  difference between the quark and antiquark momentum fractions.

  We  discuss four types of elastic mesonic form factors: (a) the $vector$ ones, associated with hard scattering
  of a photon; (b) the $scalar$ ones, associated with scattering via a Higgs boson exchange.
    (Of course, those are not
in practice doable, but  scalar form factors were calculated numerically via lattice gauge theory simulations);
(c-d) the $gravitational$ ones, associated with scattering via a {\em graviton} or {\em  dilaton} exchange.
 In this section we report results for  three contributions to each of them, of diagrams (a,c,d). 
 Plots are provided only for the pseudoscalar and vector mesons, and only for the vector and scalar formf actors.
 For other cases we present the expressions for the scattering amplitudes.

 \subsection{ Vector form factors of the pseudoscalar mesons}
  \label{sec_results_pion}

 We will keep the notations of the contributions as explained in Fig.\ref{fig_diags}. For example,
the (photon-induced)  {\em vector scattering amplitude} on the pion, with
 perturbative one-gluon exchange  will be referred to as $V^\pi_a$, which is

 \begin{eqnarray} \label{eqn_Vapi}
&&V^\pi_a(Q^2)=\epsilon_\mu(q)(p^\mu+p^{\prime\mu})\,(e_u+e_{\overline d})\,
\bigg[\bigg(\frac{2C_F\pi\alpha_sf_{\pi}^2}{N_cQ^2}\bigg) \nonumber\\
&&\times   \int  dx_1 dx_2 
\bigg({ 1\over \bar x_1\bar x_2+m_{\rm gluon}^2/Q^2}\bigg)\bigg( \varphi_{\pi}(x_1)\varphi_{\pi}(x_2)  \nonumber\\
&&+2{\chi_{\pi}^2 \over Q^2}    \bigg(
\varphi_{\pi}^P(x_1)\varphi^P_{\pi}(x_2)\bigg( {1 \over \bar x_2 +E_\perp^2/Q^2}-1\bigg)
+\frac 16\varphi_{\pi}^P(x_1)\varphi^{\prime \, T}_\pi(x_2)\bigg( {1 \over \bar x_2 +E_\perp^2/Q^2}+1\bigg)\bigg)\bigg)
\bigg]\nonumber\\
\end{eqnarray}
Here we show explicitly the electromagnetic charges $e_u=2/3,e_{\overline d}=1/3$, although of course 
the total charge of a positive pion is $e_u+e_{\overline d}=1$. The color matrices give the factor
$C_F=(N_c^2-1)/2N_c=4/3$ with $N_c=3$ number of colors.  The large spacelike photon momentum is $q^\mu$ and  $q_\mu q^\mu=-Q^2<0$. The photon polarization
vector is $\epsilon_\mu(q)$, with $\epsilon_\mu q^\mu=0$.  The momenta of the initial and final mesons are called $p$ and $p'$. The pion decay constant is $f_{\pi}\approx 133\, {\rm MeV}$,  it 
characterizes 
the wave 
function at the origin in the transverse plane, $r_\perp=0$. 
For the pion distribution we use the expression (\ref{WF1}) 
	which includes not only the chirally diagonal part of the
 distribution  $\varphi_\pi(x)$ 
 but also the chirally non-diagonal ones, such as $\varphi_{\pi}^P(x)$.  
The DA's  depend only on one longitudinal momentum fraction $x$,  which 
  correspond to the 2-body sector of the full wave function.  The  regulators of the divergent 
  integrals by extra terms in the denominators are  discussed  in  section  
  \ref{sec_regulated}, where the relative magnitude
 of both chiral contributions are compared.  
 Here the bar indicates that the momentum fractions are those of  antiquarks, $\bar x_i\equiv 1-x_i$. 
In terms of the  asymmetry parameters, these variables read as $x_i=(1+\xi_i)/2,\bar x_i=(1-\xi_i)/2$.
 The regulators are the gluon mass and quark ``transverse energy".

\begin{figure}[h]
\begin{center}
\includegraphics[width=8cm]{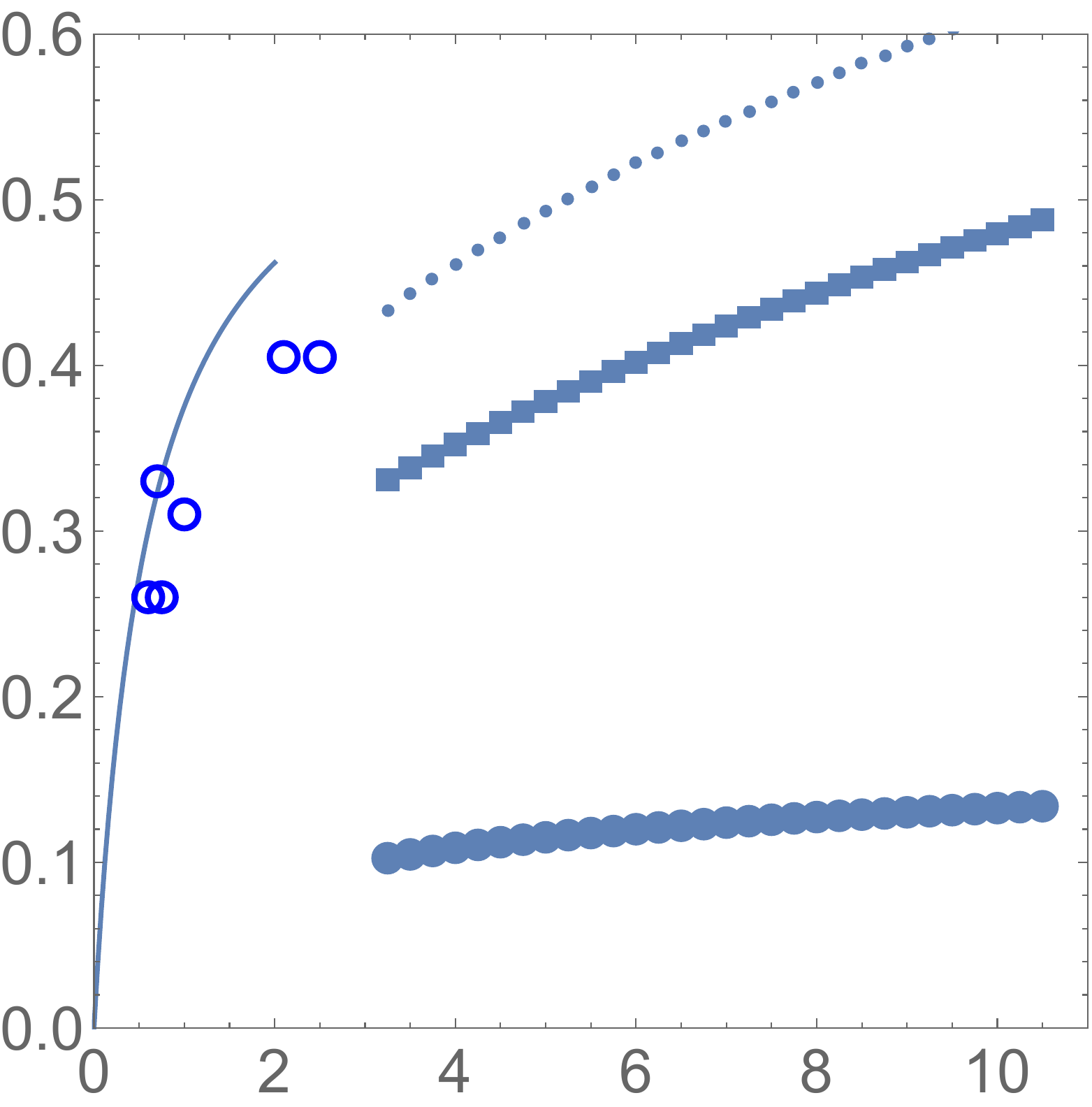}
\caption{The vector 
	 form factors of the pion times the squared momentum transfer, $Q^2 F_\pi(Q^2)\,\, ({\rm GeV}^2)$ versus $Q^2 ({\rm GeV}^2)$.
The closed  discs show the total 
perturbative contribution.  The squares correspond to the
instanton contribution from  the non-zero mode propagators $S_{NZ}$.  The dotted line above is their sum.
The  solid line  is the usual dipole fit with the rho meson mass, and the open points are from the experimental measurements.
We do not show the multiple data points at smaller $Q^2$, where
there is good agreement with  the dipole formula. 
}
\label{fig_pionV_all}
\end{center}
\end{figure}

 Note that  the last two terms $\sim \chi_\pi^2/Q^2$ are kept  because $\chi_\pi$
is large, unlike the masses and transverse momenta squared of quarks which are ignored.
We note further that the sum of them is shown  in the 
summary plot in Fig.~\ref{fig_pionV_all}. 
 
The contribution we call 
the {\em Born-like instanton contribution} $V_b^\pi$ has the same Dirac traces and,
as explained in section \ref{sec_inst_Born},  and 
is obtained by  substituting   in  $V_a^\pi$
 the Fourier transforms of the instanton gauge field (\ref{eqn_G}) instead of  the  gluon propagator, with
 
 \begin{eqnarray} \label{eqn_substitution}
  \pi \alpha_s(Q/2) \rightarrow  \kappa \left<\mathbb G^2(Q\rho\sqrt{\overline{x}_1\overline{x}_2})
 \right> \end{eqnarray}
and is therefore

  \begin{eqnarray} \label{eqn_Vbpi}
&&V^\pi_b(Q^2)=\epsilon_\mu(q)(p^\mu+p^{\prime\mu})\,(e_u+e_{\overline d})\,
\bigg[\bigg(\frac{2C_F\kappa f_\pi^2}{N_cQ^2}\bigg)   \nonumber\\
&&\times    \int  dx_1 dx_2 \,\left<\mathbb G^2(Q\rho\sqrt{\overline{x}_1\overline{x}_2})\right>
\bigg({ 1\over \bar x_1\bar x_2+m_{\rm gluon}^2/Q^2}\bigg)\bigg( \varphi_\pi(x_1)\varphi_\pi(x_2)  \nonumber\\
&&+2{\chi_\pi^2 \over Q^2}    \bigg(
\varphi_{\pi}^P(x_1)\varphi^P_\pi(x_2)\bigg( {1 \over \bar x_2 +E_\perp^2/Q^2}-1\bigg)
+\frac 16\varphi_{\pi}^P(x_1)\varphi^\prime_T(x_2)\bigg( {1 \over \bar x_2 +E_\perp^2/Q^2}+1\bigg)\bigg)\bigg)
\bigg]\nonumber\\
\end{eqnarray}
The  instanton induced form factor  $\mathbb G$ is given in (\ref{eqn_G}). The angular brackets indicate averaging over the instanton size.

The contribution $V_c^\pi$, from three $non$-zero mode propagators, are
discussed in section \ref{sec_LSZ}, it leads to  the following result

\begin{eqnarray} \label{eqn_Vcpi}
V_c^\pi= && \epsilon_\mu(q)(p^\mu+p^{\prime\mu})\,(e_u+e_{\overline d})\, \bigg[
{\kappa \pi^2f_{\pi}^2 \chi_\pi^2 \over N_cM_Q^2}
\langle\rho^2  {\mathbb G}_V(Q \rho) \rangle  \nonumber  \\
&&\times\int dx_1 dx_2 \overline x_1  
 \bigg(\varphi_{\pi}^P(x_1)\varphi_{\pi}^P(x_2)
-\frac 1{36}\varphi_{\pi}^{'T}(x_1)\varphi_{\pi}^{'T}(x_2)\bigg)\bigg]
\label{eqn_Vcpi}
\end{eqnarray}
The function $\mathbb G_V$ is given  in (\ref{eqn_GV}).
Again, the  angular brackets indicate that
it is averaged over the instanton size distribution,
as explained in section \ref{sec_averaging}.
Note that the partonic integrand involves a single 
momentum fraction $\bar x_1$
(or $\bar x_2$ in the symmetric term not shown).
The other integral is simply over the function itself.
For all of them except $\varphi_{\pi}^{'T}$ it is
the normalization integral equal to 1. Yet for this one
it is an integral of the derivative, and therefore
it vanishes due to the quark-antiquark symmetry $x\leftrightarrow \bar x$
\begin{equation} 
\label{eqn_intT'}
\int_0^1 \varphi_{\pi}^{'T}(x) dx=\varphi_{\pi}^{T}(1)-\varphi_{\pi}^{T}(0)\rightarrow 0
\end{equation}
so the last term in  (\ref{eqn_Vcpi}) does not actually contribute to the form factor.

The contribution of the mixed zero mode and non-zero mode ($^\prime$t Hooft vertex)  derived in
  section~\ref{sec_hooft},  to the pion vector form factor is
 
\begin{eqnarray} \label{eqn_Vdpi}
&&V_d^\pi= -\epsilon_\mu(q)(p^\mu+p^{\prime\mu})\,(e_u+e_{\overline d})\, \nonumber\\
&&\times \bigg[\bigg(\frac {1}{N_c^2(N_c+1)}\bigg) \frac {4\kappa \pi^2f_\pi^2\chi_\pi^2}{3M_Q^2}\left<\rho^2\frac{K_1(Q\rho)}{Q\rho}\right>
\int dx_1 dx_2 \varphi_{\pi}^P(x_1)\varphi_{\pi}^{'T}(x_2) \bigg]
\end{eqnarray}  
As  we noted in (\ref{eqn_intT'}), this contribution vanishes after the $x$ integration is carried.

The summary plot of the pion vector form factor is shown
in Fig.~\ref{fig_pionV_all}, taking all three DA's as flat distributions $\varphi_\pi(x)= \varphi^P_\pi(x)=1$ and $  \varphi^{'T}_\pi=0$. 
This selection is motivated by our view that the  flat distributions
represent  an upper bound on the DA, with the asymptotic form providing  a lower bound.

The
perturbative contributions $V_a^\pi$ 
(closed circles)
 is the {\em sum of all} chiral structures of the pion density matrices.
The corresponding integrals for each
of them separately  are  shown in Fig.\ref{fig:i1i2}, from which it is seen that
the  chiral-nondiagonal term is about twice
larger than the  chiral-diagonal one in the range of momenta considered. This feature was anticipated already
in \cite{Geshkenbein:1982zs}.

The instanton Born-style contributions to $V_b^\pi$ 
is relatively close to $V_a^\pi$ if the instanton diluteness parameter is $\kappa=1$.
 (For a discussion of its value see the end of section \ref{sec_instantons}.) To avoid any misunderstanding, we 
 note that the $V^\pi_b$ conribution does $not$ really constitute a consistent account for the instanton effects, as are $V^\pi_c,\,\,V^\pi_d$, and
 therefore is  $not$ shown 
 in the summary plot.

The instanton-induced contribution $V_c^\pi$ (squares) 
at $\kappa=1$ is comparable to the perturbative $V_a^\pi$
in magnitude, but has a different 
dependence on $Q^2$. The instanton form  factor is of course a
decreasing function of momentum transfer, but on the plot it is multiplied by an extra $Q^2$ and is therefore
slowly increasing.

 Taken together (dotted line at the top)
they account for the pion form factor for the corresponding
values of  $Q^2$, reasonably well joining the experimental data at the lower end. We stress that 
no parameters were fitted for this to happen.
Our main focus is still a  comparison between all plots,
with the same set of parameters.

 \subsection{Scalar form factors of the pion} \label{sec_S_pi}

 One may  think of a point-like scalar quantum, hitting
 one of the quarks with momentum transfer $q^\mu$ to be the Higgs boson. If so, the corresponding couplings are Yukawa couplings $\lambda_q, q=u,d,...$
 of the standard model.
 However,  these couplings are unimportant for  the form factors. (E.g. lattice groups use for convenience $\lambda_u=1,\lambda_d=0$).
 The corresponding amplitude of 
  the elastic scattering on a pion, with a
 perturbative one-gluon exchange between quarks, leads to the following scattering amplitude

\begin{eqnarray}\label{eqn_Sapi}
S_a^\pi=-(\lambda_u+\lambda_d)M_Q
&&\bigg[\bigg(\frac{2\pi C_F\alpha_sf_\pi^2 \chi_\pi}{N_c Q^2 M_Q}\bigg)\,
\int_0^1 
{ dx_1dx_2\over \overline{x}_1\overline{x}_2+m_{gluon}^2/Q^2}
\nonumber\\
&&\times\bigg(
	\varphi_{\pi}(x_1)\varphi_{\pi}^P(x_2)  \bigg( 1+ {2 \over \overline{x}_2+E_\perp^2/Q^2}\bigg)
+{1 \over 6}\varphi_{\pi}(x_1)\varphi_{\pi}^{T\prime}(x_2)\bigg)
\bigg]
\end{eqnarray}
Note first that we included outside the square bracket the
quark constituent mass $M_Q$,  which is balanced by the same constituent 
mass in the  denominator. We did it
to facilitate the comparison with the instanton-induced expressions to follow.

In the previous section, on the vector form factor, it was
obvious that the factors with charges and momenta in the amplitude
do not belong to form factors, as they are also present
in the forward scattering at $Q=0$. The situation with the scalar
formf actors is a bit different. The forward scattering amplitude on a hadron $h$ is proportional to 

\begin{equation}
\sum_q\lambda_q \langle h | \bar q q | h \rangle=-\sum_q \lambda_q {\partial M^2_h \over \partial m_q}
\end{equation}
thanks to the Feynman-Hellman theorem.
The derivative appearing here is known for  the  pion from the Gell-Mann-Oaks-Renner relation, and for most hadrons from
lattice chiral extrapolations. 

In the scalar plots  to follow, we will the show square brackets in the amplitudes times $Q^2$ without  the factors in front,
as we did for the vector cases. Indeed, our main focus is on  the
relative magnitude of different contributions. However
the reader should be cautioned that the true scalar formf actor $F_S(Q^2)$ requires multiplication by an
 additional
factor 

\begin{equation} \label{eqn_factor_in_S}
K_S= {\sum_q \lambda_q \partial M^2_h / \partial m_q  \over  M_Q \sum_q\lambda_q}
\end{equation} 
to enforce the standard form factor normalization $F_S(Q=0)=1$.

An additional  contribution proportional to the quark mass $M_Q$, instead of $\chi_\pi$,  is explicitly given in (\ref{eqn_MSapi}), 
but, being subleading, it is not mentioned here. 

Note also  that 
the scalar amplitudes have negative overall sign, which really
does not matter as the  couplings $\lambda_q$ are arbitrary.
This sign of course does not affect the contribution to the form factor as captured by the square bracket.  
 The {\em Born-like instanton contribution} to the scalar pion scattering, $S_b^\pi$, is obtained by the same
 substitution (\ref{eqn_substitution}) to $S_a^\pi$ and is therefore not shown here.

 \begin{figure}[h]
\begin{center}
\includegraphics[width=8cm]{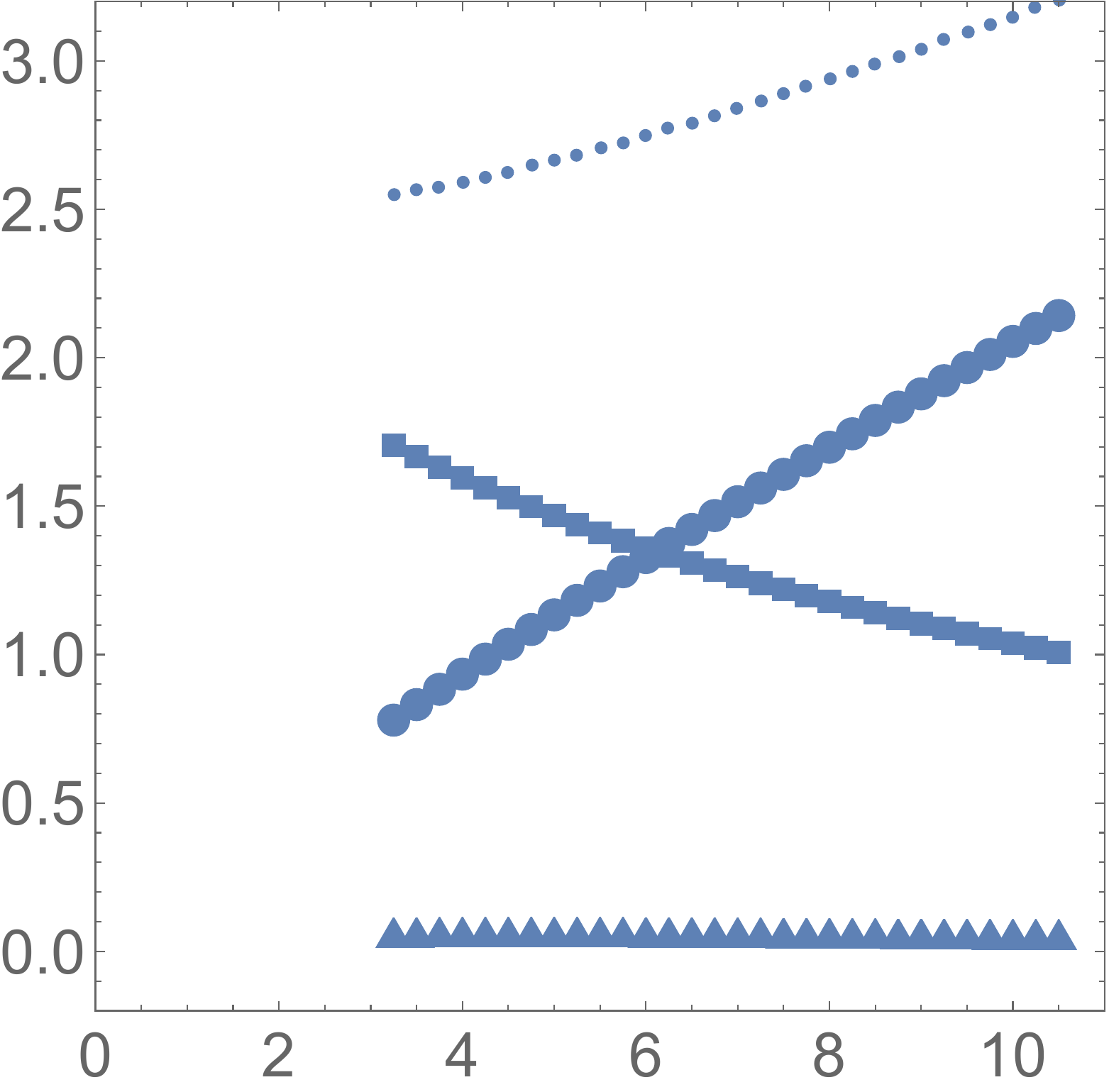}
\caption{ We show the square bracket in the 
scalar scattering amplitudes of the pion as in (\ref{eqn_Sapi}) etc, times the momentum transfer squared  $({\rm GeV}^2)$, versus $Q^2, ({\rm GeV}^2)$. 
As in the previous plot,  the
black closed disks correspond to the one-gluon exchange contribution,
the black  squares  to the instanton
contribution from three non-zero mode propagators $S_{NZ}$,  the black triangles correspond to the zero-mode terms $S_Z$ in two propagators.
The dotted line above is their sum. (We recall  that the full normalized scalar form factor is
obtained by multiplying these results by the extra factor
(\ref{eqn_factor_in_S}).)
}
\label{fig_Spi_all}
\end{center}
\end{figure}

 The contribution of the instanton-induced diagram (c)
(with three nonzero mode propagators) is

 \begin{eqnarray} \label{eqn_Scpi}
S_c^\pi(Q^2)=&&-(\lambda_u+\lambda_{\overline d})\,M_Q\,\bigg[\bigg(\frac{\kappa\pi^2 \chi_\pi f_\pi^2}{N_cM_Q^3}\bigg)\,
\left<(Q\rho)^2\mathbb G_S(Q\rho)\right>\,\nonumber\\
&&\times\int_0^1dx_1dx_2\,  \bar x_2 \varphi_{\pi}(x_1)
\bigg(\varphi_{\pi}^P(x_2)
-\frac 16 \varphi_{\pi}^{T\prime}(x_2)\bigg)
\bigg]
\end{eqnarray}
Unlike  the vector form factor $V_c^\pi$ (\ref{eqn_Vcpi}),  the scalar form factor here involves 
the form factor ${\mathbb G}_S $ in (\ref{eqn_GS}), which is part of  ${\mathbb G}_V $.

The contribution from the mixed zero modes and non-zero modes or  {\em 't Hooft vertex}  is

\begin{eqnarray}
\label{eqn_Sdpi}
S_d^\pi(Q^2)=&&-(\lambda_u+\lambda_{\overline d})\,M_Q\,\bigg[\bigg(\frac {1}{N_c^2(N_c+1)}\bigg)
\bigg(\frac{\kappa\pi^2 f_\pi^2\chi_\pi}{M_Q^3}\bigg)\,
\left<Q\rho K_1(Q\rho)\right>\nonumber\\
&&\times\,\int_0^1dx_1dx_2\,\bigg( x_1\varphi_\pi(x_2)\bigg(\varphi_\pi^P(x_1)+\frac 16\varphi_\pi^{T\prime}(x_1)\bigg)+x_2\varphi_\pi(x_1)\bigg(\varphi_\pi^P(x_2)+\frac 16\varphi_\pi^{T\prime}(x_2)\bigg)\bigg)\bigg]\nonumber\\
\end{eqnarray}

The perturbative and instanton contributions 
to the  scalar form factor of the pion 
(with flat DA's) are shown  in Fig.\ref{fig_Spi_all}  versus $Q^2$. Again, one finds 
them to be comparable in magnitude but quite
different in their $Q$ dependence. Moreover, their
sum is roughly independent of $Q^2$.

\subsection{ Form factors of transversely polarized vector mesons}    
\label{sec_results_rho}

The transversely polarized rho meson form
factors are both electric and magnetic (see below). 
For simplicity, we quote here the contribution to the electric or charge form factor  by choosing the transverse
polarization $\epsilon_T(p,p^\prime)$ of the $\rho$ with momentum $p, p^\prime$ to be also transverse to $q$,  or $\epsilon_T(p, p^\prime)\cdot q=0$.

 The perturbative contribution (a) for the transversely polarized rho vector form factor  is formally subleading (containing an extra factor of $m_\rho^2/Q^2$),
like the $\chi_\pi^2/Q^2 $  contribution in the second term of $V_a^\pi$, and is found to be

\begin{eqnarray}
\label{VECTOR1}
V^\rho_a(Q^2)=&&\epsilon_\mu(q)(p^\mu+p^{\prime\mu})\,(e_u+e_{\overline d})\,(-\epsilon^{\prime*}_T\cdot\epsilon_T)\, 
\bigg(\frac{2\pi C_F\alpha_sf_\rho^2m_\rho^2}{N_cQ^4}\bigg) \int_0^1 {dx_1dx_2 \over 
	\bar x_1\bar x_2  + m_{\rm gluon}^2/Q^2}  \nonumber\\
&&\times 
\bigg[\bigg(
\varphi_{\rho}(x_1)\varphi_{\rho}(x_2)
	-\frac 1{16} \varphi^{A\prime}_{\rho}(x_1)\varphi^{A\prime}_{\rho}(x_2) \bigg)
\bigg( {1\over \bar x_1 +E_\perp^2/Q^2}+{1\over \bar x_2 +E_\perp^2/Q^2}-2\bigg) \nonumber \\
&& +{1 \over 2} \varphi^{A\prime}_{\rho}(x_1)\varphi_{\rho}(x_2)
	\bigg({1\over \bar x_1 +E_\perp^2/Q^2} -{1\over \bar x_2 +E_\perp^2/Q^2}  \bigg)
\bigg]
\end{eqnarray}
Note that the  minus sign in the product of polarization vectors in fact means that this contribution is $positive$,  
since $\epsilon^*_T\cdot\epsilon_T^\prime<0$ in  the Minkowski metric used here.  The Born-like instanton contribution to the rho vector form factor $V_b^\rho$
is also given by the substitution (\ref{eqn_substitution}) to $V_a^\rho$, and will not be given.

The one-gluon exchange to the scalar form factor
of the transverse rho meson is

\begin{eqnarray}
&&S^\rho_a(Q^2)=-(\lambda_u+\lambda_{\overline d})\,M_Q\, (-\epsilon^{\prime*}_T\cdot\epsilon_T)  
\bigg[\bigg(\frac{\pi C_F\alpha_s}{N_c}\,\frac{m_\rho f_\rho f^T_\rho}{M_Q Q^2}\bigg)\nonumber\\
&&\times  \int {dx_1 dx_2 \over \overline{x}_1\overline{x}_2+m^2_{\rm gluon}/Q^2}
\bigg(\frac {\varphi^T_{\rho}(x_1)(\varphi_{\rho}(x_2)-\varphi^{A\prime}_\rho(x_2)/4)}{\overline{x}_1+E_\perp^2/Q^2}
+\frac {(\varphi_{\rho}(x_1)-\varphi_\rho^{A\prime}(x_1)/4)\varphi^T_{\rho}(x_2)}{\overline{x}_2+E_\perp^2/Q^2}\bigg)\bigg]\nonumber\\
\end{eqnarray}

 \begin{figure}[h]
	\begin{center}
		\includegraphics[width=8cm]{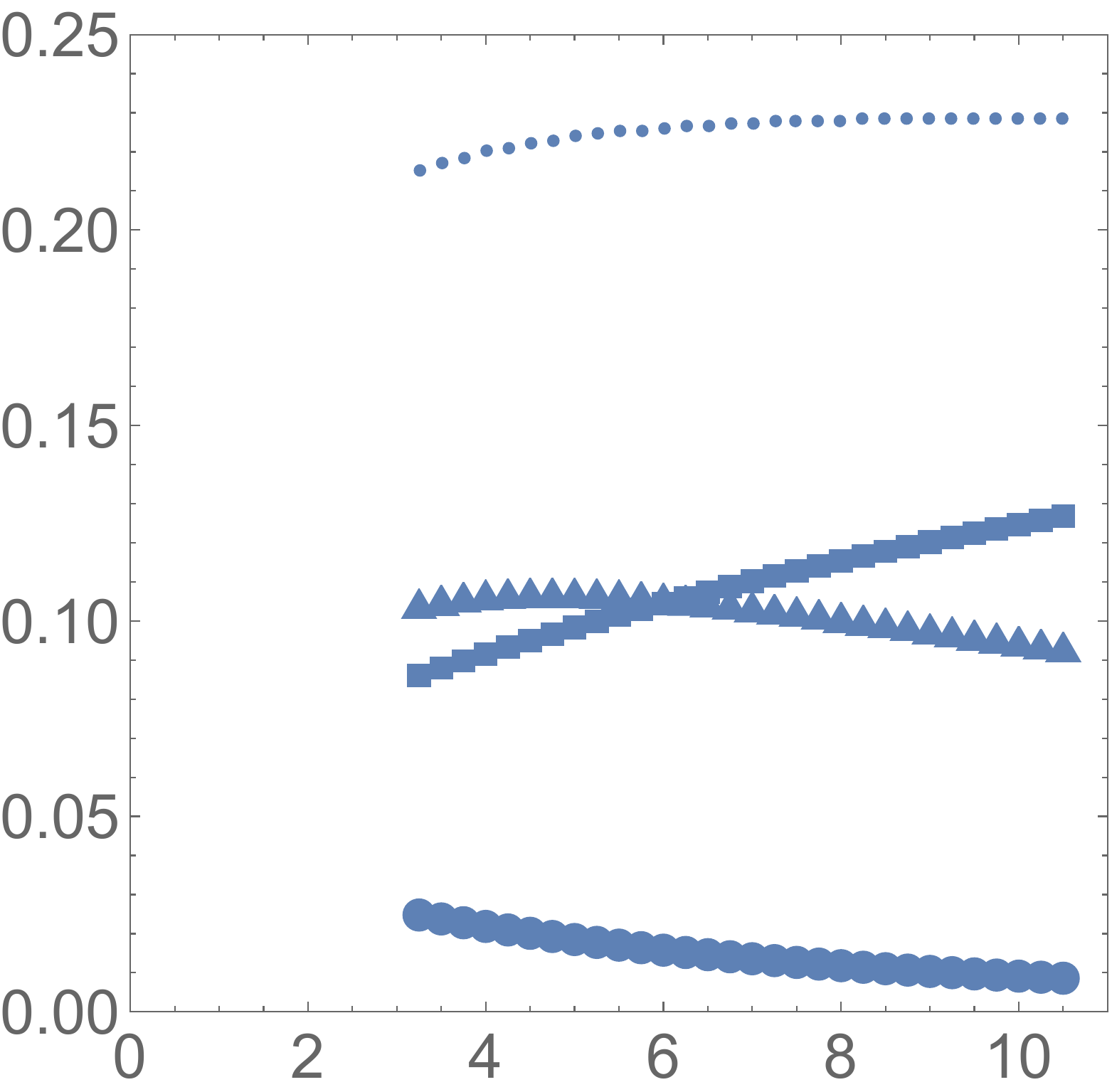}
		\caption{The  vector  form factors of the transversely polarized rho meson, times the momentum transfer squared , $Q^2 F^S_\pi(Q^2)\,\, ({\rm GeV}^2)$ versus $Q^2, ({\rm GeV}^2)$.
			The black closed points show the perturbative contribution,
			the black triangles correspond to the instanton zero mode
			('t Hooft vertex) contribution, and the squares are the 
			contribution of the non-zero mode propagators $S_{NZ}$.	
The dotted line above is their sum.		}
		\label{fig_Vrhoall}
	\end{center}
\end{figure}

The contribution of the the mixed zero mode and non-zero mode  to the transverse rho meson vector form factor is

\begin{eqnarray}
\label{TRANSVX}
V_c^\rho(Q^2)=&&\epsilon_\mu(q)(p^\mu+p^{\prime\mu})\,(e_u+e_{\overline d})\,(-\epsilon^{\prime*}_T\cdot\epsilon_T)
\bigg[ {\kappa\pi^2 f_\rho^2 m_\rho^2 \over N_cM_Q^2}\,\left<\rho^2\mathbb G_V(Q\rho)\right>\,\nonumber\\ 
&&\times\int_0^1 dx_1 dx_2
\,\bar x_1\bigg(\varphi_\rho(x_1)-\frac{ \varphi_\rho^{A\prime}(x_1)}4\bigg)\bigg(\varphi_\rho(x_2)+\frac{ \varphi_\rho^{A\prime}(x_2)}4\bigg)\bigg]
\end{eqnarray}


The contribution of  the mixed zero mode and non-zero mode to the rho meson scalar form factor is


\begin{eqnarray}
&& S_c^\rho(Q^2)=-(\lambda_u+\lambda_{\overline d})\,M_Q\,(-\epsilon^{\prime*}_T\cdot\epsilon_T)\,
\bigg[\bigg(\frac{\kappa \pi^2f_\rho f^{T2}_\rho m_\rho}{4N_c M_Q^4}\bigg)\,\left<(Q\rho)^2{\mathbb G}_S(Q\rho)\right>\nonumber\\
&&\times\int dx_1dx_2\bigg(\bar x_1\bigg(\varphi_{\rho}(x_1)-\frac{\varphi^{A\prime}_\rho(x_1)}4\bigg)\varphi^T_{\rho}(x_2)
+\bar x_2\bigg(\varphi_{\rho}(x_2)-\frac{\varphi^{A\prime}_\rho(x_2)}4\bigg)\varphi^T_{\rho}(x_1)\bigg)
\bigg] 
\end{eqnarray}

The contribution of the {\em 't Hooft vertex} to the vector form factor of the transversely polarized rho is detailed
in (\ref{TVR}) with the result

\begin{equation}
V_d^\rho(Q^2)=-(e_u+e_{\overline d})\bigg(\epsilon_\mu(q)(p^\mu+p^{\prime \mu})\,(\epsilon^{\prime*}_T\cdot\epsilon_T)\bigg)\,
\bigg[\left<Q\rho\,K_1(Q\rho)\right>\,\bigg(\frac{2\kappa\pi^2 }{N_c^2(N_c+1)}\frac{ {f^{T2}_\rho} }{M^2_Q}\bigg)\bigg]
\end{equation}

The contribution of the mixed non-zro mode and zero-mode contribution ({\em 't Hooft vertex}) to the  scalar form factor of the transversely polarized rho 
is

\begin{eqnarray}
S_d^\rho(Q^2)=&&+(\lambda_u+\lambda_{\overline d})\,M_Q\,(-\epsilon^{\prime*}_T\cdot\epsilon_T)
\bigg[\bigg(\frac {1}{N_c^2(N_c+1)}\bigg)\bigg(\frac{\kappa\pi^2f_\rho f_\rho^T m_\rho}{M_Q^3}\bigg)\,
\left<Q\rho K_1(Q\rho)\right>\nonumber\\
&&\times\,\int_0^1dx_1dx_2\,\bigg( x_1\varphi^T(x_2)\bigg(\varphi_\rho(x_1)+\frac 14\varphi_\rho^{A\prime}(x_1)\bigg)
+x_2\varphi_\rho^T(x_1)\bigg(\varphi_\rho(x_2)+\frac 14\varphi_\rho^{A\prime}(x_2)
\bigg)\bigg)\bigg]\nonumber\\
\end{eqnarray}


 \begin{figure}[h]
	\begin{center}
		\includegraphics[width=8cm]{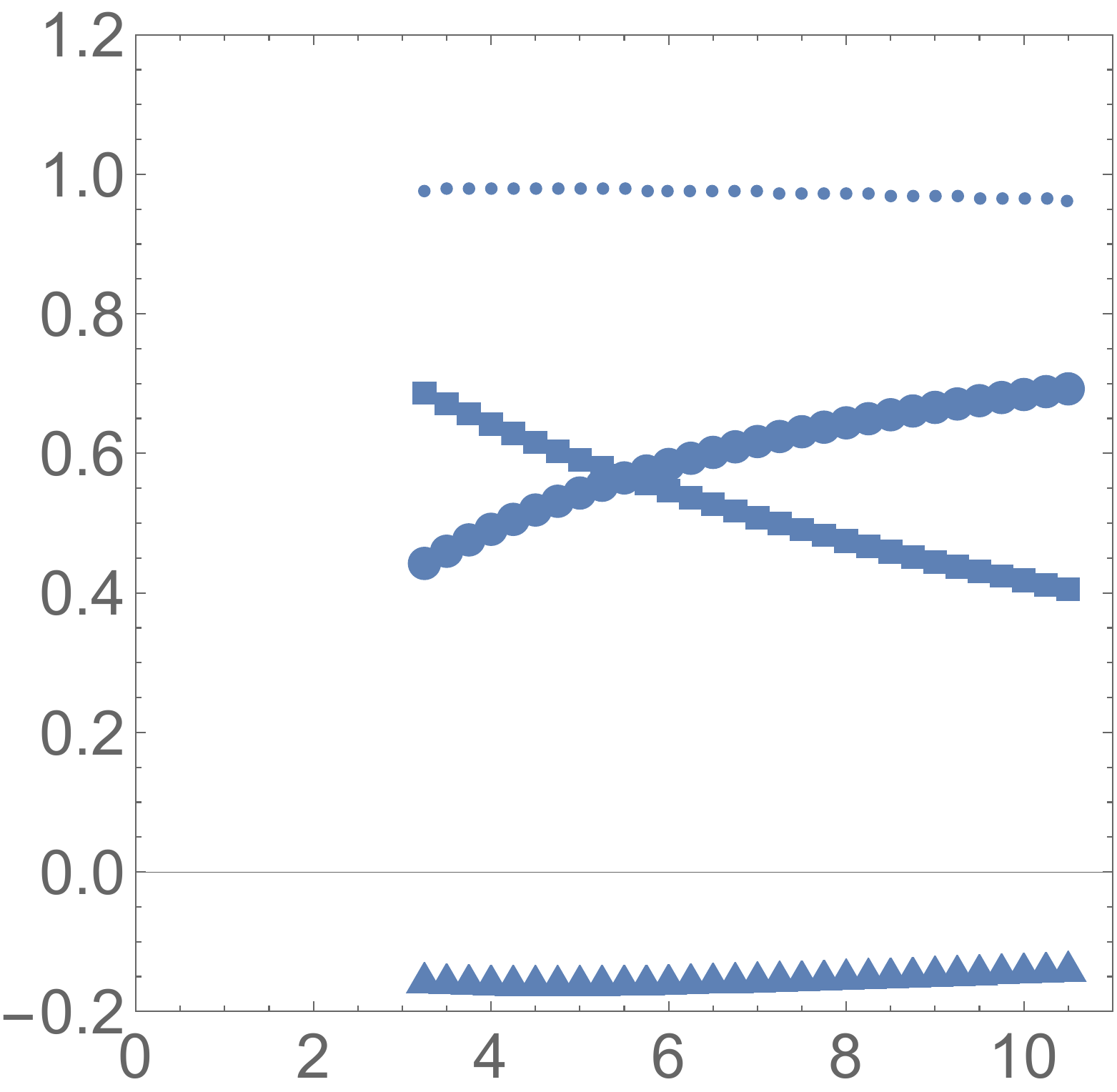}
		\caption{		
			The square bracket quoted in  the 
			 scalar scattering amplitudes of the rho given in the text, times the momentum transfer squared  $Q^2 [...] \,\,\, ({\rm GeV}^2)$ versus $Q^2, ({\rm GeV}^2)$. 
			The black closed points correspond to the one-gluon exchange contribution,
			the black  squares  to the instanton
			contribution of three non-zero mode propagators $S_{NZ}$, and the black triangles to the zero-mode terms $S_Z$ 
			 in two propagators.	The dotted line above is their sum.
			(Recall that the normalized scalar form factor is
			obtained by multiplication by the extra factor in
			(\ref{eqn_factor_in_S}).)	
		}
		\label{fig_Srhoall}
	\end{center}
\end{figure}

Completing this section, let us summarize the lessons from these  four
plots: a/  the first conclusion stemming from all of them,  is that the instanton effects are indeed
{\em comparable to the perturbative ones in magnitude}.; b/
the second  conclusion is that, {\em while separate contributions have 
different $Q$-dependence, the total sums tend to be flat}. 
While this conclusion seems to correspond to lattice data, we still need to remind the reader
 that  the $absolute$ normalization of the instanton-induced 
effects (squares and triangles in the previous 4 plots) remains relatively uncertain. The value $\kappa=1$
is  motivated (as for the DAs) to represent the ``maximal
but reasonable" value.
The quark mass used $M_{Q}=0.4\, {\rm GeV}$ may be
strongly modified in the denominators. With better knowledge of the gauge topology and DAs, these  curves may be modified.

We end this section by noting that the hard block is not sensitive to the current quark mass $m_q$, as
it was assumed that $Q^2 \gg m_q^2$. Therefore, going from the pion to $\eta_s$ and perhaps even
$\eta_c$ (in the appropriate kinematic range) one only needs to change the wave functions and the 
decay constant $f_\pi\rightarrow f_{\eta s}$. We will discuss the comparison to lattice data in section~\ref{sec-lat}.

\subsection{Form factors of the scalar meson $a_0^+$}\label{FFa0}

The form factors of the scalar meson $a_0^+$ 
results from the same hard blocks, but with completely
different DA's.
The vector form factor of the $a_0^+$ meson is

 \begin{eqnarray} \label{eqn_Vadel}
&&V^{a0}_a(Q^2)=\epsilon_\mu(q)(p^\mu+p^{\prime\mu})\,(e_u+e_{\overline d})\,
\bigg[\bigg(\frac{2C_F\pi\alpha_sf_{a0}^2}{N_cQ^2}\bigg)
   \int 
{ dx_1 dx_2 
 \over \bar x_1\bar x_2+m_{\rm gluon}^2/Q^2}\bigg( \varphi^V_{a0}(x_1)\varphi^V_{a0}(x_2)  \nonumber\\
&&+{ 2\over Q^2}    \bigg((\chi_{a0}^S)^2 
\varphi_{a0}^S(x_1)\varphi^S_{a0}(x_2)\bigg( {1 \over \bar x_2 +E_\perp^2/Q^2}-1\bigg)
+{\chi_{a0}^S\chi_{a0}^T\over 6}\varphi_{a0}^S(x_1)\varphi_{a0}^{\prime\, T}(x_2)\bigg( {1 \over \bar x_2 +E_\perp^2/Q^2}+1\bigg)
\bigg) \bigg]\nonumber\\
\end{eqnarray}
which is to be compared with (\ref{eqn_Vapi}).  All constants are different than in the pion case. The repulsive 
character of the interaction in this channel should penalize the wavefunction at the origin, leading to smaller
values for the parameters in comparison to the pion parameters. 

The scalar form factor of the $a_0^+$ reads 
\begin{eqnarray}\label{eqn_Saplus}
&&S_a^{a_0}(Q^2)=-(\lambda_u+\lambda_d)\,M_Q\,
\bigg[\bigg(\frac{2C_F\pi\alpha_sf_{a_0}^2 }{N_c M_QQ^2}\bigg)\,\nonumber\\
&&\times\int_0^1 
{ dx_1dx_2\over \overline{x}_1\overline{x}_2+m_{gluon}^2/Q^2}
\bigg(\chi_{a_0}^S
\varphi^S_{a_0}(x_1)\varphi^V_{a_0}(x_2)\bigg(1+\frac 2{\bar x_2+E_\perp^2/Q^2}\bigg)+\chi_{a_0}^T\varphi^V_{a_0}(x_1)\frac{\varphi_{a_0}^{T\prime}(x_2)}6\bigg)
\bigg]\nonumber\\
\end{eqnarray}
which is structurally similar to the pion result (\ref{eqn_Sapi}).

The non-zero mode propagators contribute to the vector
and scalar form factors of the $a_0^+$ as follows 
\begin{eqnarray} \label{eqn_Vaplus}
&&V_c^{a_0}=  \epsilon_\mu(q)(p^\mu+p^{\prime\mu})\,(e_u+e_{\overline d})\, 
\bigg[\bigg({\kappa \pi^2  f_{a_0}^2  \over  N_cM_Q^2}\bigg)
\langle  \rho^2{\mathbb G}_V(Q \rho) \rangle  \nonumber  \\
&&\times\int dx_1 dx_2 \bigg( 
\overline x_1  (\chi_{a_0}^S)^2 \varphi_{a_0}^S(x_1)\varphi_{a_0}^S(x_2)-(\overline x_2-\overline x_1)\varphi_{a_0}^S(x_1)\frac{\chi_{a_0}^S\chi_{a_0}^T\varphi_{a_0}^{T\prime}(x_2)}6
-\bar x_1\frac{(\chi_{a_0}^T)^2\varphi_{a_0}^{T\prime}(x_1)}6\frac{\varphi_{a_0}^{T\prime}(x_2)}6\bigg)\bigg]\nonumber\\
\label{eqn_Vaplus}
\end{eqnarray}
 is similar
to (\ref{eqn_Vcpi})

 \begin{eqnarray} \label{eqn_Saplus}
S_c^{a_0}(Q^2)=&&-(\lambda_u+\lambda_{\overline d})\,M_Q\,\bigg[\bigg(\frac{\kappa\pi^2  f_{a_0}^2}{N_cM_Q^3}\bigg)\,
\left<(Q\rho)^2\mathbb G_S(Q\rho)\right>\bigg]\,\nonumber\\
&&\times\int_0^1dx_1dx_2\,\bigg[  \bar x_2 \varphi^V_{a_0}(x_1)
\bigg(\chi_{a_0}^S\varphi^S_{a_0}(x_2)
-\frac 16 \chi_{a_0}^T\varphi_{a_0}^{T\prime}(x_2)\bigg)
\bigg]
\end{eqnarray}

The contribution of the mixed zero mode and non-zero mode propagators ($^\prime$t Hooft vertex) 
 to the $a_0^+$ vector form factor is

\begin{eqnarray} \label{eqn_Vaplus}
&&V_d^{a_0}(Q^2)=\epsilon_\mu(q)(p^\mu+p^{\prime\mu})\,(e_u+e_{\overline d})\, \nonumber\\
&&\times \bigg[\bigg(\frac {1}{N_c^2(N_c+1)}\bigg)\frac {4\kappa\pi^2f_{a_0}^2\chi_{a_0}^S\chi_{a_0}^T}{3M_Q^2}\left<\rho^2\frac{K_1(Q\rho)}{Q\rho}\right>
\int dx_1 dx_2 \varphi_{a_0}^S(x_1)\varphi_{a_0}^{T\prime}(x_2) \bigg]
\end{eqnarray}  
As  we noted  for the pion  in (\ref{eqn_Vdpi}), this contribution vanishes after the partonic integration is carried.

The contribution from the mixed zero modes and non-zero modes or  {\em 't Hooft vertex}  to the scalar form factor of the 
$a_0^+$ meson  is

\begin{eqnarray}
\label{eqn_Sdaplus}
S_d^{a_0}(Q^2)=&&+(\lambda_u+\lambda_{\overline d})\,M_Q\,\bigg[\bigg(\frac {1}{N_c^2(N_c+1)}\bigg)
\bigg(\frac{\kappa\pi^2 f_{a_0}^2}{M_Q^3}\bigg)\,\int_0^1dx_1dx_2\,
\left<Q\rho K_1(Q\rho)\right>\nonumber\\
&&\times\,\bigg( x_1\varphi^V_{a_0}(x_2)\bigg(\chi_{a_0}^S\varphi_{a_0}^S(x_1)+\frac 16\chi_{a_0}^T\varphi_{a_0}^{T\prime}(x_1)\bigg)
+x_2\varphi^V_{a_0}(x_1)\bigg(\chi_{a_0}^S\varphi_{a_0}^S(x_2)+\frac 16\chi_{a_0}^T\varphi_{a_0}^{T\prime}(x_2)\bigg)\bigg)\bigg]\nonumber\\
\end{eqnarray}
We note the overall sign flip in comparison to the pion contribution in (\ref{eqn_Sdpi}), but otherwise a similar structural result.

\subsection{Graviton and dilaton  form factors of the pion}

The energy-momentum form factor  of the pion follows  follows from the replacement of the vector vertex by the 
symmetrized tensor vertex (\ref{VERTEX}).  The form factor decomposition is detailed in~\ref{GRAVITON} with the $00$ (graviton)
and $\mu\mu$ (dilaton) perturbative contributions

 \begin{eqnarray}
 \label{eqn_T00PT}
T^\pi_{00a}(Q^2)=&&\bigg(\frac{4C_F\pi\alpha_sf_\pi^2}{N_c}\bigg)\int dx_1dx_2\,
\frac 1{\bar x_1\bar x_2+m_{\rm gluon}^2/Q^2}\bigg(x_1\varphi_\pi(x_1)\varphi_\pi(x_2)\nonumber\\
&& +\frac{2\chi_\pi^2}{Q^2}\bigg(\bigg(\frac 1{\bar x_1+E_\perp^2/Q^2}+\bar x_1-2\bigg)\varphi_\pi^P(x_1)\varphi_\pi^P(x_2)
+\bigg(\frac 1{\bar x_1+E_\perp^2/Q^2}-\bar x_1\bigg)\varphi_\pi^P(x_2)\frac{\varphi_\pi^{T\prime}(x_1)}6\bigg)\nonumber\\
T^\pi_{\mu\mu a}(Q^2)=&&\bigg(\frac{4C_F\pi\alpha_sf_\pi^2}{N_c}\bigg)\int dx_1dx_2\,
\frac 1{\bar x_1\bar x_2+m_{\rm gluon}^2/Q^2}\bigg(\varphi_\pi(x_1)\varphi_\pi(x_2)\nonumber\\
&& +\frac{2\chi_\pi^2}{Q^2}\bigg(\bigg(\frac 1{\bar x_1+E_\perp^2/Q^2}-3\bigg)\varphi_\pi^P(x_1)\varphi_\pi^P(x_2)
+ \bigg(\frac 1{\bar x_1+E_\perp^2/Q^2}-1\bigg)\varphi_\pi^P(x_2)\frac{\varphi_\pi^{T\prime}(x_1)}6\bigg)\nonumber\\
 \end{eqnarray}

The  non-zero mode contribution follows  the same reasoning as that for
the vector contribution through  the substitution (\ref{VERTEX}),  with the result

  \begin{eqnarray}
 \label{eqn_T00c}
T^\pi_{00c}(Q^2)=&&\bigg(\frac{\kappa\pi^2f_\pi^2\chi_\pi^2}{N_cM_Q^2}\bigg)\left<(Q\rho)\mathbb G_V(Q\rho)\right>\int dx_1dx_2\,(\bar x_1-\bar x_2)\nonumber\\
&&\times\bigg(\bar x_1\varphi_\pi^P(x_1)\varphi_\pi^P(x_2)+(\bar x_2-\bar x_1)\varphi_\pi^P(x_2)\frac{\varphi_\pi^{T\prime}(x_1)}6
-\bar x_1\frac{\varphi_\pi^{T\prime}(x_1)}6\frac{\varphi_\pi^{T\prime}(x_2)}6\bigg)\nonumber\\
T^\pi_{\mu\mu c}(Q^2)=&&\bigg(\frac{2\kappa\pi^2f_\pi^2\chi_\pi^2}{N_cM_Q^2}\bigg)\left<(Q\rho)\mathbb G_V(Q\rho)\right>\int dx_1dx_2\,\bar x_1 x_2\nonumber\\
&&\times\bigg(\varphi_\pi^P(x_1)\varphi_\pi^P(x_2)-\varphi_\pi^P(x_2)\frac{\varphi_\pi^{T\prime}(x_1)}6
-\frac{\varphi_\pi^{T\prime}(x_1)}6\frac{\varphi_\pi^{T\prime}(x_2)}6\bigg)
 \end{eqnarray}
The  mixed-zero mode and non-zero mode contribution (\ref{NZVEM}) contribute equally to $00$ and $\mu\mu$
in the Breit frame,  with the result
 
 \be
 \label{eqn_T00d}
 T^\pi_{00d}(Q^2)=T^\pi_{\mu\mu d}(Q^2)=-\frac {1}{N_c^2(N_c+1)}\bigg(\frac{16\kappa \pi^2 f_\pi^2\chi_\pi^2}{3M_Q^2}\bigg)
 \left<(Q\rho)K_1(Q\rho)\right>
 \int dx_1dx_2\,\varphi_\pi(x_1)\frac{\varphi_\pi^{T\prime}(x_2)}6
 \ee
 which is seen to vanish after $x$ integration.

\section{Hard block from one-gluon exchange  and its possible extensions}

\subsection{ The one-gluon exchange contributions }\label{sec_pert}

After we presented the results, we  now turn to their derivation starting from the hard block induced by the lowest order perturbative diagram, 
for completeness.

The one-gluon exchange contribution to the mesonic form factor is illustrated in Fig.~\ref{fig_diags}(a), where
also the definition of  the momenta involved is  also given
(see also Appendix \ref{sec_notations}).
 Of course,  Fig.~\ref{fig_diags}(a)  is one of four diagrams,  with a photon insertion appearing on 
the upper line of the $u$-quark before the gluon vertex. 
In the Breit frame, the space-like photon carries
$q=(0,0,Q,0)$, with the energy as the 4-th component.
The incoming pion carries $p=( 0,0,-Q/2,\sqrt{m_\pi^2+(Q/2)^2})$ and the outgoing pion carries ${p}^\prime=(0,0,+Q/2,\sqrt{m_\pi^2+(Q/2)^2})$.
We will however ignore the	pion mass $m_\pi$ in the energy,
by approximating the latter by $Q/2$ in the hard momentum limit.

The quark momenta should not be directed strictly along the direction of the meson momentum, as they carry 
  some nonzero transverse momenta $\vec k_\perp\neq 0$ in the wave functions. In principle, one needs to integrate over their distribution in hadrons.
  This brings in a question discussed e.g. in  \cite{Chibisov:1995ss}, who pointed out that
the smallness of the mean transverse momenta $\langle k_\perp^2 \rangle \ll Q^2$ does not in general exclude the existence 
and importance of a wave function component with a larger $k_\perp^2\sim Q^2$. In particular, it can also be induced by instantons, as momenta and field strength
are simply related by the equations of motion. In general, such a component, when present, would
violate factorization and produce an additional contribution to the exclusive processes.

 Nevertheless, in this paper we
will for now  ignore such contributions. The wave functions depending on $k_\perp$ will
appear only  in the integrated DA's, times
 the probability to find both quarks {\em at the same transverse location}. Those are constants like $f_\pi^2$.
Therefore, we will approximate the quark momenta as 
simply proportional to the mesonic ones
$k_1^\mu=x_1 p^\mu$ etc.  In the two-body sector
of the mesonic wave function, the longitudinal momentum fraction of the anti-quark is just $ \overline x_{1,2}=1-x_{1,2}\leq 1$. 

The contributions  of four perturbative diagrams of type (a)  are

\begin{eqnarray}
\label{HARD1}
&&e_u\,\bigg(\frac{-\,g_{\mu\nu}}{(\underline{k}_1-\underline{k}_2)^2}\bigg)
\bigg(\overline{u}(k_2)g_sT^a\,\gamma^\mu S_u(\underline{k}_1+p^\prime)\,\epsilon(q)\cdot\gamma u(k_1)\bigg)
\bigg(\overline{d}(\underline{k}_1)g_sT^a\gamma^\nu d(\underline{k}_1)\bigg)\nonumber\\
+&&e_u\,\bigg(\frac{-\,g_{\mu\nu}}{(\underline{k}_1-\underline{k}_2)^2}\bigg)
\bigg(\overline{u}(k_2)\epsilon(q)\cdot\gamma\,S_u(\underline{k}_2+p)\, g_sT^a\,\gamma^\mu u(k_1)\bigg)
\bigg(\overline{d}(\underline{k}_1)g_sT^a\gamma^\nu d(\underline{k}_1)\bigg)\nonumber\\
+&&e_{\overline d}\,\bigg(\frac{-\,g_{\mu\nu}}{(k_1-k_2)^2}\bigg)
\bigg(\overline{u}(k_2)g_sT^a\gamma^\mu u(k_1)\bigg)
\bigg(\overline{d}(\underline{k}_1)g_sT^a\gamma^\nu S_d(k_2-p)\epsilon(q)\cdot\gamma u(\underline{k}_2)\bigg)\nonumber\\
+&&e_{\overline d}\,\bigg(\frac{-\,g_{\mu\nu}}{(k_1-k_2)^2}\bigg)
\bigg(\overline{u}(k_2)g_sT^a\gamma^\mu u(k_1)\bigg)
\bigg(\overline{d}(\underline{k}_1)\epsilon(q)\cdot\gamma S_d(k_1-p^\prime)g_sT^a\gamma^\nu (\underline{k}_2)\bigg)
\end{eqnarray}
with the usual free quark propagators $S_f(p)=1/(\slashed{p}-m_f)$.  Note that the diagram \ref{fig_diags}(a)  corresponds to the second line.
Here  $\epsilon(q)\cdot\gamma$ is  the convolution of the photon polarization vector $\epsilon_\mu$ with gamma matrices, for brevity indicated by a slash. 
The propagator denominators  of the exchanged gluon 
simplify in the hard momentum limit as follows

\begin{eqnarray}
&&(k_1-k_2)^2=-x_1x_2Q^2-\sqrt{2}q(x_1k_1^-+x_2k_2^+)-2k_1^-k_2^+-(k_{1\perp}-k_{2\perp})^2\approx -x_1x_2Q^2\nonumber\\
&&(\underline{k}_1-\underline{k}_2)^2=-\bar x_1\bar x_2q^2-\sqrt{2}Q(k_1^-\bar x_1+k_2^+\bar x_2)-2k^+_2k^-_1-(k_{1\perp}-k_{2\perp})^2
\approx -\bar x_1\bar x_2Q^2\nonumber\\
\end{eqnarray}
Similarly, 
 the free fermion propagators 
simplify as

\begin{eqnarray}
\label{PROP}
S_f(\underline k_1+{p}^\prime)=&&\frac{{\slashed{\underline k}}_1 +{\slashed p}^\prime+m_f}{-\overline x_1 Q^2-\sqrt{2}q\overline{x}_1\underline{k}_1^--\underline{k}_{1\perp}^2-m_f^2}\approx
\frac{{\slashed p}^\prime-\overline{x}_1\slashed p}{-\bar x_1 Q^2}\nonumber\\
S_f(\underline{k}_2+p)=&&\frac{{\slashed {\underline k}}_2+\slashed p+m_f}{-\bar x_2 Q^2-\sqrt{2}q\overline{x}_2k_2^+-k_{2\perp}^2-m_f^2}\approx
\frac{{\slashed p}-\overline{x}_2\slashed p^\prime}{-\bar x_2 Q^2}\nonumber\\
\end{eqnarray}

Since there are two denominators, from the quark and gluon propagators,  
one encounters  certain negative powers of 
$x_i$ in the answer, with  potentially divergent integrals of the distributions. To keep it from happening, one should keep the
``regulating" masses and other subleading terms {\em  only in the denominator}. The magnitude of these
``regularized" integrals is discussed in section \ref{sec_regulated}.
When two parts of the hard blocks are sandwiched between two pion  DA amplitudes 
for  the outgoing and incoming pion 
 (both defined in  (\ref{WF1}-\ref{WF1X1}), each term becomes a single color-Dirac trace.
 The final expression for $V_a^\pi$ were  reported in the results section
(\ref{eqn_Vapi}).

\subsection{Convolutions with the wave functions and regularization of the $x$-integrals} \label{sec_regulated}

After  substitution of the DA's into  expressions for  form factors (and other exclusive processes)
one immediately finds that the integrands contains factor that diverge  at the end points,
$\xi=\pm 1$ or $ x,\bar x =0,1$. Therefore, some of the wave functions so far mentioned
(flat, semicircular and asymptotic ones)   lead to divergent
integrals. 
When $Q$ is taken to infinity, the integrals over momentum fractions obtain
 end-point singularities ($x\rightarrow 0,1$), up to quadratic ones
$$ \int  {dx}\,\frac {\varphi(x)}{x^2}$$
 For some 
wave functions, including the asymptotic one, 
such integrals are divergent. 

However, the  very derivations of the corresponding expressions provide a natural way out of this problem, 
by  keeping $subleading$ terms in the denominators.
In particular, the well known twist-two  perturbative contribution to the pion  vector form factor in (\ref{eqn_Vapi})
can be written as  

\begin{equation}
I_1= \int dx_1 dx_2\, {\varphi_\pi(x_1) \varphi_\pi(x_2) \over 
	\bar x_1\bar x_2 + m_{\rm gluon}^2/Q^2 
}  \label{eqn_I1}
\end{equation}
with a nonzero gluon mass used as an IR regulator.
More generally, one may view it as an effective parameter, representing a sum of higher-twist operators
which would appear if one expands the integral in powers of $1/Q$. For estimates below we will use
a value of $m^2_{\rm gluon}\sim 1 \, {\rm GeV}^2$.

\begin{figure}
	\centering
	\includegraphics[width=0.5\linewidth]{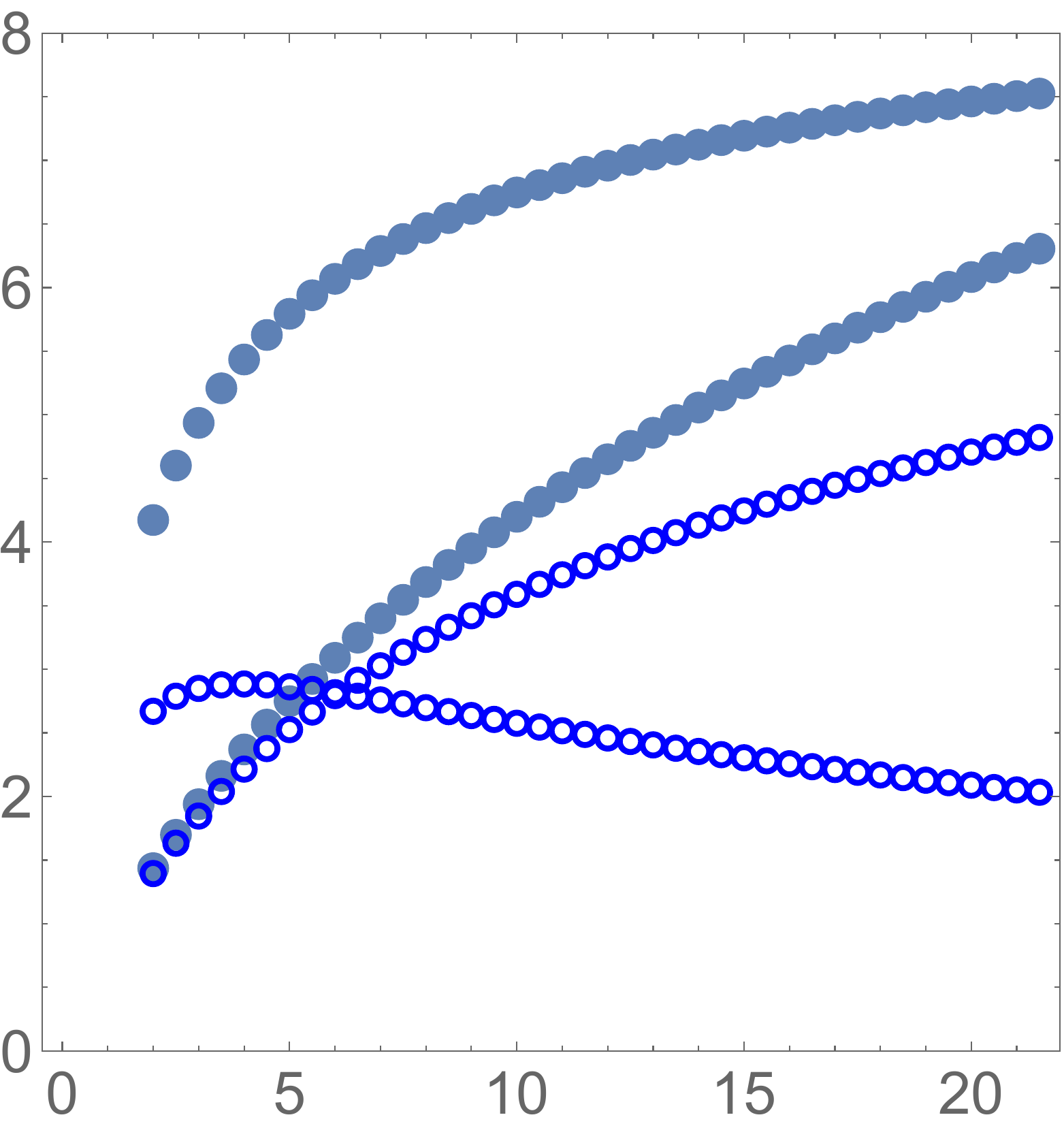}
	\caption{The regulated
		integrals $I_1$ (\ref{eqn_I1})
		(lower curves at right ) and
		$I2$
		(\ref{eqn_I2})
		(upper curves)	 versus the momentum transfer $Q^2 \,\, ({\rm GeV}^2)$. The closed points are for  the flat distribution (p=0), the open points are for the asymptotic distribution (p=1). The values of the gluon mass and quark transverse energy used are $m_{\rm gluon}^2=1\, {\rm GeV}^2, E_\perp^2=0.3\, {\rm GeV}^2$.	
	}
	\label{fig:i1i2}
\end{figure}

The twist-three contributions  have  a higher singularity, stemming  from
the denominator of the quark propagator. For instance, by combining the 
quark transverse momentum and quark mass into a
``transverse energy"
$E_\perp^2=\vec k_\perp^2+M^2$, the first twist-three contribution can be recast in the  ``regulated" form

\begin{equation} \label{eqn_I2}
I_2=\bigg({\chi_\pi^2 \over Q^2}\bigg) \int dx_1 dx_2\,{\varphi_\pi^P(x_1) \varphi_\pi^P(x_2) \over 
	\bar x_1\bar x_2  + m_{\rm gluon}^2/Q^2}
\bigg(2- {1\over \bar x_1 +E_\perp^2/Q^2}-{1\over \bar x_2 +E_\perp^2/Q^2}
\bigg)
\end{equation}
The dependence of these integrals on $Q$ is shown in Fig
\ref{fig:i1i2}, for flat (closed points)
and asymptotic wave functions (open points).

We recall that for the ``flat" distribution both un-regularized integrals are divergent, while for the asymptotic one only the second one is divergent.
However, they cannot be compared. Remarkably,  the regulated versions of the traditional part, $I_1$, and
new one $I_2$, turned out to be comparable. Moreover, although $I_2$ has  a $1/Q^2$ upfront
as a twist-three contribution, its regulated version  shows quite a weak  $Q$  dependence! Only asymptotically, the full twist-three
contribution in the pion vector form factor asymptotes $1/Q^4$ as it should.

 Note finally, that  the DA $\varphi_\pi(x)$, $\varphi_\pi^P(x)$  and $\varphi_\pi^T(x)$ 
 are distributions of independent chiral components 
 of the pion, and there is no general reasons for them to be the same. Moreover, we do know that the constants in
 front are quite different, so the distributions over the transverse momenta $must$ be different.
For instance, 
 $\varphi_\pi^P(x)$ is more compact -- has a larger probability to find the pair of quarks  at the same point in the transverse plane -- 
 so it is perhaps closer to flat than  $\varphi_\pi(x)$.  It is possible that the distributions over  the longitudinal momenta are  also
 different.

\subsection{Scalar pion form factor from one-gluon exchange}

The hard scalar (Higgs) block follows from
the same diagrams (\ref{HARD1}) with the substitutions $$(\epsilon(q)\cdot\gamma)\rightarrow 1,\,\,\, e_q\rightarrow \lambda_q$$
with the Yukawa couplings instead of the electric charges.  To understand the nature of the scalar  form factor in perturbation theory,
and for simplicity, let us set the tensor DA $\varphi^T(x)=0$. 
With the same regulation procedure as used in the previous subsection, the result for the form factor is

\begin{eqnarray}
S_a^\pi(Q^2)\rightarrow &&-(\lambda_u+\lambda_d)
\bigg[\bigg(\frac{\pi C_F\alpha_sf_\pi^2 \chi_\pi}{N_c Q^2}\bigg)\,
\int_0^1dx_1dx_2\,\varphi_{\pi}(x_1)\varphi_{\pi}^P(x_2)\nonumber\\
&&\times \bigg(\frac {1}{\overline{x}_1\overline{x}_2+m_{gluon}^2/Q^2}\bigg)\bigg(1+ {1 \over \overline{x}_1+E_\perp^2/Q^2} +\frac 1{\overline{x}_2+E_\perp^2/Q^2}\bigg)\bigg]
\end{eqnarray}
Note that this term appears from the product of two different
chiral components of the density matrix. It must be so because,  unlike the  interaction with the photon, the scalar vertex flips chirality,
and therefore needs to be complemented by $another$ chirality flip.

Another contribution to the scalar form factor 
stems from the mass  term in the quark propagators (\ref{PROP}),  $\sim M_Q/Q^2$, which we usually neglect.
Since it also  flips chirality, it  generates a subleading contribution (as we assume $M_Q\ll \chi_\pi$)

\begin{eqnarray} \label{eqn_MSapi}
S_a^\pi\rightarrow -(\lambda_u+\lambda_d)
&&\bigg[\bigg(\frac{\pi C_F\alpha_sf_\pi^2 M }{N_c Q^2}\bigg)\,
\int_0^1dx_1dx_2\,
\bigg( \varphi_{\pi}(x_1)\varphi_{\pi}(x_2)-
{4\chi_\pi^2 \over Q^2} \varphi_{\pi}^P(x_1)\varphi_{\pi}^P(x_2)\bigg) \nonumber \\
&& \times \bigg(\frac {1}{\overline{x}_1\overline{x}_2+m_{gluon}^2/Q^2}\bigg)\bigg( {1 \over \overline{x}_1+E_\perp^2/Q^2} +\frac 1{\overline{x}_2+E_\perp^2/Q^2}\bigg)\bigg]
\end{eqnarray}
which we have not included  in  the results quoted above.

 \subsection{Gravitaton and dilaton pion form factors  from one-gluon exchange}\label{GRAVITON}

 QCD is characterized by the symmetric energy momentum tensor

\bea
\label{EM1}
T^{\mu\nu}=&&\frac 2{\sqrt{-g}}\frac{\delta S_{QCD}}{\delta{g_{\mu\nu}}}
=F^{a\mu\lambda}F^{a\nu}_\lambda+\frac 14 g^{\mu\nu}F^2+\frac 12 \overline\psi \gamma^{[\mu} i\overleftrightarrow D^{\nu]_+}\psi
\eea
with the symmetric and long derivative $\overleftrightarrow{D}=\overrightarrow{D}-\overleftarrow{D}$,  and $[]_+$ denotes symmetrization.
It is conserved $\partial_\mu T^{\mu\nu}=0$, with a non-vanishing trace

\be
\label{TT}
T^\mu_\mu=\frac{\beta(g)}{2g} F_{\mu\nu}^aF^{a\mu\nu}+m\overline\psi\psi
\ee
due to the conformal anomaly, with the  one-loop beta function $\beta(g)=-(11N_c/3-2N_f/3)g^3/16\pi^2$.

The form factor of the energy-momentum tensor in a pion (or any pseudoscalar) state 
  is contrained by Lorentz symmetry, parity  and energy-momentum conservation.
Under  these strictures it takes the general form

\begin{align}
\label{THETA123}
\left<p_2\left|T^{\mu\nu}\right|p_1\right>=\frac 12 (g^{\mu\nu}q^2-{q^\mu q^\nu})\Theta_1(q^2)
+\frac 12 {p^\mu p^\nu}\Theta_2(q^2)
\end{align}
with $q^\mu=p^\mu_2-p_1^\mu$ and $p^\mu=p_1^\mu+p_2^\mu$. 
The two form factors correspond to the spin representations
$1\otimes 1=0\oplus 1\oplus 2$ with  $1$ excluded by parity. They reflect on the tensor exchange or graviton ($2$), and
the scalar exchange or dilaton (0). The graviton form factor is described by $\Theta_2$, and the dilaton form factor is  described by the trace

 \be
\label{TT}
\left<p_2\left| T^\mu_\mu\right|p_1\right>=\frac{3}2q^2\Theta_1(q^2)+\frac 12(4m_\pi^2-q^2)\Theta_2(q^2)
\ee
The normalization $\Theta_2(0)=1$ is fixed by recalling that $H=\int dx\, T^{00}$ is the Hamiltonian, with
$\left<p_1\left|H\right|p_1\right>=p_1^0[2p_1^0(2\pi)\delta_p(0)]$.  At low energy, the Goldstone nature of
the pion allows  to organize (\ref{EM1}) in a momentum expansion

\begin{align}
\label{MOM}
\left<p_2\left|T^{\mu\nu}\right|p_1\right>=p_2^\mu p_1^\nu+p_2^\nu p_1^\mu+\frac 12 g^{\mu\nu} q^2+{\cal O}(p^4)
\end{align}
which shows that the two invariant form factors normalize to 1,  $\Theta_1(0)=\Theta_2(0)=1$. Note that in two-dimensions,
there is only one invariant form factor for the dilaton, and an exact non-perturbative result can be derived in the context of 
the large $N_c$ limit~\cite{Ji:2020bby}.

In general, the invariant form factors $\Theta_{1,2}$ are fixed by the energy density
$T^{00}$ and the trace identity (\ref{TT})

\begin{align}
\label{MOM2}
-\frac{Q^2}2\Theta_1(Q^2)+2\bigg(m_\pi^2+\frac{Q^2}4\bigg)\Theta_2(Q^2)&=\left<p_2|T^{00}(0)|p_1\right>\nonumber\\
-\frac{3Q^2}2\Theta_1(Q^2)+2\bigg(m_\pi^2+\frac{Q^2}4\bigg)\Theta_2(Q^2)&=\left<p_2|T^\mu_\mu(0)|p_1\right>
\end{align}
 in the Breit frame  with  $q^\mu=(0,Q)$ and $p^\mu_{1,2}=(E_p,\mp Q/2)$.   More specifically,  we have

 \begin{align}
 \label{MOM3}
 \Theta_1(q^2)&=\frac 1{q^2}\left<p_2\left|\big(T^{\mu}_\mu(0)-T^{00}(0)\big)\right|p_1\right>\equiv \frac 1{q^2}\big(T_{\mu\mu}^\pi(q^2)-T_{00}^\pi(q^2)\big)
 \nonumber\\
 \Theta_2(q^2)&=\frac 1{q^2-4m^2_\pi}\left<p_2\left|\big(T^{\mu}_\mu(0)-3T^{00}(0)\big)\right|p_1\right>
 \equiv \frac 1{q^2-4m_\pi^2}\big(T_{\mu\mu}^\pi(q^2)-3T_{00}^\pi(q^2)\big)
\end{align} 
with $q^2=-Q^2$.
 We identify the strength of the graviton coupling  as  $-3T^{00}+T^\mu_\mu$ from (\ref{MOM3}), and the
strength of the dilaton coupling as $T^\mu_\mu$ from (\ref{TT}).

 The hard graviton and dilaton blocks follow from the same diagrams (\ref{HARD1})  using the elementary quark vertices 
 in (\ref{EM1}) 
 
 \begin{align}
 \label{VERTEX}
 \left<f(k_1+q)\left|\frac 12 \overline\psi \gamma^{[\mu} i\overleftrightarrow \partial^{\nu]_+}\psi(0)\right|f(k_1)\right> &= \frac 12 \gamma^{[\mu}(2k_1+q)^{\nu]_+}
 \nonumber\\
  \left<f(k_2)\left|\frac 12 \overline\psi \gamma^{[\mu} i\overleftrightarrow \partial^{\nu]_+}\psi(0)\right|f(k_2-q)\right> &= \frac 12 \gamma^{[\mu}(2k_2-q)^{\nu]_+}
 \end{align}
 for $f=u,d$. The $00$-coupling corresponds to the vertex $\gamma^0(2k_{1,2}\pm q)^0$, and the $\mu\mu$-coupling corresponds to
 the vertex $(2\slashed{k}_{1,2}\pm \slashed q)$. With this in mind, the  corresponding perturbative contributions to the pion 
 after regularization  are   listed in~(\ref{eqn_T00PT}). The non-perturbative contributions in the context of semi-classics will follow.

\subsection{The form factor of the transversely polarized\\rho meson  from one-gluon exchange }

The  perturbative and unregulated contribution to the vector form factor for the rho meson with transverse polarization,
can be obtained through similar contractions. For instance, if we set in this case the axial DA amplitude $\varphi_\rho^A=0$,
then  a typical perurbative contribution to the rho vector form factor  follows from the vector insertion


\begin{eqnarray}
\label{WF2}
V_a^\rho(Q^2)\rightarrow &&-\frac{e_uC_F}{N_c}\int_0^1\,dx_1\,dx_2\,\bigg(\frac {-1}{- \bar x_1\bar x_2Q^2}\bigg)\nonumber\\
&&\times {\rm Tr}\bigg[\epsilon_T(q)\cdot \gamma\, \bigg(\frac{{\slashed p}^\prime-\overline{x}_1\slashed p}{-\bar x_1Q^2}\bigg)
\,g_s\gamma^\alpha\,\gamma^0
\bigg(-\frac i4\slashed{\epsilon}^\prime_T({f_\rho} m_\rho\, \varphi_{\rho}(x_1)
+{f^T_\rho} \slashed{p}^\prime\, \varphi^T_{\rho}(x_1))\bigg)^\dagger\gamma^0\nonumber\\
&&\qquad\times g_s\gamma_\alpha \bigg(-\frac i4 \slashed{\epsilon}_T({f_\rho} m_\rho\, \varphi_{\rho}(x_2)+{f^T_\rho} \slashed{p}\, \varphi^T_{\rho}(x_2))\bigg)
\bigg]
\end{eqnarray}

Since $\gamma^\alpha \slashed{\epsilon}^\prime_T\slashed{p}^\prime\gamma_\alpha=4\epsilon_{T\mu}^\prime p^{\prime\mu}=0$,
the tensor contribution $\slashed{\epsilon}_T\slashed{p}$ drops out in the spin trace. The  final result with all insertions combined
is subleading in $m_\rho^2/Q^2$

\begin{eqnarray}
\label{VECTOR1}
V_a^\rho(Q^2)\rightarrow &&-\epsilon_\mu(q)(p^\mu+p^{\prime\mu})\,(e_u+e_{\overline d})\,\epsilon^{\prime *}_T\cdot\epsilon_T\,
\bigg[\bigg(\frac{2\pi C_F\alpha_sf_\rho^2m_\rho^2}{N_cQ^4}\bigg)\nonumber\\
&&\times \int_0^1dx_1dx_2\,\varphi_{\rho}(x_1)\varphi_{\rho}(x_2)
\bigg(\frac {1}{\overline{x}_1\overline{x}_2}\bigg(\frac 12 \bigg(\frac 1{\overline{x}_1} +\frac 1{\overline{x}_2}\bigg)-1\bigg)\bigg]
\end{eqnarray}
Similar arguments applied to the scalar form factor of the transversely polarized vector meson yield  the unregulated result

\begin{eqnarray}
S_a^\rho(Q^2)\rightarrow &&(\lambda_u+\lambda_{\overline d})\,M_Q\,\epsilon^{\prime *}_T\cdot\epsilon_T
\bigg[\bigg(\frac{\pi C_F\alpha_s}{N_c}\,\frac{m_\rho f_\rho f^T_\rho}{M_QQ^2}\bigg)\nonumber\\
&&\times \int dx_1dx_2\,\frac 1{\overline{x}_1\overline{x}_2}\bigg(\frac 1{\overline{x}_2}\varphi_{\rho}(x_1)\varphi^T_{\rho}(x_2)+\frac 1{\overline{x}_1}\varphi_{\rho}(x_2)\varphi^T_{\rho}(x_1)\bigg)\bigg]
\end{eqnarray}
The full perturbative results including the axial DA $\varphi_\rho^A$ are quoted in the results section above.

\subsection{Including other NJL-type local 4-fermion operators?} \label{sec_including_NJL}

In the spirit of the effective scattering theory for quarks,
one may think of introducing  $all$  local 
operators of the type
$${\cal O}_{\Gamma}\equiv (\bar q \Gamma q)(\bar q \Gamma q) $$
where the matrices $\Gamma$ include all possible Dirac, color and flavor structures. 
Naively, including any of them  is rather straightforward. The obvious practical problem is due to the fact that  total basis
of all operators is way too large for meaningful applications.
One needs  an organizational principle for the selection of only relevant ones, to be kept in the hard block.

From short distances, one-gluon exchange corresponds to the product of two color currents, with $\Gamma=\gamma_\mu \tau^a/2$.
 Colorless  exchanges start from two gluons, or perhaps scalar and tensor glueballs (in the discussion section
we will explain why the latter seems to be especially important, based on high energy scattering phenomenology). If so, $\Gamma=1$ or
the stress tensor $\Gamma_T= i\partial_\mu \gamma_\nu$.  

From large distance perspective, one
may  think about mesonic exchanges, as 
is done for nuclear forces. If this is the case, colorless scalar, pseudoscalar and vector $\Gamma$
 should be used, with or without  
flavor matrices. Still, the basis is too large for this approach to be practical.

Instantons generate a very specific effective quark-quark interactions. The most prominent is the one
discovered by 't Hooft~\cite{tHooft:1976snw}. It provides a unique  nontrivial selection of matrices, and so, in this work, we have 
focused on this particular choice. The organizational principle is the use of semi-classics in the hard block supplemented by a perturbative correction (one-loop).

\subsection{Born-style estimates of the instanton effects} 
 \label{sec_inst_Born}

This section is
devoted to estimate of the diagram (b) of Fig.~\ref{fig_diags}. Note that the point in which the hard
photon (scalar) is absorbed  is separated, by a quark propagator,
 from the location of the quark-antiquark scattering.
(On general grounds, one may question why such a separation is always  possible,
and in fact we will not assume it  in the next section.)

In this warm-up section, we   include the instanton field in the ``naive Born-like approximation", 
just by substituting the  gluon propagator by  the (Fourier transform) of the {\em instanton field}.
The reader must be warned that such approach is  a ``naive estimate" of the effect,
similar in spirit to our treatment of the NJL vertex above. 
 However, we note that the instanton field is nonperturbative, $gA_\mu\sim O(1)$  in  the weak coupling $g\ll 1$ regime.
Therefore, a consistent treatment should include  the
instanton field in the  full quark propagator  to {\em all orders}, with all zero and nonzero modes, through the instanton, a task
relegated to later sections below.

Before we start, let us mention the issue of gauge selection. Historically, the instantons were discovered in the
so called ``regular" gauge, in which the topological singularity is at infinity. In contrast, the so called "singular" gauge
put it at the origin. The difference between them  became apparent in any discussion of multi-instanton
configurations (and ensembles): only the singular ones can be used, since there is only one infinity for all of them.
This is important for our estimate, since the point-like gauge singularity will show up in the Fourier transform.

With this in mind, the instanton field has the form
\begin{eqnarray}
&&(A_\mu(x))^i_j(x)=-\frac i{2g}U^i_\alpha(\sigma_\mu\overline x-x\overline{\sigma}_\mu)^\alpha_\beta U^{\dagger \beta}_j\frac {\rho^2}{x^2(x^2+\rho^2)}\nonumber\\
&&(A_\mu(x))^i_j(x)=+\frac i{2g}U^i_\alpha(\sigma_\mu\overline x-x\overline{\sigma}_\mu)^\alpha_\beta U^{\dagger \beta}_j\frac {\rho^2}{(-x^2+i0)(-x^2+i0+\rho^2)}
\end{eqnarray}
in Euclidean and Minkowski space respectively. In Euclidean space, the Fourier transform of the instanton is

\begin{eqnarray}
\label{AF1}
A_\mu^a(k)=\frac{(2\pi\rho)^2}{2g}{\rm Tr}(T^aU(\sigma_\mu\overline{k}-k\overline{\sigma}_\mu)U^\dagger)\frac {\mathbb{G}(\rho\sqrt{q^2})}{k^2}
=\frac{i(2\pi \rho)^2}{g}{D^{ab}(U)\overline{\eta}^b_{\mu\nu}k^\nu}\frac{\mathbb{G}(\rho\sqrt{q^2})}{k^2}\nonumber\\
\end{eqnarray}
with the field form factor

\begin{equation} \label{eqn_G}
\mathbb{G}(\rho\sqrt{k^2})=\bigg(\frac 4{k^2\rho^2}-2K_2(\rho \sqrt{k^2})\bigg)
\end{equation}
which is normalized to 1, $\mathbb{G}(0)=1$. 
(No minus sign under the root because here we use Euclidean notations.)
The D-function is
$D^{ab}(U)={\rm Tr}(T^aU\tau^bU^\dagger)$ with  the normalization ${\rm Tr}(T^aT^b)=\delta^{ab}/2$. 
In particular, the analytical continuation of (\ref{AF1}) to Minkowski space  with amputation gives

\begin{eqnarray}
\lim_{k^2\rightarrow 0} (-k^2)\,\epsilon^\mu(k) A_\mu^a(k)=&&-\frac{2\pi^2\rho^2}{g}{\rm Tr}\bigg(T^a\,U(\epsilon(k)\overline k-k\overline{\epsilon}(k))U^\dagger\bigg)\,\mathbb{G}(0)\nonumber\\
\rightarrow&&-i \frac{(2\pi \rho)^2}{g}\,{D^{ab}(U)\overline{\eta}^b_{\mu\nu}\epsilon^\mu(k)k^\nu}
\end{eqnarray}
Note that the  2-point gluon correlator in both spaces read 

\begin{equation}
\label{TWO}
A_\mu^a(k) A_\nu^b(-k)=\frac{(2\pi\rho)^4}{g^2}D^{ac}(U)D^{bd}(U)\frac {\overline{\eta}^c_{\mu\alpha}\overline{\eta}^d_{\nu\beta}k^\alpha k^\beta}{k^4}\,\mathbb{G}^2(\rho\sqrt{k^2})
\end{equation}

The non-perturbative contribution to the mesonic form factor in the Born approximation illustrated in Fig.~\ref{fig_diags}b, 
can be evaluated using the single instanton contribution in (\ref{TWO}). This contribution  corresponds to the instanton 
(anti-instanton) effect on a pair of non-zero quark modes. For light quarks, it  is subleading in diluteness with the contribution shown in 
Fig.~\ref{fig_diags}c which involves non-zero mode contributions. For heavy quarks it is the sole and dominant non-perturbative contribution. 

The non-perturbative gluon propagator (\ref{TWO}) when averaged over an instanton plus anti-instanton contribution gives

\begin{eqnarray}
\label{TWOX}
\Delta_{\mu\nu}^{ab}(k)=&&\bigg<A_\mu^a(k) A_\nu^b(-k)\bigg>_{I+\overline I}\nonumber\\
=&&\frac n2 
\frac{(2\pi\rho)^4}{g_s^2}\bigg<D^{ac}(U)D^{bd}(U)\bigg>_U\,\bigg(\overline{\eta}^c_{\mu\alpha}\overline{\eta}^d_{\nu\beta}
+{\eta}^c_{\mu\alpha}{\eta}^d_{\nu\beta}\bigg)
\frac {k^\alpha k^\beta}{k^4}\,\mathbb{G}^2(\rho\sqrt{k^2})\nonumber\\
\rightarrow &&-\delta^{ab}\,\frac{n(2\pi\rho)^4}{g_s^2}\frac{g_{\mu\nu}k^2-k_\mu k_\nu}{k^4}\,\mathbb{G}^2(\rho\sqrt{-k^2})\equiv \delta^{ab}\Delta_{\mu\nu}(k)
\end{eqnarray}
with the last relation following in Minkowski space.
The contribution of Fig.~\ref{fig_diags}b follows that  in Fig.~\ref{fig_diags}a in the form (\ref{HARD1}) with the
substitution of (\ref{TWOX}) for the gluon propagator, namely

\begin{equation}
\frac{-g_{\mu\nu}}{({k}_1-{k}_2)}\rightarrow \Delta_{\mu\nu}({k}_1-{k}_2)\qquad
\frac{-g_{\mu\nu}}{(\underline{k}_1-\underline{k}_2)}\rightarrow \Delta_{\mu\nu}(\underline{k}_1-\underline{k}_2)
\end{equation}

\section{Instanton-induced effects} \label{sec_inst}

\subsection{From non-zero-mode propagators to hard block operators} \label{sec_LSZ}

As we already discussed, the instanton field is nonperturbative, or  strong  $A_\mu \sim 1/g$. Therefore
even if the coupling $g$ is small, it cancels out.
The propagation in such field  cannot be calculated in powers of $g$. Instead, one should  use
 the fully  dressed (re-summed) propagators. With this in mind,  the next step is the
 identification of the hard block, via  the
 ``amputation" of the free propagators
also known as Lehmann-Symanzik-Zimmermann 
  (LSZ) reduction.  We start explaining how this procedure
works in the coordinate representation, starting from the simpler case of spinless (scalar) quarks, 
as discussed in~\cite{Shuryak:1994ay}.

The propagator for a  massless scalar  particle in an instanton field has the form~\cite{Brown:1977eb}

\begin{equation}
\label{PROSCALAR}
\Delta(x,y) =\Delta_0(x-y)\bigg(1+ \rho^2 \frac{[Ux \bar y U^\dagger]}{x^2 y^2}\bigg) \frac 1{(\Pi(x) \Pi(y))^{\frac 12}}
\end{equation} 
with $\Delta_0(x)=1/(2\pi x)^2$ the free scalar propagator, and $\Pi(x)=1+\rho^2/x^2$, and 
$x,\bar y$ are convoluted with (Euclidean 4d) sigma matrices (\ref{eqn_4dsigma}). 
To see how the LSZ reduction operates on (\ref{PROSCALAR}), we 
consider the limit  $x,y \gg \rho$,  which is dominated by the asymptotic of 
$1/\sqrt{\Pi(x)}\approx (1-\rho^2/2x^2+...)$. For a single quark line, the color averaging 
in (\ref{PROSCALAR})  yields

\begin{equation} 
\label{AVERAGEX}
\left<[Ux \bar y U^\dagger] \right>_U=\frac{x\cdot y}{N_c}
\end{equation}
Inserting (\ref{AVERAGEX}) and keeping only the asymptotic contributions, give

\begin{equation}
\Delta(x-y)\approx \Delta_0(x-y)\bigg[1-\frac{\rho^2}{2x^2y^2}\bigg((x-y)^2+2x\cdot y\bigg(1-\frac 1{N_c}\bigg)\bigg)\bigg]
\end{equation}
one finds that the
term of order $\rho^2$  in the numerator becomes exactly the combination $(x-y)^2$ 
 in the denominator, so that it is canceled out. 
Subtracting the free propagator, one observes  that the $O(\rho^2)$ lowest-order instanton contribution
is proportional to $1/x^2 y^2=(4\pi^2)^2 D_0(x)D_0(y)$,
just  the product of Green function describing free propagation to and from the instanton.
So, in this case  the LSZ procedure is just an ``amputation" of  these  free propagators.

This result can be generalized to  an  ``amputated line operator", in the momentum representation
with arbitrary in- and out-momenta 

\begin{equation} {\cal T}(k,k^\prime) =\int d^4 x d^y e^{ikx-ik^\prime y} \big(\partial_x^2 \Delta(x,y) \partial_y^2\big) \end{equation}
where the second derivatives over $x$ and $y$ stand for the  ``amputation" of the 
trivial large distance part of the Green function.
Out of those one can construct n-body scattering amplitudes
by taking their powers, averaging over the positions of the instanton center $z_\mu$ and tracing over the color indices
\begin{equation}{\cal A}(k_i,k^\prime_i)={\rm Tr}\bigg[\prod_{i=1}^n {\cal T}(k_i,k^\prime_i)\bigg] \end{equation} 
The simplest of them, n=1, leads to the forward scattering amplitude on the instanton
\begin{equation}T(k,k)=\frac{4\pi^2 \rho^2}{N_c} \end{equation}
used by one of us  long ago, in \cite{Shuryak:1982hk}. 
This result explains the instanton suppression term at finite temperatures previously
calculated in~\cite{Pisarski:1980md}, and allowed its generalization to the case of finite temperature and density.
The n=2 case corresponds to two-by-two scattering, n=3 to three-by-three scattering, and so on. 
Averaging over the instanton position leads to momentum conservation $\sum_i k_i
=\sum k'_i$. The former case is important for meson form factors, the latter for baryon ones. 

The remaining  important detail is that in Euclidean calculations  $k=\sqrt{k_\mu^2}$ where all coordinates appear with plus sign. Going to Minkowski kinematics with ``on-shell"  $k\rightarrow 0$,
 partons can only mean here all components going to zero, or $x,y$ go to large distances . The scattering
 amplitude one gets from this procedure is just a constant, corresponding to low energy local interaction. There is
 no correlation between $k,k'$ momenta, or any angular distribution.  There is no nonlocality or explicit form factors in this procedure, and thus no dependence 
 on the momentum transfer $k-k^\prime$ in quark-antiquark scattering.

The extension of the (massless) scalar case to the (massless) spinor   case is done by using the full quark non-zero
mode propagator in the chiral-split form~\cite{Brown:1977eb}

\begin{eqnarray} 
\label{BROWN}
S_{NZ}(x,y)= &&\overrightarrow{\Dslash_x}\Delta(x,y) {1+\gamma_5 \over 2} +  \Delta(x,y) \overleftarrow{ \Dslash}_y {1-\gamma_5 \over 2} \nonumber\\
=&& \overline{S}(x,y) \frac{1+\gamma_5}2+{S}(x,y)\frac{1-\gamma^5}2
\end{eqnarray}
with the free Weyl  propagators $S_0=1/\overline{\partial}$ and $\overline{S}_0=1/{\partial}$, in the notations detailed in Appendix~\ref{app_defs}. 
The long derivative $\slashed{D}=\slashed{\partial}-i\slashed{A}$ acts on the left and right respectively of the (massless) scalar 
propagator, with each explicit contribution

\begin{eqnarray}
\label{BROWNCHIRAL}
\overline{S}(x,y)=\bigg(\overline{S}_0(x-y)\bigg(1+\rho^2\frac {[Ux\bar yU^\dagger]}{x^2y^2}\bigg)+\frac{\rho^2\overline{\sigma}_\mu}{4\pi^2}
\frac{[Ux\overline{\sigma}_\mu( x- y)\bar yU^\dagger]}{\Pi_xx^4(x-y)^2y^2}\bigg)\,\frac 1{(\Pi_x\Pi_y)^{\frac 12}} \nonumber\\
S(x,y)=\bigg(S_0(x-y)\bigg(1+\rho^2\frac {[Ux\bar yU^\dagger]}{x^2y^2}\bigg)+\frac{\rho^2\sigma_\mu}{4\pi^2}
\frac{[Ux(\bar x-\bar y)\sigma_\mu\bar yU^\dagger]}{x^2(x-y)^2y^4\Pi_y}\bigg)\,\frac 1{(\Pi_x\Pi_y)^{\frac 12}}
\end{eqnarray}
and with $U$ valued in $SU(N_c)$.
When a mixture of color and spinor indices occurs, the spinor matrices act on the upper left corner of the
$N_c\times N_c$ color matrices. 
Recall that the terms without and with the bar here correspond to Weyl notations with two-by-two matrices. 
They do not correspond to quarks and antiquarks -- the diagonal of $\gamma_0$ -- but to the left and right quark polarizations, diagonal of $\gamma_5$. These notations
are compatible with other Weyl-style notations used.

In the case of a scalar (Higgs) probe on a $q\bar q$ meson pair, the chirality of the quark is flipped, and therefore one part of the diagram 
contributes

$$ \bar S(x,z)   S(z,y) +S(x,z) \bar  S(z,y)\,, $$
in which case the endpoints $x,y$ should be taken to large distances while the intermediate
point $z$ is still residing  inside the instanton field. 
In the former term the covarient derivatives, acting from both sides, create free fermionic propagators,
which can be readily amputated. What is left, depending on the point $z$ is just the factor $1/\Pi_z$. Its Fourier transform
with momentum transfer $q_\mu$ is 

\begin{equation}
\int {d^4 z }\, {e^{i q\cdot z} \over \Pi_z}
\end{equation}

Unfortunately, this  is not so simple in the second part of the diagram. The second term of $S(x,z)$ at large $x$ is of order $1/x^2$,
with a power not matching the free fermion propagator  $S_0 \sim x/x^4$. It means that  the LSZ reduction in coordinate
representation is not local. Let us use the following trick: multiply and divide by $\Dslash$. The $\Dslash$ in the numerator now
reproduces the free propagator, which we can amputate. The $\Dslash$ in the denominator  will become
the negative power of momentum in the amplitude when taken to the momentum representation, generating 
a $negative$ moment of the wave function by convolution to the wavefunction.

Now let us focus on the line in which there is no external probe. There is a single $S(x,y)$ in which $both$ coordinates
are taken to infinity. Again, in each term one dependence leads to a straightforward LSZ procedure, and the other lacks one
power of the distance. We use the same trick and represent it as $\Dslash_x \Delta \Dslash_y (1/\Dslash_y)$. The effective
amplitude takes the form ${\cal A}\sim \rho^2({1\over x}+{1\over x'})$. We now proceed to give a more quantitative derivation
of these results.

\begin{figure}[h!]
	\begin{center}
		\includegraphics[width=8cm]{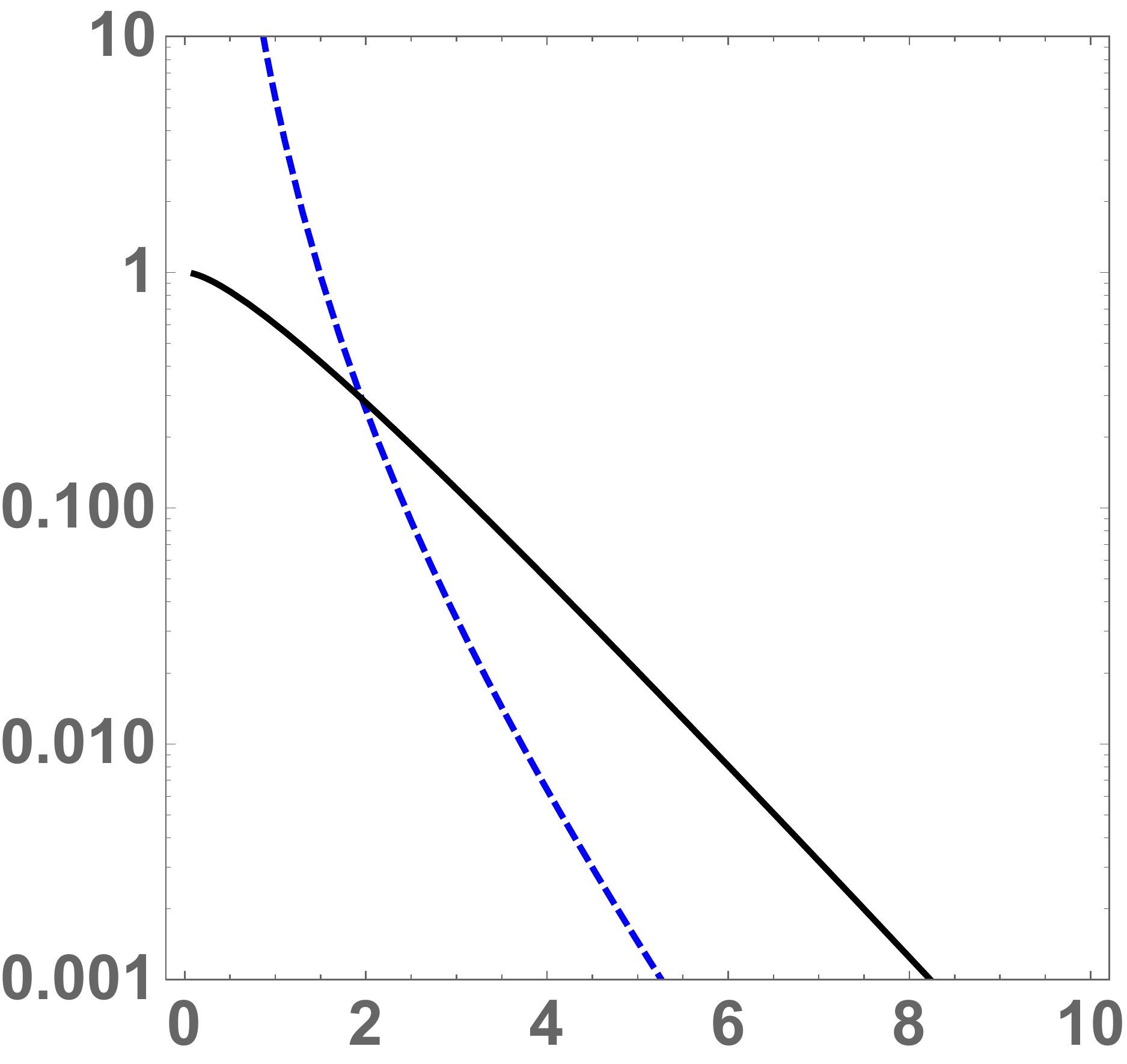}
		\caption{The blue dashed  and the solid black lines correspond to the 
			 functions $F_1(Q\rho)$ and $F_2(Q\rho)$ versus $Q$ as given in (\ref{eqn_F2}),  respectively.
		}
		\label{fig_F1_F2}
	\end{center}
\end{figure}

The LSZ reduced non-zero mode contributions to  the  $q\overline{q}$ vector vertex with polarization $\epsilon_\mu(q)$, 
can be formally written in the chiral  split  form as 

\begin{eqnarray}
\label{NZ1}
\left<\left<k_2\left|{\partial}\,\overline{S}\,\overline{\epsilon}(q)\,\overline{S}\,{\partial}\,\frac{1+\gamma_5}2
+\overline{\partial}\, {S}\,{\epsilon}(q)\,{S}\,\overline{\partial}\,\frac{1-\gamma_5}2
\right|k_1\right>\otimes
\left<\underline{k}_1\left|{\partial}\,\overline{S}\,{\partial}\,\frac{1+\gamma_5}2
+\overline{\partial}\, {S}\,\overline{\partial}\,\frac{1-\gamma_5}2
\right|\underline{k}_2\right>\right>_U\nonumber\\
\end{eqnarray}
where the  overall averaging over color is indicated by the subscript $U$, and the mass shell conditions $k_{1,2}^2\rightarrow 0$ and $\underline{k}_{1,2}^2\rightarrow 0$ are subsumed. 
Using the results in Appendix~\ref{app_51}, 
the  final vector vertex  follows by  adding (\ref{REDUCED}) to the color averaged (\ref{REDUCED3}) 
and combining  the result with (\ref{NZ8})  to finally obtain

\begin{eqnarray}
\label{NZV}
&& \bigg[2\kappa\mathbb G_V(q\rho)\overline{u}(k_2)\slashed{\epsilon}(q)u(k_1)\bigg]
\times\bigg[\bigg({{2\pi^2\rho^2}}\bigg) \overline{d}(\underline k_1)
\bigg(\frac 1{\slashed{\underline k}_1}+\frac 1{\slashed{\underline k}_2}\bigg)d(\underline k_2)\bigg]\nonumber\\
\end{eqnarray}
The induced vector form factor ${\mathbb G}_V$ consists  of two parts

\begin{equation}{\mathbb G}_V(\xi)=F_1(Q\rho)+{1 \over N_c M^2\rho^2}F_2(Q\rho)
\label{eqn_GV}\end{equation}
with specifically

\begin{eqnarray}
F_1(x)\equiv  \bigg(\frac {K_1(x)}{x}\bigg)^{\prime\prime}=&&
{1 \over 4 x^3}(4 x K_0(x) + (8 + 3 x^2) K_1( x) + 
x (4 K_2( x) + x K_3(x)))\nonumber\\
F_2(x)\equiv x \bigg(\frac{(x K_1)^\prime}{x}\bigg)^\prime=&&
{1 \over 4 x}(-2 x K_0(x) + (-4 + 3 x^2) K_1(x) + 
x (-2 K_2(x) + x K_3(x)))\equiv xK_1(x)\nonumber\\
\label{eqn_F2}
\end{eqnarray}
We have summed over $n/2$ instantons plus $n/2$ anti-instantons,  analytically continued to Minkowski signature, and
dropped the extra factor of $i$ since (\ref{NZ1}) follows from  $S=1+iT$ with $T$ identified with the vector vertex. Overall
momentum conservation follows from the Z-integration over the instanton and anti-instanton positions leading to
$q+k_1+\underline{k}_1=k_2+\underline{k}_2$ for the 2-body vertex (\ref{NZV}). In Fig.~\ref{fig_F1_F2} the behavior
of $F_{1,2}(Q\rho)$ in (\ref{eqn_F2}) is shown, with $F_2(Q\rho)$ dominant at large $Q$.

After the hard block is defined, one carries  the trace
with the pion (or rho) density matrices. The propagators remaining in the second bracket of 
(\ref{NZV}) are  treated as follows
$$ {1\over \slashed{\underline k}_1} \rightarrow {\slashed{\underline k}_1 \over M_Q^2} $$
with $M_Q$ being the constituent quark mass. 
The final  expression is (\ref{eqn_Vcpi}).

\begin{figure}[h!]
	\begin{center}
		\includegraphics[width=8cm]{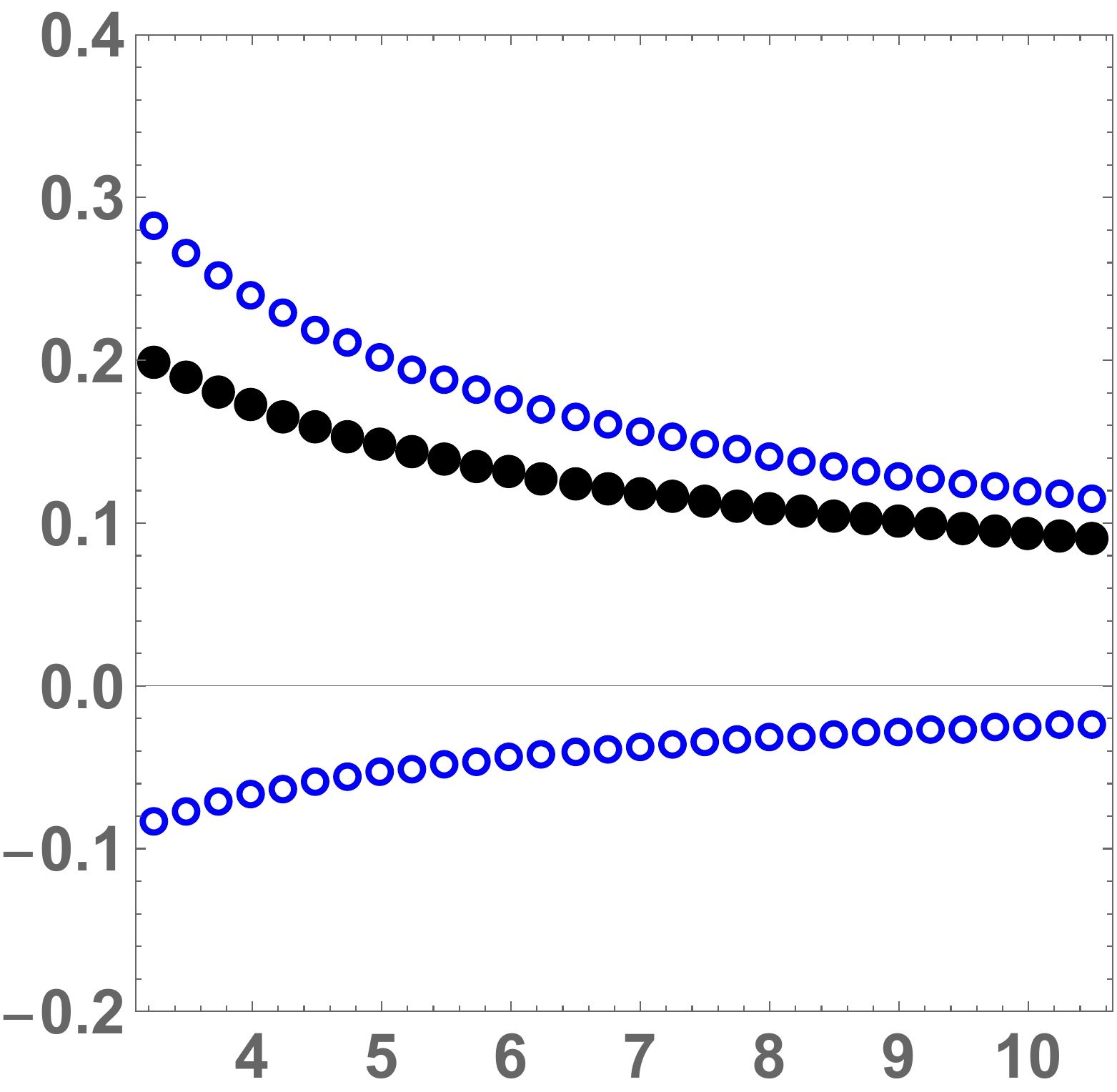}
		\caption{
			The nonzero mode contributions to the vector pion form factor times $Q^2 \, ({\rm GeV}^2)$,
			 versus  $Q^2 \, ({\rm GeV}^2)$.
			The blue circle above and below
			show the contributions of $\langle F_1 \rangle_\rho$ and $\langle  F_2\rangle_\rho $, respectively.
		The black closed circles are their sum.
		}
		\label{fig_SSSV}
	\end{center}
\end{figure}

We have checked that similar arguments apply to the scalar form factor, which is seen to mix 
chirality through $S\overline S$ and $\overline S S$ contributions, but the result is found to be identical to (\ref{NZ2}) with the substitution $\slashed{\epsilon}(q)\rightarrow 1$
and no additional contribution. Hence, 
the same result holds for the scalar vertex with the substitution $\slashed\epsilon(q)\rightarrow 1$ in (\ref{NZV}) and the induced scalar form factor

\begin{equation}\label{eqn_GS}
\mathbb G_V(\xi)\rightarrow {\mathbb G}_S(\xi)=\bigg(\frac {K_1(\xi)}{\xi}\bigg)^{\prime\prime}
\end{equation}
Note that the two terms in (\ref{eqn_GV})
have opposite signs. So their sum is sensitive
to the averaging over the instanton size (see section \ref{sec_averaging}). Fig.\ref{fig_SSSV} displays the 
contribution of each of them, as well as their sum.  After convolution of the hard block with the pion density matrices we get the final result for $V_c^\pi$, 
as given in (\ref{eqn_Vcpi}).

\subsection{The non-zero mode  contributions to  the rho  vector form factors}

The general decomposition of the vector form factor of the rho meson compatible with parity, time-reversal symmetry  and Lorentz symmetry is of the form 

\begin{eqnarray}
\label{GENERAL}
\left<\rho(p^\prime, \epsilon^\prime)\left|J_\mu(0)\right|\rho(p, \epsilon)\right>&&=
F_V(q^2)\,\epsilon^{\prime *}\cdot \epsilon\,(p_\mu+p_\mu^\prime)\nonumber\\
&&+\frac{G_V(q^2)}{2m_\rho}\,\left(\epsilon^{*\prime}_\mu\epsilon\cdot q-\epsilon_\mu\epsilon^{*\prime}\cdot q\right)
+\frac{H_V(q^2)}{4m_\rho^2}\,\epsilon^{\prime *}\cdot q \epsilon\cdot q\,(p_\mu+p_\mu^\prime)
\end{eqnarray}
with $F_V,H_V$ contributing to the electric form factor and $G_V$ to the magnetic form factor of the rho.

The contribution to the vector form factor of the transverse rho meson in the large momentum limit, involves all three form factors
in (\ref{GENERAL}) in general. For simplicity, we focus  on the contribution to the electric part $F_V$ by choosing the transverse
polarization $\epsilon_T(p,p^\prime)$ of the $\rho_\perp$ with momentum $p, p^\prime$ to be also transverse to $q$,  or $\epsilon_T(p, p^\prime)\cdot q=0$.
In the large momentum limit  and for the axial DA $\varphi^A(x)=0$ for simplicity, the unregulated contribution to $F_V$ is

\begin{eqnarray}
\label{TRANSV}
V_c^\rho(Q^2)\rightarrow &&-\epsilon_\mu(q)(p^\mu+p^{\prime\mu})\,\epsilon^{\prime *}_T\cdot\epsilon_T\,(e_u+e_{\overline d})\,
\bigg[\frac{\kappa\pi^2{f}^{T2}_\rho}{N_c}\,\left<\rho^2 \mathbb G_V(Q\rho)\right>\int_0^1dx_1dx_2\,\nonumber\\
&&\times\bigg(\bigg(\frac 1{\overline{x}_1}+\frac 1{\overline{x}_2}\bigg)\varphi^T_{\rho}(x_1)\varphi^T_{\rho}(x_2)
+\frac 12 \frac{f_\rho^2m_\rho^2}{{f}_\rho^{T2}M_Q^2}\varphi_{\rho}(x_1)\varphi_{\rho}(x_2)\,
(\overline{x}_1+\overline{x}_2)\bigg)\bigg]\nonumber\\
\end{eqnarray}

\subsection{The non-zero mode  contribution to the rho  scalar form factors}

The general decomposition of the scalar form factor of the rho meson compatible with parity, time-reversal symmetry  and Lorentz symmetry is of the form 

\begin{eqnarray}
\label{GENERAL1}
\left<\rho(p^\prime, \epsilon^\prime)\left|S(0)\right|\rho(p, \epsilon)\right>=
F_S(q^2)\,\epsilon^{\prime *}\cdot \epsilon
+\frac{H_S(q^2)}{4m_\rho^2}\,\epsilon^{\prime *}\cdot q \epsilon\cdot q
\end{eqnarray}
Similarly to   the
pion scalar form factor,  the contribution to the scalar form factor of the longitudinal rho meson vanishes because of a poor spin trace. As a result, the invariant scalar form
factors in (\ref{GENERAL1}) satisfy

\begin{equation}
F_S(q^2)-\frac{q^4}{16 m_\rho^4}H_S(q^2)\approx 0
\end{equation}
in the large momentum limit. We can extract $F_S(q^2)$ from the transversely polarized $\rho$ by choosing $\epsilon_T(p, p^\prime )\cdot q=0$.  
For simplicity, if we set the axial DA amplitude $\varphi^A(x)=0$, the unregulated result is

\begin{eqnarray}
\label{FSCALAR}
S_c^\rho(Q^2)\rightarrow &&(\lambda_u+\lambda_{\overline d})\,\epsilon_T^{\prime *}\cdot \epsilon_T\,M_Q\,
\bigg[\frac{\kappa\pi^2 f_\rho  f^T_\rho }{2N_c}\frac {m_\rho}{M_Q}\,\left<\rho^2\mathbb G_S(Q\rho)\right>\int_0^1dx_1dx_2\,\nonumber\\
&&\bigg(\varphi^T_{\rho}(x_1)\varphi_{\rho}(x_2)\bigg(\frac 1{\overline{x}_1}+\frac{q^2}{M_Q^2}\overline{x}_2\bigg)
+\varphi^T_{\rho}(x_2)\varphi_{\rho}(x_1)\bigg(\frac 1{\overline{x}_2}+\frac{q^2}{M_Q^2}\overline{x}_1\bigg)\bigg)
\bigg]\nonumber\\
\end{eqnarray}

 \subsection{The non-zero mode contribution to the graviton and dilaton\\ form factor of the pion }

The instanton and anti-instanton contributions  to the hard block with the  energy-momentum tensor vertex, follows a similar reasoning as that
for the vector insertion  with the substitutions 
$$\epsilon^\mu(q)\rightarrow (2k_{1,2}^\mu+q^\mu),\,\,\, e_q\rightarrow 1$$
and symmetrization. 
 In particular, the  non-zero mode contributions  to the energy-momentum vertex follow from (\ref{NZV}) in the form

\begin{eqnarray}
\label{NZVEM}
&& \bigg[2\kappa\mathbb G_V(Q\rho)\overline{u}(k_2)(k_1+k_2)^{[\mu}\gamma^{\nu]_+}u(k_1)\bigg]
\times\bigg[\bigg({{2\pi^2\rho^2}}\bigg) \overline{d}(\underline k_1)
\bigg(\frac 1{\slashed{\underline k}_1}+\frac 1{\slashed{\underline k}_2}\bigg)d(\underline k_2)\bigg]
\end{eqnarray}
with the   induced vector form factor ${\mathbb G}_V$ given in (\ref{eqn_GV}-\ref{eqn_F2}).
The non-zero mode vertex  (\ref{NZVEM}) when sandwiched betwen the pion DA yields (\ref{eqn_T00c}).

\subsection{Quark zero modes and 't Hooft effective Lagrangian} \label{sec_hooft}

The quark propagator in the instanton background when expressed in the eigenmode basis, is a sum over all modes. In this section we focus on the specific term of this sum containing the zero modes.
For  a single instanton, this contribution takes  the form
\begin{equation} \label{eqn_SZ}
	S_{Z}(x,y)= {\psi_0(x)  \psi^*_0(y) \over i m_q} \end{equation}
with zero eigenvalue plus the quark mass in the denominator. This appears singular in the chiral
limit $m_q\rightarrow 0$, but, as explained by 't Hooft, since the amplitude for a single instanton
is itself proportional to the product of masses of all light quark flavors, $\sim \prod_{q=u,d,s} m_q$, the Green functions and vertices with
$N_f$ fermions are finite. This is how  the famous   't Hooft effective  Lagrangian was derived.

In ``empty" (perturbative) vacuum the mass here is
that from QCD Lagrangian. However,
when an instanton is not in the perturbative but rather in $physical$ QCD vacuum, the problem is more complex.
A nonzero quark condensate makes  the  instanton amplitude nonzero even in the chiral limit. 
The current quark mass $m$ is supplemented by the so called ``determinantal mass" $M^*$  \cite{Shifman:1979uw}
\begin{equation} M^*\equiv {2\pi^2 \over 3} |\langle \bar q q \rangle | \rho^2 \approx  200 \, {\rm MeV}\bigg({\rho \over \rho_0 }\bigg)^2\end{equation}
(Note that  this is  $not$ the on-shell quark mass at zero momentum, which is 
about twice larger.)  This mass was used in the first mean-field-style 
treatment of the instanton ensemble  \cite{Shuryak:1981ff}, appending the quark masses both
in the instanton determinant and in the denominator of the quark propagator.

After the formalism of the interacting instanton liquid model (IILM) was further developed, 
the so called ``single instanton approximation" (SIA)   for treating effects produced by a single member of the ensemble was
 further discussed in Ref.\cite{Faccioli:2001ug}. It was pointed out there that the OPE expression from \cite{Shifman:1979uw} was derived assuming factorization of the VEVs of 4-fermion 
operators in the QCD vacuum, which is also a version of the mean-field treatment. However, 
the instanton ensemble is highly correlated, and the expectation values of different multi-quark operators
 are highly inhomogeneous, and therefore  the mean-field-style approximations are quite inaccurate. 
In particular,  the operators 
of the type of 't Hooft Lagrangian  under consideration

$$\langle (\bar u u)(\bar d d) \rangle \gg  \langle (\bar u u)\rangle  \langle (\bar d d) \rangle   $$  
have strongly enhanced VEVs. 
The quark propagator in the QCD vacuum, is  approximated by  the form 

\begin{equation}S(x,y)=S_Z(x,y)+\sum_{I,J} \psi_{0I}^*(x) \bigg( {1 \over T} \bigg)_{I,J} \psi_{0J}(y)
\end{equation}
where $T_{IJ}$ denotes the so called ``instanton hopping" matrix,  constructed out of the Dirac
zero modes overlaps between neighboring  instantons $I,J$. Note that
here enters the $inverse$ matrix, as propagators are inverse to Dirac operators.
So, when one discusses a process in which
both points $x,y$ are inside one instanton $I^*$, like when defining the hard block here, 
we  can restrict the sum to only the term with the zero 
mode of this very instanton. This leads to the following
redefinition of the  ``determinantal mass"  

\begin{equation}{1 \over M_u}\equiv \bigg< \bigg( {1 \over T} \bigg)_{I^*I^*}    \bigg>  \end{equation}
Furthermore, in the diagrams containing $two$ quark propagators of {\em different flavors} one has a different
averaging

\begin{equation}
{1 \over M^2_{uudd}}\equiv \bigg< \bigg( {1 \over T} \bigg)^2_{I^*,I^*}    \bigg>
 \end{equation}
These two quantities were calculated in the random and interacting instanton liquid models,
and in all calculations one finds that
\begin{equation}1/M_u^2 \ll  1/M^2_{uudd}\end{equation}
In the interacting instanton liquid these quantities are
\begin{equation}{1 \over M_u^2}={1 \over (177 \, MeV)^2}, \,\,\,\,
	{1 \over M^2_{uudd}}\approx {1\over (103\, MeV)^2} \label{eqn_Muudd}
\end{equation}
The chief consequence of these substantial  deviations from mean field can be captured by 
a ``'t Hooft operator enhancement factor"

\begin{equation}
f_{\rm tHooft}\equiv {M^{-2}_{uudd} \over M_u^{-2}} \approx 3 
    \label{eqn_hooft_inhancement} 
\end{equation}

Ending this section, we briefly explain the values used to generate the plots in the "results" section. 
Since we decided to take a round $maximal$ value for the instanton diluteness parameter $\kappa\rightarrow 1$,
we have  $not$ included this additional enhancement factor (\ref{eqn_hooft_inhancement}). When the quark 
effective mass appears, in  the numerator or denominator, we use a round value of $M_Q= 400\, {\rm MeV}$.
This uniform but simplified approximation in all  our numerical plots, does not 
exclude the need for further  systematic lattice studies of the VEVs and their averages over 
the mesons of $all$  4-quark operators. 
To our knowledge the only such work, reporting the enhacement just mentioned on the lattice
is a rather old study in \cite{Faccioli:2003qz}. Since those
operators are widely used in hadronic phenomenology,
such studies are, in our opinion, long overdue.

\subsection{The zero mode contributions to the  vector form factor}

The zero mode part of the propagator (\ref{eqn_SZ})
can be schematically shown as two disconnected 
quark lines, with different chirality, ending in the instanton shown with the labels +, see
Fig.~\ref{fig_fr3}, with the  rules for these diagrams  given in Appendix~\ref{RULESX}. 
The corresponding contributions to a hard 
block  are

\begin{figure}[h!]
\begin{center}
\includegraphics[width=10cm]{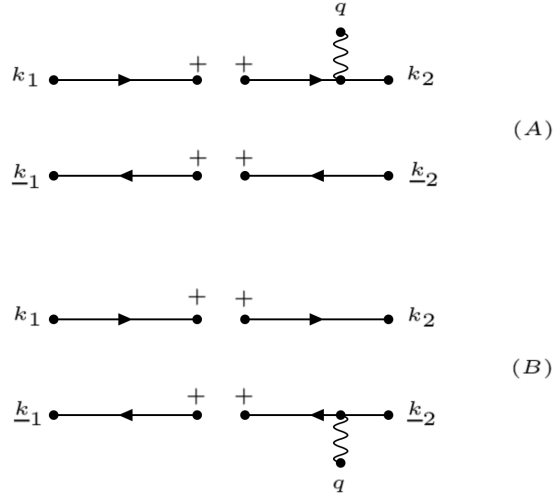}
\caption{A quark-antiquark pair absorbing or emitting a   vector photon in an instanton background labeled by $+$.}
\label{fig_fr3}
\end{center}
\end{figure}

\begin{eqnarray}
\label{VERTEX1}
{\rm Fig}.~\ref{fig_fr3}A=&&e_u\times \bigg<
\bigg(\bigg(u_R^\dagger(k_2)(-ik_2)\epsilon(q)\cdot\overline{V}(q,-k_2)\bigg)\bigg(\overline{\phi}(k_1)(i\overline{k}_1)u_L(k_1)\bigg)\bigg)\nonumber\\
&&\,\,\,\,\,\,\otimes
\bigg(\big(d_R^\dagger(\underline{k}_1)(i\underline{k}_1)K(\underline{k}_1)\bigg)
\bigg(\overline{\phi}(-\underline{k}_2)(-i\underline{k}_2)d_L(\underline{k}_2)\bigg)\bigg)\bigg>_U\nonumber\\
{\rm Fig}.~\ref{fig_fr3}B=&&e_{\overline d}\times\bigg<
\bigg(\bigg(u_R^\dagger(k_2)(-ik_2)K(-k_2)\bigg)\bigg(\overline{\phi}(k_1)(i\overline{k}_1)u_L(k_1)\bigg)\bigg)\nonumber\\
&&\,\,\,\,\,\,\otimes
\bigg(\big(d_R^\dagger(\underline{k}_1)(i\underline{k}_1)K(\underline{k}_1)\bigg)
\bigg(\epsilon(q)\cdot V(q, -\underline{k}_2)(-i\overline{\underline{k}}_2)d_L(\underline{k}_2)\bigg)\bigg>_U
\end{eqnarray}
where we have now made explicit the different flavors running in the vector vertex in Fig.~\ref{fig_fr3}, with the generic notation
$u_\alpha^i(k_{1,2})$ with flavor charge $e_u$, and ${d}_\alpha^{\dagger i}(\underline{k}_{1,2})$ with flavor charge $e_{\overline d}$,
and $\alpha=1,2$ for spin and $i=1, .., N_c$ for color. Note that we
have now attached a color index to each incoming and outgoing quark-antiquark line which is contracted with the pertinent $U$-matrix 
in the corresponding bracket. To carry the color averaging in (\ref{VERTEX1}) we use the identity

\begin{equation}
\label{NCOLOR}
\int dU\, U^i_\alpha U^{\dagger\beta}_jU^k_\gamma U^{\dagger \delta}_l=
\frac 1{N_c^2-1}\bigg(\delta^i_j\delta^k_l\delta^\beta_\alpha\delta^\delta_\gamma+\delta^i_l\delta^k_j\delta^\beta_\gamma\delta^\delta_\alpha\bigg)
-\frac 1{N_c(N_c^2-1)}\bigg(\delta^i_j\delta^k_l\delta^\beta_\gamma\delta^\delta_\alpha+\delta^i_l\delta^k_j\delta^\beta_\alpha\delta^\delta_\gamma\bigg)
\end{equation}
The result for the symmetrized vector insertion on the quark line  in  Fig.~\ref{fig_fr3}A  is

\begin{eqnarray}
\label{VV1}
&&{\rm Fig}.~(\ref{fig_fr3}A+\ref{fig_fr3}\tilde A)_{I}=i \frac{(2\pi\rho^{\frac 32})^4}{(-iM_Q\rho)^2}\times e_u\times\nonumber\\
&&\bigg(\frac 1{N_c^2-1}\bigg(u_R^\dagger(k_2)(\mathbb F_V(q,k_1)+\mathbb F_V(q,-k_2))u_L(k_1)\bigg)\bigg(d_R^\dagger (\underline{k}_1)d_L(\underline{k}_2)\bigg)\nonumber\\
&&+\frac 1{N_c^2-1}\bigg(u_R^\dagger(k_2)(\mathbb F_V(q,k_1)+\mathbb F_V(q,-k_2))d_L(\underline{k}_2)\bigg)\bigg(d_R^\dagger (\underline{k}_1)u_L(k_1)\bigg)\nonumber\\
&&-\frac 1{N_c(N_c^2-1)}  \bigg(u_{Ri}^\dagger(k_2)(\mathbb F_V(q,k_1)+\mathbb F_V(q,-k_2))u^j_L(k_1)\bigg)\bigg(d_{Rj}^\dagger (\underline{k}_1)d^i_L(\underline{k}_2)\bigg)\nonumber\\
&&-\frac 1{N_c(N_c^2-1)} \bigg( u_{Ri}^\dagger(k_2)(\mathbb F_V(q,k_1)+\mathbb F_V(q,-k_2))d^j_L(\underline{k}_2)\bigg)\bigg(d_{Rj}^\dagger (\underline{k}_1)u^i_L(k_1)\bigg)\bigg)\nonumber\\
\end{eqnarray}
with $M_Q$ the  constituent quark mass discussed earlier. 
To avoid cluttering the spin-color indices in the Weyl spinors have been omitted. 
Each of the L,R-Weyl spinor in the 
bracket is contracted over the dummy spin $\alpha=1,2$ and color $i=1, ..., N_c$ indices, unless the contraction is out-of-bracket in which case the pertinent
(color) index contraction is displayed. 
The I-subscript refers to the instanton contribution. The anti-instanton contribution follows from (\ref{VV1}) through the substitution $L\leftrightarrow R$.
The corresponding result for the symmetrized vector insertion on the anti-quark line  in  Fig.~\ref{fig_fr3}B  is

\begin{eqnarray}
\label{VV2}
&&{\rm Fig}.~(\ref{fig_fr3}B+\ref{fig_fr3}\tilde B)_{I}=i \frac{(2\pi\rho^{\frac 32})^4}{(-iM_Q\rho)^2}\times e_{\overline d}\times\nonumber\\
&&\bigg(\frac 1{N_c^2-1}\bigg(u_R^\dagger(k_2)u_L(k_1)\bigg)\bigg(d_R^\dagger (\underline{k}_1)
\mathbb (\mathbb F_V(q,\underline{k}_1)+\mathbb F_V(q,-\underline{k}_2))d_L(\underline{k}_2)\bigg)\nonumber\\
&&+\frac 1{N_c^2-1}\bigg(u_R^\dagger(k_2)d_L(\underline{k}_2)\bigg)\bigg(d_R^\dagger (\underline{k}_1)
\mathbb (\mathbb F_V(q,\underline{k}_1)+\mathbb F_V(q,-\underline{k}_2))u_L(k_1)\bigg)\nonumber\\
&&-\frac 1{N_c(N_c^2-1)} \bigg( u_{Ri}^\dagger(k_2)u^j_L(k_1)\bigg)\bigg(d_{Rj}^\dagger (\underline{k}_1)
\mathbb (\mathbb F_V(q,\underline{k}_1)+\mathbb F_V(q,-\underline{k}_2))d^i_L(\underline{k}_2)\bigg)\nonumber\\
&&-\frac 1{N_c(N_c^2-1)} \bigg( u_{Ri}^\dagger(k_2)d^j_L(\underline{k}_2)\bigg)\bigg(d_{Rj}^\dagger (\underline{k}_1)
\mathbb (\mathbb F_V(q,\underline{k}_1)+\mathbb F_V(q,-\underline{k}_2))u^i_L(k_1)\bigg)\bigg)\nonumber\\
\end{eqnarray}
with the substitution $L\leftrightarrow R$ for the anti-instanton contribution. 
The  spin-valued induced form factor

\begin{eqnarray}
\label{VV3}
\mathbb F_V(q, k)=\frac{\epsilon(q)\overline{k}-k\overline{\epsilon}(q)}{2k\cdot q}F(\rho\sqrt{q^2})
+\bigg(\frac{\epsilon(q)(\overline q+\overline k)-(q+k)\overline{\epsilon}(q)}{(k+q)^2}-
\frac{\epsilon(q)\overline{k}-k\overline{\epsilon}(q)}{2k\cdot q}\bigg)F(\rho\sqrt{(k+q)^2})\nonumber\\
\end{eqnarray}
simplifies when the quark line is taken on  mass-shell ($F(x)=xK_1(x)$)

\begin{equation}
\label{VV4}
\lim_{k^2\rightarrow 0}\,\mathbb F_V(q, k)=\frac {\epsilon(q)\overline{q}-q\overline{\epsilon}(q)}{q^2}\,F(\rho\sqrt{q^2})
=\epsilon_\mu(q)q_\nu(\sigma^\mu\overline{\sigma}^\nu-\sigma^\nu\overline{\sigma}^\mu)\frac{1}{q^2}\,F(\rho\sqrt{q^2})
\end{equation}


The full contribution to the {\it hard vector form factor} is (\ref{VV1}) plus (\ref{VV2}) weighted by the instanton averaged density $n/2$,
plus the corresponding anti-instanton contribution. The analytically continued result is

\begin{eqnarray}
\label{VFULL}
&&\epsilon_\mu(q)\mathbb V^\mu (k_1, k_2; \underline{k}_1, \underline{k}_2; q)=-\frac{8\kappa\pi^2}{M_Q^2}\times\nonumber\\
&&\bigg[e_u\times\bigg\{\frac 1{N_c^2-1}\bigg(\overline{u}_R(k_2)\mathbb (\mathbb F_V(q,k_1)+\mathbb F_V(q,-k_2))u_L(k_1)\bigg)\bigg(\overline{d}_R (\underline{k}_1)d_L(\underline{k}_2)\bigg)\nonumber\\
&&+\frac 1{N_c^2-1}\bigg(\overline{u}_R(k_2)\mathbb (\mathbb F_V(q,k_1)+\mathbb F_V(q,-k_2))d_L(\underline{k}_2)\bigg)\bigg(\overline{d}_R (\underline{k}_1)u_L(k_1)\bigg)\nonumber\\
&&-\frac 1{N_c(N_c^2-1)}  \bigg(\overline{u}_{Ri}(k_2)\mathbb (\mathbb F_V(q,k_1)+\mathbb F_V(q,-k_2))u^j_L(k_1)\bigg)\bigg(\overline{d}_{Rj}(\underline{k}_1)d^i_L(\underline{k}_2)\bigg)\nonumber\\
&&-\frac 1{N_c(N_c^2-1)} \bigg( \overline{u}_{Ri}(k_2)\mathbb (\mathbb F_V(q,k_1)+\mathbb F_V(q,-k_2))d^j_L(\underline{k}_2)\bigg)\bigg(\overline{d}_{Rj}(\underline{k}_1)u^i_L(k_1)\bigg)\bigg\}\nonumber\\
&&+e_{\overline d}\times\bigg\{\frac 1{N_c^2-1}\bigg(\overline{u}_R(k_2)u_L(k_1)\bigg)\bigg(\overline{d}_R (\underline{k}_1)\mathbb (\mathbb F_V(q,\underline{k}_1)+\mathbb F_V(q,-\underline{k}_2))d_L(\underline{k}_2)\bigg)\nonumber\\
&&+\frac 1{N_c^2-1}\bigg(\overline{u}_R(k_2)d_L(\underline{k}_2)\bigg)\bigg(\overline{d}_R(\underline{k}_1)\mathbb (\mathbb F_V(q,\underline{k}_1)+\mathbb F_V(q,-\underline{k}_2))u_L(k_1)\bigg)\nonumber\\
&&-\frac 1{N_c(N_c^2-1)} \bigg( \overline{u}_{Ri}(k_2)u^j_L(k_1)\bigg)\bigg(\overline{d}_{Rj}(\underline{k}_1)\mathbb (\mathbb F_V(q,\underline{k}_1)+\mathbb F_V(q,-\underline{k}_2))d^i_L(\underline{k}_2)\bigg)\nonumber\\
&&-\frac 1{N_c(N_c^2-1)} \bigg( \overline{u}_{Ri}(k_2)d^j_L(\underline{k}_2)\bigg)\bigg(\overline{d}_{Rj} (\underline{k}_1)\mathbb (\mathbb F_V(q,\underline{k}_1)+\mathbb F_V(q,-\underline{k}_2))u^i_L(k_1)\bigg)\bigg\}\bigg]
+L\leftrightarrow R\nonumber\\
\end{eqnarray}
We dropped a factor of $i$ in switching from the S-matrix to the T-matrix element in the identification of the vector vertex.
For the free spinors, we made the substitutions  $u_{L,R}^\dagger\rightarrow \overline{u}_{L,R}$ and $d_{R,L}^\dagger\rightarrow \overline{d}_{R,L}$
when analytically going to  Minkowski space  as in (\ref{FREEX}), with the standard Minkowski relation between Dirac and Weyl spinors

\begin{equation}
u=\frac {1+\gamma_5}2 \,u+\frac {1-\gamma_5}2 \,u\equiv u_R+u_L
\end{equation}
and similarly for $d\equiv d_R+d_L$. 
More explicitly, the first contribution in (\ref{VFULL})  due to the instanton can be recast in the form 

\begin{equation}
\label{FIRST}
\bigg(\frac{N_c}{N_c^2-1}\bigg)\bigg[e_u\mathbb F_P(\rho\sqrt{-q^2})\,
 {\overline{u}(k_2){\frac{i\epsilon_\mu(q)\sigma^{\mu\nu}q_{\nu}}{2M_Q}\frac {(1-\gamma_5)}2u(k_1)}}\bigg]
\bigg[\frac {(2\pi\rho)^2}{M_Q}\overline{d} (\underline{k}_1)\frac{(1-\gamma_5)}2d(\underline{k}_2)\bigg]
\end{equation}
with the spin-valued  matrix $\sigma^{\mu\nu}=\frac i{2} [\gamma^\mu, \gamma^\nu]$.
The first bracket in (\ref{FIRST}) shows the vector interaction
with a chirally flipped u-quark which is purely magnetic. The  corresponding  Pauli form factor  is 

\begin{equation}
\label{FIRST1}
\mathbb F_P(\rho\sqrt{-q^2})=\frac{8\kappa}{N_c}\frac{K_1(\rho\sqrt{-q^2})}{\rho\sqrt{-q^2}}\,
\end{equation}
(We again recall  that $-q^2=Q^2>0$.)
The second bracket is the chirality flipped d-quark through the instanton zero mode. All contributions in (\ref{VFULL}) are of this type. 
(Note that if the amplitude  is evaluated  at near-zero $Q$, this instanton term contributes 
to the constituent quark magnetic moment, see~\cite{Kochelev:2003cp}.)

\subsection{The zero mode contributions to the scalar form factor}

This  contribution to the hard scalar form factor follows a similar reasoning as in the previous subsection, with two modifications:
1/ in the form factors (\ref{VV3}-\ref{VV4}) the polarization $\epsilon(q)\rightarrow 1$; 2/ in the contributions (\ref{VV1}-\ref{VV2}) 
there is no chirality flip on the leg with the scalar form factor insertion.  With this in mind, and making use of the LSZ amputations
(\ref{ZM-ZM}-\ref{ZM-ZM-X}) we have 

\begin{eqnarray}
\label{SS1}
&&{\rm Fig}.~(\ref{fig_fr3}A+\ref{fig_fr3}\tilde A)_{S,I}=-i\frac{(2\pi\rho^{\frac 32})^4}{(-iM_Q\rho)^2}\times \lambda_u\times \nonumber\\
&&\bigg[\frac 1{N_c^2-1}\bigg(u_L^\dagger(k_2){\mathbb{F}}_S(q,-k_2)u_L(k_1)+u_R^\dagger(k_2)\overline{\mathbb F}_S(q,k_1)u_R(k_1)
\bigg)\bigg(d_R^\dagger (\underline{k}_1)d_L(\underline{k}_2)\bigg)\nonumber\\
&&+\frac 1{N_c^2-1}
\bigg(\bigg(u_L^\dagger(k_2)\mathbb F_S(q,-k_2)d_L(\underline{k}_2)\bigg)\bigg(d_R^\dagger (\underline{k}_1)u_L(k_1)\bigg)+
\bigg(d_R^\dagger(\underline{k}_1)\overline{\mathbb F}_S(q,k_1)u_R({k}_1)\bigg)\bigg(u_R^\dagger ({k}_2)d_L(\underline{k}_2)\bigg)\bigg)
\nonumber\\
&&-\frac 1{N_c(N_c^2-1)} 
\bigg(\bigg( u_{Li}^\dagger(k_2)\mathbb F_S(q,-k_2)u^j_L(\underline{k}_2)\bigg)\bigg(d_{Rj}^\dagger (\underline{k}_1)u^i_L(k_1)\bigg)
+\bigg( d_{Rj}^\dagger(\underline{k}_1)\overline{\mathbb F}_S(q,k_1)u^i_R({k}_1)\bigg)\bigg(u_{Ri}^\dagger ({k}_2)d^j_L(\underline{k}_2)\bigg)\bigg)\nonumber\\
&&-\frac 1{N_c(N_c^2-1)} \bigg( u_{Li}^\dagger(k_2)\overline{\mathbb {F}}_S(q,-k_2)u^j_L(k_1)
+u_{Ri}^\dagger(k_2)\overline{\mathbb F}_S(q,k_1)u^j_R(k_1)\bigg)\bigg(d_{Rj}^\dagger (\underline{k}_1)d^i_L(\underline{k}_2)\bigg)
\bigg]\nonumber\\
\end{eqnarray}
with a scalar charge $\lambda_u$, and  the scalar form factors

\begin{eqnarray}
\label{SS4}
{\mathbb{F}}_S(q,-k_2)&=&\frac{\overline{k}_2}{M_Q^2}F(\rho\sqrt{(q+k_2)^2})\rightarrow \frac{\overline{k}_2}{M_Q^2}F(\rho\sqrt{-q^2})\nonumber\\
{\overline{\mathbb{F}}}_S(q,k_1)&=&\frac{{k_1}}{M_Q^2}F(\rho\sqrt{(q-k_1)^2})\rightarrow \frac{{k_1}}{M_Q^2}F(\rho\sqrt{-q^2})
\end{eqnarray}
 The  S-subscript refers to the scalar vertex, and the 
I-subscript referring to the instanton contribution. The anti-instanton contribution follows from (\ref{VV1}) through the substitution $L\leftrightarrow R$.
The rightmost identity in (\ref{SS4}) is the leading contribution in the  $k^2\rightarrow 0$ limit, with $F(x)=xK_1(x)$ after analytical continuation to Minkowski space.
The corresponding result for the symmetrized scalar insertion on the anti-quark line  in  Fig.~\ref{fig_fr3}B  is

\begin{eqnarray}
\label{SS2}
&&{\rm Fig}.~(\ref{fig_fr3}B+\ref{fig_fr3}\tilde B)_{S,I}=-i\frac{(2\pi\rho^{\frac 32})^4}{(-iM_Q\rho)^2}\times \lambda_{\overline d}\times \nonumber\\
&&\bigg[\frac 1{N_c^2-1}\bigg(u_R^\dagger(k_2)u_L(k_1)\bigg)
\bigg(\bigg(d_L^\dagger (\underline{k}_1)\mathbb F_S(q,-\underline{k}_2)d_L(\underline{k}_2)\bigg)
+\bigg(d_R^\dagger (\underline{k}_1)\overline{\mathbb F}_S(q,\underline{k}_1)d_R(\underline{k}_2)\bigg)\bigg)
\nonumber\\
&&+\frac 1{N_c^2-1}\bigg(\bigg(u_L^\dagger(k_2)\mathbb F_S(q, -\underline{k}_2)d_L(\underline{k}_2)\bigg)\bigg(d^\dagger_R(\underline{k}_1)u_L(k_1)\bigg)
+\bigg(d^\dagger(\underline{k}_1)\overline{\mathbb F}_S(q, \underline{k}_1)u_L(k_1)\bigg)\bigg(u^\dagger_R(k_2)d_L(\underline{k}_2)\bigg)\bigg)\nonumber\\
&&-\frac 1{N_c(N_c^2-1)} \bigg( u^\dagger_{Li}(k_2)\mathbb F_S(q, -\underline{k}_2)d^j_{L}(\underline{k}_2)\bigg)
\bigg(d^\dagger_{Rj}(\underline{k}_1)u_L^i(k_1)\bigg)
+\bigg(d^\dagger_{Lj}(\underline{k}_1)\overline{\mathbb F}_S(q, \underline{k}_1)u^i_{L}(k_1)\bigg)
\bigg(u^\dagger_{Ri}(k_2)d_{L}^j(\underline{k}_2)\bigg)
\nonumber\\
&&-\frac 1{N_c(N_c^2-1)} \bigg( u^\dagger_{Ri}(k_2)u_L^j(k_1)\bigg)
\bigg(d_{Rj}^\dagger(\underline{k}_1)\mathbb F_S(q, -\underline{k}_2) d^i_{L}(\underline{k}_2)
+d^\dagger_{Lj}(\underline{k}_1)\overline{\mathbb F}_S(q, \underline{k}_1)d^i_{L}(\underline{k}_2)\bigg)
\bigg]\nonumber\\
\end{eqnarray}
The anti-instanton contribution follows through the substitution $L\leftrightarrow R$.

The full contribution  to the {\it  hard scalar form factor} is (\ref{SS1}) plus (\ref{SS2}) weighted again by the instanton averaged density
$n/2$, plus the corresponding anti-instanton contribution. The analytically continued result to Minkowski space  is

\begin{eqnarray}
\label{SFULL}
&&\mathbb S (k_1, k_2; \underline{k}_1, \underline{k}_2; q)= +\frac{8\kappa\pi^2}{M_Q^2}\times \nonumber\\
&&\bigg[\lambda_u\times\bigg\{
\frac 1{N_c^2-1}\bigg(\overline{u}_L(k_2){\mathbb{F}}_S(q,-k_2)u_L(k_1)+\overline{u}_R(k_2)\overline{\mathbb F}_S(q,k_1)u_R(k_1)
\bigg)\bigg(\overline{d}_R(\underline{k}_1)d_L(\underline{k}_2)\bigg)\nonumber\\
&&+\frac 1{N_c^2-1}
\bigg(\bigg(\overline{u}_L(k_2)\mathbb F_S(q,-k_2)d_L(\underline{k}_2)\bigg)\bigg(\overline{d}_R (\underline{k}_1)u_L(k_1)\bigg)+
\bigg(\overline{d}_R(\underline{k}_1)\overline{\mathbb F}_S(q,k_1)u_R({k}_1)\bigg)\bigg(\overline{u}_R({k}_2)d_L(\underline{k}_2)\bigg)\bigg)
\nonumber\\
&&-\frac 1{N_c(N_c^2-1)} 
\bigg(\bigg( \overline{u}_{Li}(k_2)\mathbb F_S(q,-k_2)d^j_L(\underline{k}_2)\bigg)\bigg(\overline{d}_{Rj}(\underline{k}_1)u^i_L(k_1)\bigg)
+\bigg( \overline{d}_{Rj}(\underline{k}_1)\overline{\mathbb F}_S(q,k_1)u^i_R({k}_1)\bigg)\bigg(\overline{u}_{Ri}({k}_2)d^j_L(\underline{k}_2)\bigg)\bigg)\nonumber\\
&&-\frac 1{N_c(N_c^2-1)} \bigg( \overline{u}_{Li}(k_2)\overline{\mathbb {F}}_S(q,-k_2)u^j_L(k_1)
+\overline{u}_{Ri}(k_2)\overline{\mathbb F}_S(q,k_1)u^j_R(k_1)\bigg)\bigg(\overline{d}_{Rj}(\underline{k}_1)d^i_L(\underline{k}_2)\bigg)\bigg\}\nonumber\\
&&+\lambda_{\overline d}\times\bigg\{\frac 1{N_c^2-1}\bigg(\overline{u}_R(k_2)u_L(k_1)\bigg)
\bigg(\bigg(\overline{d}_L (\underline{k}_1)\mathbb F_S(q,-\underline{k}_2)d_L(\underline{k}_2)\bigg)
+\bigg(\overline{d}_R(\underline{k}_1)\overline{\mathbb F}_S(q,\underline{k}_1)d_R(\underline{k}_2)\bigg)\bigg)
\nonumber\\
&&+\frac 1{N_c^2-1}\bigg(\bigg(\overline{u}_L(k_2)\mathbb F_S(q, -\underline{k}_2)d_L(\underline{k}_2)\bigg)\bigg(\overline{d}_R(\underline{k}_1)u_L(k_1)\bigg)
+\bigg(\overline{d}(\underline{k}_1)\overline{\mathbb F}_S(q, \underline{k}_1)u_L(k_1)\bigg)\bigg(\overline{u}_R(k_2)d_L(\underline{k}_2)\bigg)\bigg)\nonumber\\
&&-\frac 1{N_c(N_c^2-1)} \bigg( \overline{u}_{Ri}(k_2)\mathbb F_S(q, -\underline{k}_2)d^j_{R}(\underline{k}_2)\bigg)
\bigg(\overline{d}_{Rj}(\underline{k}_1)u_L^i(k_1)\bigg)
+\bigg(\overline{d}_{Lj}(\underline{k}_1)\overline{\mathbb F}_S(q, \underline{k}_1)u^i_{L}(k_1)\bigg)
\bigg(\overline{u}_{Ri}(k_2)d_{L}^j(\underline{k}_2)\bigg)
\nonumber\\
&&-\frac 1{N_c(N_c^2-1)} \bigg( \overline{u}_{Ri}(k_2)u_L^j(k_1)\bigg)
\bigg(\overline{d}_{Rj}(\underline{k}_1)\mathbb F_S(q, -\underline{k}_2) d^i_{R}(\underline{k}_2)
+\overline{d}_{Lj}(\underline{k}_1)\overline{\mathbb F}_S(q, \underline{k}_1)d^i_{L}(\underline{k}_2)\bigg)
\bigg\}\bigg]
+L\leftrightarrow R
\end{eqnarray}
again after dropping a factor of $i$ in going from the S-matrix to the T-matrix element in the identification of the scalar vertex.
More explitly, using the limiting form factors (\ref{SS4})  the first contribution in (\ref{SFULL})  can be recast in the compact form

\begin{equation}
\label{FIRST2}
 \bigg(\frac{N_c}{N_c^2-1}\bigg)\bigg[\lambda_u \tilde{\mathbb F}_P(\rho\sqrt{-q^2})\,
 {\overline{u}(k_2)\bigg({\frac{k_{+\mu}\gamma^\mu+k_{-\mu}\gamma^\mu\gamma_5}{2M_Q}\bigg)u(k_1)}}\bigg]
\bigg[\frac {(2\pi\rho)^2}{M_Q}\overline{d} (\underline{k}_1)\frac{(1-\gamma_5)}2d(\underline{k}_2)\bigg]
\end{equation}
with $k_{\pm}=(k_1\pm k_2)/2$ (not to be confused with the light cone momenta) and the scalar form factor

\be
\tilde{\mathbb F}_P(x)=\frac{x^2\,\mathbb F_P(x)}{(M_Q\rho)^2}=\frac{8\kappa}{N_c}\frac{F(x)}{(M_Q\rho)^2}=\frac{8\kappa}{N_c}\frac{xK_1(x)}{(M_Q\rho)^2}
\ee
Similar reductions hold for the other contractions.  In Appendix~\ref{redux}  we  give an alternative but simplified derivation of
 (\ref{FIRST2}) before analytical continuation and color averaging.

The zero mode instanton plus anti-instanton contribution to the pion vector form factor  vanishes 
for a vanishing tensor DA amplitude $\varphi^T_\pi(x)=0$,

\begin{eqnarray}
\label{TVP}
&&-(e_u+e_{\overline d})\,\mathbb F_P(\rho\sqrt{-q^2})\,\bigg(\frac{N_c}{N^2_c(N_c+1)}\bigg)\int_0^1dx_1dx_2\nonumber\\
&&\times{\rm Tr}\bigg[\bigg(\frac{i\epsilon_\mu(q)\sigma^{\mu\nu}q_{\nu}}{2M_Q}\frac {(1-\gamma_5)}2\bigg)
\bigg(\frac{if_\pi}4 \gamma^5(\slashed{p}\,\varphi_{\pi}(x_1)-\chi_\pi\varphi_{\pi}^P(x_1))\bigg)\nonumber\\
&&\qquad\times\bigg(\frac {(2\pi\rho)^2}{M_Q}\frac{(1-\gamma_5)}2\bigg)
\gamma^0\bigg(\frac{if_\pi}4 \gamma^5(\slashed{p}^\prime\,\varphi_{\pi}(x_2)-\chi_\pi\varphi_{\pi}^P(x_2))\bigg)^\dagger\gamma^0
\bigg]+{\rm L}\rightarrow {\rm R}=0\nonumber\\
\end{eqnarray}
which is seen to spin trace to zero.  The color factor follows directly from the color contraction of (\ref{NCOLOR}) in a  colorless meson state

\be
\frac {\delta^{ik}}{N_c}\bigg[\bigg(\frac{\delta^i_j\delta^k_l}{N_c^2-1}-\frac{\delta^i_l\delta^k_j}{N_c(N_c^2-1)}\bigg)\delta_\alpha^\beta\delta_\gamma^\delta\bigg]
\frac {\delta^{jl}}{N_c}=\frac{\delta_\alpha^\beta\delta_\gamma^\delta}{N_c^2(N_c+1)}
\ee
The  non-vanishing result with the tensor DA amplitude is given in the results section above.
Similarly, the zero mode  instanton plus anti-instanton contribution to the pion scalar form factor  is given by

\begin{eqnarray}
\label{TSP}
&&-(\lambda_u+\lambda_{\overline d})\,\tilde{\mathbb F}_P(\rho\sqrt{-q^2})\,\bigg(\frac{N_c}{N^2_c(N_c+1)}\bigg)\int_0^1dx_1dx_2\nonumber\\
&&\times{\rm Tr}\bigg[\bigg(\frac{k_{+\mu}\gamma^\mu+k_{-\mu}\gamma^\mu\gamma_5}{2M_Q}\bigg)
\bigg(\frac{if_\pi}4 \gamma^5\bigg(\slashed{p}\,\varphi_{\pi}(x_1)
-\chi_\pi\varphi_{\pi}^P(x_1)+i\chi_\pi\sigma_{\alpha\beta}\frac{p^\alpha p^{\prime \beta}}{p\cdot p^\prime}\frac{\varphi_\pi^{T\prime}(x_1)}6\bigg)\bigg)\nonumber\\
&&\times\bigg(\frac{(2\pi\rho)^2}{M_Q}\frac {(1-\gamma_5)}2\bigg)
\gamma^0\bigg(\frac{if_\pi}4 \gamma^5\bigg(\slashed{p}^\prime\,\varphi_{\pi}(x_2)
-\chi_\pi\varphi_{\pi}^P(x_2)+i\chi_\pi\sigma_{\alpha\beta}\frac{p^{\prime\alpha} p^{ \beta}}{p\cdot p^\prime}\frac{\varphi_\pi^{T\prime}(x_2)}6\bigg)\bigg)^\dagger\gamma^0\bigg]
\nonumber\\
&&+{\rm L}\rightarrow {\rm R}
\end{eqnarray}
with the result of all tracing given in the results section above.

 \subsection{The zero mode  contribution to the graviton and dilaton\\ form factor of the pion }
 
 The mixed zero-mode and  non-zero mode contribution ($^\prime$t Hooft vertex)  follows from (\ref{VFULL}) with (\ref{FIRST}) now reading

\begin{equation}
\label{FIRSTEM}
\bigg(\frac{N_c}{N_c^2-1}\bigg)\bigg[\mathbb F_P(Q\rho)\,
 {\overline{u}(k_2){\frac{i (k_1+k_2)^{[\mu}\sigma^{\nu]_+\tau}q_{\tau}}{2M_Q}\frac {(1-\gamma_5)}2u(k_1)}}\bigg]
\bigg[\frac {(2\pi\rho)^2}{M_Q}\overline{d} (\underline{k}_1)\frac{(1-\gamma_5)}2d(\underline{k}_2)\bigg]
\end{equation}
The  mixed-zero mode and non-zero mode  vertex  (\ref{FIRSTEM}) contributes equally to $00$ and $\mu\mu$
in the Breit frame,  with the result
 
 \be
 T^\pi_{00d}(Q^2)=T^\pi_{\mu\mu d}(Q^2)=\frac 1{N_c^2(N_c+1)}\bigg(\frac{16\kappa \pi^2 f_\pi^2\chi_\pi^2}{3M_Q^2}\bigg)
 \left<(Q\rho)K_1(Q\rho)\right>
 \int dx_1dx_2\,\varphi_\pi(x_1)\frac{\varphi_\pi^{T\prime}(x_2)}6
 \ee
 which is seen to vanish.

\subsection{The zero mode contribution to the transverse rho  form factors}

The instanton plus anti-instanton contribution to the transversely polarized  vector form factor is

\begin{eqnarray}
\label{TVR0}
&&V_d^\rho(Q^2)=-(e_u+e_{\overline d})\,\mathbb F_P(\rho Q)\,\bigg(\frac{N_c}{N_c^2(N_c+1)}\bigg)\int_0^1dx_1dx_2\nonumber\\
&&\times{\rm Tr}\bigg[\bigg(\frac{i\sigma^{\mu\nu}q_{\nu}}{2M_Q}\frac {(1-\gamma_5)}2\bigg)
\bigg(\frac  i4 \slashed\epsilon_T\bigg(f_\rho m_\rho\varphi_\rho(x_1) -f_\rho^T\slashed{p}\varphi_\rho^T(x_1)\bigg)
+  \frac{f_\rho m_\rho}{4p\cdot p^\prime}\epsilon_{\mu\nu\rho\sigma}\gamma^\mu\gamma_5\epsilon^\nu p^\rho p^{\prime\sigma}\frac{\varphi_\rho^{A\prime}(x_1)}4\bigg)
\nonumber\\
&&\times\bigg(\frac {(2\pi\rho)^2}{M_Q}\frac{(1-\gamma_5)}2\bigg)
\gamma^0\bigg(\frac  i4 \slashed\epsilon_T\bigg(f_\rho m_\rho\varphi_\rho(x_2) -f_\rho^T\slashed{p}^\prime\varphi_\rho^T(x_2)\bigg)
+  \frac{f_\rho m_\rho}{4p\cdot p^\prime}\epsilon_{\mu\nu\rho\sigma}\gamma^\mu\gamma_5\epsilon^\nu p^{\prime\rho} p^{\sigma}\frac{\varphi_\rho^{A\prime}(x_2)}4\bigg)^\dagger\gamma^0
\nonumber\\
&&+{\rm L}\rightarrow {\rm R}\nonumber\\
&&=(e_u+e_{\overline d})\,\mathbb F_P(\rho\sqrt{-q^2})\,\bigg(\frac{N_c}{N_c^2(N_c+1)}\bigg)\nonumber\\
&&\times {\rm Tr}\bigg[\bigg(\frac{i\sigma^{\mu\nu}q_{\nu}}{2M_Q}\bigg)
\bigg(+\frac i4 { f^T_\rho} \slashed{p}\,\slashed{\epsilon}_T\bigg)
\left(\frac {(2\pi\rho)^2}{M_Q}\right)
\bigg(-\frac i4 \slashed{\epsilon}^{\prime *}_T\,{f^T_\rho} \slashed{p}^\prime\bigg)\bigg]\nonumber\\
\end{eqnarray}
Unwinding the last trace gives

\begin{eqnarray}
\label{TVR}
V_d^\rho(Q^2)=-(e_u+e_{\overline d})\bigg((p^\mu+p^{\prime \mu})\,\epsilon_T^{\prime *}\cdot\epsilon_T\bigg)\,
\bigg[\left<(\rho Q)\,{K_1(\rho Q)}\right>\,\bigg(\frac{2\kappa\pi^2 }{N_c^2(N_c+1)}\frac{ {\tilde f_\rho}^2 }{M^2_Q}\bigg)\bigg]
\end{eqnarray}
which is to be compared to the hard perturbative contribution (\ref{VECTOR1}).
The instanton plus anti-instanton contribution to the transversely polarized rho  scalar form factor is given by

\begin{eqnarray}
\label{TSR}
&&S_d^\rho(Q^2)=-(\lambda_u+\lambda_{\overline d})\,\tilde{\mathbb F}_P(\rho Q)\,\bigg(\frac{N_c}{N_c^2(N_c+1)}\bigg)\int_0^1dx_1dx_2\nonumber\\
&&\times{\rm Tr}\bigg[\bigg(\frac{k_{+\mu}\gamma^\mu+k_{-\mu}\gamma^\mu\gamma_5}{2M_Q}\bigg)
\bigg(\frac  i4 \slashed\epsilon_\perp\bigg(f_\rho m_\rho\varphi_\rho(x_1) -f_\rho^T\slashed{p}\varphi_\rho^T(x_1)\bigg)
+  \frac{f_\rho m_\rho}{4p\cdot p^\prime}\epsilon_{\mu\nu\rho\sigma}\gamma^\mu\gamma_5\epsilon^\nu p^\rho p^{\prime\sigma}\frac{\varphi_\rho^{A\prime}(x_1)}4\bigg)
\nonumber\\
&&\times\frac{(2\pi\rho)^2}{M_Q}\bigg(\frac{(1-\gamma_5)}2\bigg)
\gamma^0\bigg(\frac  i4 \slashed\epsilon_\perp\bigg(f_\rho m_\rho\varphi_\rho(x_2) -f_\rho^T\slashed{p}\varphi_\rho^T(x_2)\bigg)
+  \frac{f_\rho m_\rho}{4p\cdot p^\prime}\epsilon_{\mu\nu\rho\sigma}\gamma^\mu\gamma_5\epsilon^\nu p^{\prime\rho} p^{\sigma}\frac{\varphi_\rho^{A\prime}(x_2)}4\bigg)^\dagger\gamma^0
\bigg]
\nonumber\\
&&+{\rm L}\rightarrow {\rm R}
\end{eqnarray}
with the full tracing result given in the results section above.

\subsection{Averaging over the instanton size distribution} \label{sec_averaging}

In the expressions above and for simplicity, we have used a single 
value of the instanton size $\rho$.
For many estimates it is sufficient to use  its r.m.s.
value of about $0.30\, {\rm fm}$. Yet in cases in which
a large momentum transfer is involved, the 
shape of the distribution over $\rho$ becomes important, especially its tail toward small sizes. Fortunately, at small $\rho$ the effective
coupling $\alpha_s(\rho)$ is small, the action is
large and semiclassical theory get more reliable.

Still, one needs the full shape of the distribution,
to get a proper averaging. The
instanton size distribution in the QCD vacuum  has been derived from
lattice works, using various degree of ``cooling"
methods, e.g. \cite{Hasenfratz:1999ng}. We will not dwell on the details of this distribution, 
and we will not get involved in the theoretical aspects of the large  large-size instantons for
which we refer to e.g. ref.\cite{Shuryak:1999fe}.  Here we make use of the interpolating formula

\begin{equation}
\label{dn_dist}
dn(\rho) \sim  {d\rho \over \rho^5}\big(\rho \Lambda_{QCD} \big)^{b_{QCD}} \, e^{-2\pi \sigma \rho^2}
\end{equation}
in which the pre-exponential  is the semi-classical contribution corrected to  one-loop with
$b_{QCD}=11N_c/3-2N_f/3\approx 9$. The exponent
is model-dependent,  with 
 $\sigma$ the QCD string tension. 
(It is  proportional to
the dual magnetic condensate, that of Bose-condensed monopoles, but we prefer the expression with the string tension which is experimentally well
determined to be  $\sigma\approx 0.42 \, {\rm GeV}^2$.)

\begin{figure}[h!]
	\begin{center}
		\includegraphics[width=8cm]{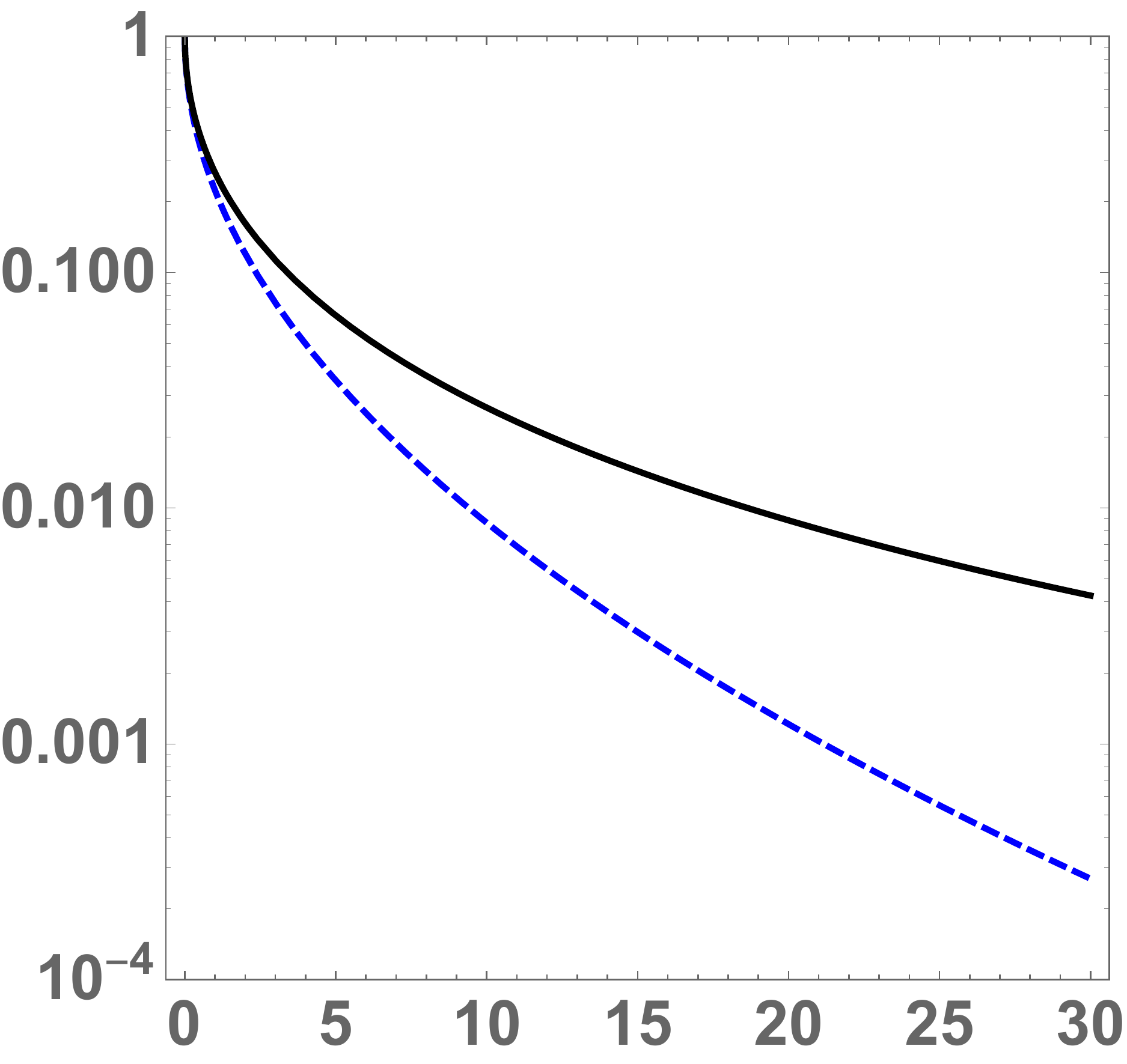}
		\caption{The blue dashed line is 
	the exponential function $e^{-Q \rho_{r.m.s.}}$ of momentum transfer $Q$, and the solid line
	is the same function averaged over the instanton size distribution, plotted versus $Q^2\, ({\rm GeV}^2)$.	
	}
		\label{fig_exp_over_sizes}
	\end{center}
\end{figure}

In Fig.\ref{fig_exp_over_sizes} we show the effect of
averaging over the instanton size distribution. We take a simple typical exponential dependence
on the momentum transfer, and compare it to 
its version after the instanton size averaging

\begin{equation}
\langle e^{-Q \rho} \rangle_\rho =
{\int dn(\rho) \,e^{-Q \rho}\over \int dn(\rho) }
\end{equation}
As one can see, at small $Q$ the two curves coincide,
but at large  $Q$ they differ significantly. The small-size instantons become more important in this limit, and the exponential decay with $Q$ changes to an inverse power.

All of the Bessel functions $K_i(Q\rho)$ appearing above in the instanton-induced form factors  behave as $\sim  e^{-Q\rho}$
at large $Q$, so the result of their averaging 
over the instanton sizes is similar to what is shown in Fig.\ref{fig_exp_over_sizes}.
Yet, when we performed  the 
instanton size averaging of the functions $F_{1,2}$,
as given in  (\ref{eqn_F2}), we found
that their corrections due to averaging differ substantially, resulting in significant changes in the outcome.

\section{Mesonic light-front  distribution amplitudes} \label{sec_wfs}

\subsection{Brief history of mesonic DA's and exclusive processes }

In the pioneering works on theory of exclusive QCD processes~\cite{Chernyak:1977as,Lepage:1979zb,Efremov:1979qk}, 
most of the general structure and observations were made. The  key element of  the theory  is  the $factorization$ into
a  hard block and a  light-front DA's, with the latter containing most of the information about the 
soft physics at  a scale much below the hard scale $Q^2$. 

The DA's are defined as the hadron wave function integrated over the transverse momentum. They are 
traditionally defined via bilocal light cone operators, 
classified in  the framework of a  {\em twist expansion}, the leading twist 2, and higher, with extra powers of $1/Q$.
 Twist is spin minus dimension. This theory originated
from DIS in which  moments of  PDFs are matrix elements of the  leading twist operators, containing only bilinear quark (or gluon)  fields. Higher twist operators, specified for unpolarized and polarized DIS in \cite{Shuryak:1981kj,Shuryak:1981pi}, contain important further
information about the nucleon structure, such as 
 correlations between quark and gluon fields, or quark-quark correlation via four-fermion operators.

However, unlike in DIS, the exclusive processes are studied at what we call  a  ``semi-hard"
 domain, in which one cannot expect the twist expansion to converge. In particular,  as pointed out early on by
 Geshkenbein and Terent'ev  \cite{Geshkenbein:1982zs},  the 
 twist-3 DA's of the pion are numerically enhanced, 
 so that their contribution may  in fact be larger than
 that of the leading twist, in the semi-hard $Q^2$ range of  interest.  
 
  Even more academic is the discussion of the asymptotic
  limit,  in which not powers of $Q^2$, but powers of log are considered to be large, i.e.  ${\rm log}(Q^2/\Lambda_{QCD}^2)\gg 1$.
When  perturbative processes of gluon radiation are included,  these logs sum into calculable {\it anomalous dimension} of various operators.  
So when the log is considered to be large, only the  leading contribution survives. 

Technically, the 
DA's  are decomposed into Gegenbauer polynomials,  and the  so called ``asymptotic wave function" corresponds to the lowest order polynomial. 
    \begin{equation}
    \varphi_\pi \rightarrow  \varphi^{\rm asymptotic}_\pi(\xi)=\frac 34 (1-\xi^2) =6x\bar x
     \label{phi_asymptotic} 
     \end{equation} 
Needless to say, this limit is very far from the 
realistic kinematic range of interest. Therefore we 
will neither use ``asymptotic wave functions", nor
restrict our analysis to the leading twist DA's.
We rather focus on the chiral structure
of the DA's, making sure that all possible and large contributions are
included.
For phenomenological purposes it is sufficient to
 consider a set of DA's approximated by the simple analytic form 
 \begin{equation}
 \varphi_\pi(\xi,p) = {\Gamma(3/2+p) \over \sqrt{\pi} \Gamma(1+p)}(1-\xi^2)^p   ={6^p\Gamma(3/2+p) \over \sqrt{\pi} \Gamma(1+p)}\,(x\bar x)^p
 \end{equation} 
The case $p=1$ is the ``asymptotic" distribution,  while the case $p=0$ is called ``flat". Several authors  have used an intermediate case $p=1/2$ called ``semicircular".

\begin{figure}[h!]
	\begin{center}
		\includegraphics[width=10cm]{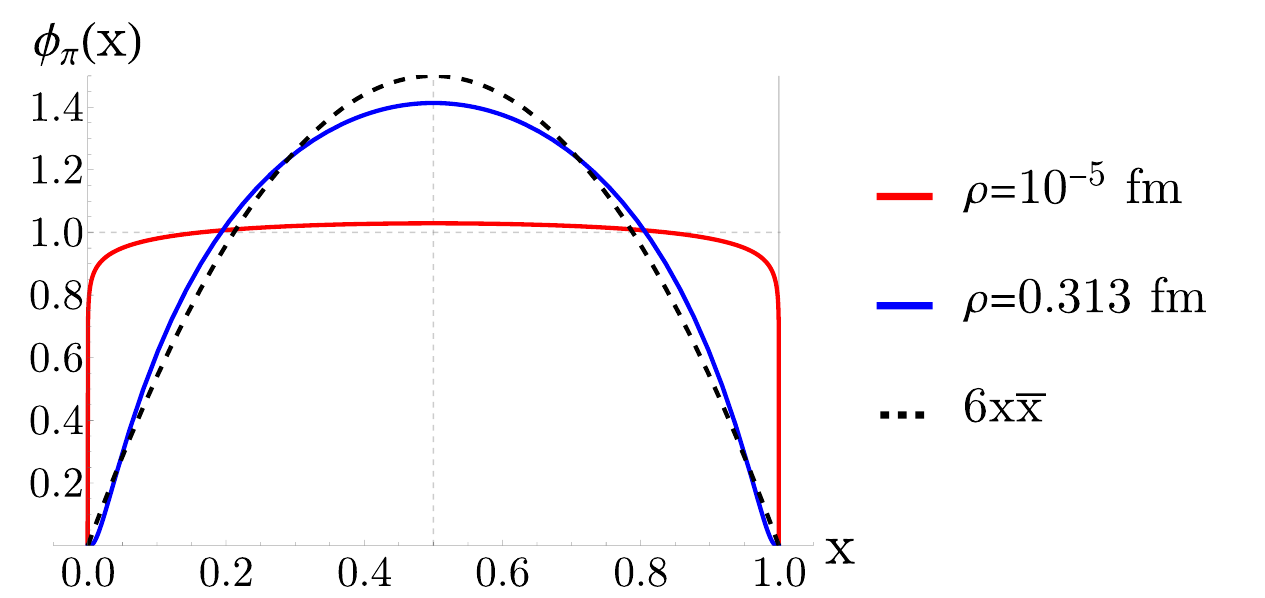}
		\includegraphics[width=8cm]{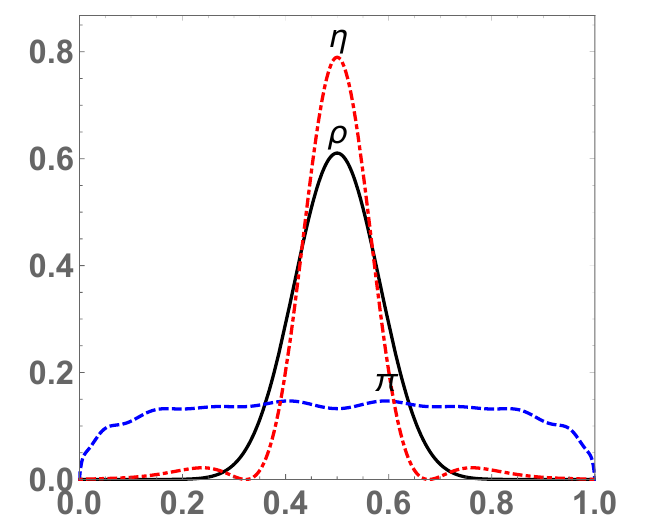}
		\caption{Upper plot: the pion light front distribution function from \cite{Kock:2020frx}. 
			Lower plot:  the light meson light front distribution functions from~\cite{Shuryak:2019zhv}.}
		\label{fig_pionwf}
	\end{center}
\end{figure}

 The pion is a particular particle, a Nambu-Goldstone mode,   and therefore
  its properties one can calculate in any theory in which  chiral symmetry gets spontaneously  broken. Historically  the NJL model
  and its nonlocal versions (some related with the  instanton liquid model) have been used to calculate the pion light-front wave function~\cite{Broniowski:2017wbr,Petrov:1997ve,Anikin:2000bn}. Before we briefly discuss the results of ``realistic" models, related to larger set of hadronic wave functions,
  let us introduce some extreme cases.    
  For example, in \cite{hep-ph/0207266} a ``flat"  pion wave function was
 used  as an ``initial condition" for radiative evolution.   
Some typical  shapes of the pion and other light meson wave functions stemming from some recent works,  are  shown in Fig.\ref{fig_pionwf}.

In  contrast, in 1980's Chernyak and collaborators \cite{Chernyak:1983ej}, using the QCD sum rules, arrived at the pion wave function 
\begin{equation}
\varphi_{CZ}(x) = 30x(1-x)(2x-1)^2 \label{eqn_double_hump}
\end{equation}
known as  ``the double-hump" one. 
But, since then no support for this shape has materialized, and it also does not
agree with lattice results on momentum fractions, so we will not discuss it. Let us state once again, that we 
see phenomenological failure of the pQCD expressions
not in the modified shape of the wave function, but in 
missing nonperturbative part of the hard blocks.

The light pseudoscalar mesons $P=\pi, K, \eta$ are related to chiral symmetry breaking and 
therefore they exist even in models without confinement, such as the NJL and ILM. 
Their wave functions and parton distribution amplitudes (PDA)
have been calculated in various approximations.  The question was addressed originally in the ILM framework
in~\cite{Petrov:1998kg}, and  more recently using the quasi-distribution proposal by one of us in~\cite{Kock:2020frx}.
The  distribution amplitude for the light  P-pseudoscalars of squared mass $m_P^2$ was derived in~\cite{Kock:2020frx}

\begin{equation}
\label{PDFRIV}
\varphi_{P}(x)=\frac{2N_c}{f^2_P}\int\frac{d^2k_\perp}{(2\pi)^3}
\frac{\theta(x\bar x)}{(k_\perp^2+M^2(0, m_q)-x\bar xm_P^2)}\,M^2\bigg(\frac{\sqrt{k^2_\perp+M^2(0,m_q)}}{\lambda_P\sqrt{x\bar x}}\bigg)
\end{equation}
where the momentum-dependent quark mass is

\begin{equation}
M(k,0)=M(0)\bigg(\bigg|z\bigg(I_0K_0-I_1K_1\bigg)^\prime\bigg|^2\bigg)_{z=\frac 12 \rho k}
\end{equation}
Here $\lambda_P$ is a cut-off parameter of order 1, e.g. $\lambda_\pi=3.41$, and  $M(0)=386\,{\rm MeV}\approx M_Q$ with an instanton size $\rho=0.3$ fm. As shown in~\cite{Kock:2020frx},
this momentum dependence has been confirmed by  lattice studies.
For a  light current quark mass $m_q$, the  running effective quark mass $M(k, m_q)\approx M(k)+m_q$. 
The corresponding  shape of the wave function is shown in the upper plot of  Fig.\ref{fig_pionwf}  as reproduced from \cite{Kock:2020frx},
is in agreement with \cite{Petrov:1998kg}. Both calculations show a wave functions rather close to the asymptotic one, and
very far from the ``double-hump"  distribution (\ref{eqn_double_hump}).

Another approach to light front wave functions is based on some model-dependent Hamiltonians.
Jia and Vary \cite{Jia:2018ary} introduced a convenient form of it, with three basic elements:  constituent quark masses
(that is, chiral symmetry breaking), plus some form of confinement, plus (NJL-type) residual quark-quark interaction.
This approach was followed by one of us~\cite{Shuryak:2019zhv}, who calculated the
wave functions for the $\pi,\rho,\eta' $ mesons,  as shown in the lower plot of Fig.\ref{fig_pionwf}. 

We do not study or discuss in this work other hadrons, such as baryons, pentaquarks and dibaryons, 
or multi-quark components of meson wave functions.  
Still, let us make here few remarks on those. Their PDF's, extracted from DIS and jets, are in this case
not sufficient to obtain the wave functions and DA's,
as they depend on more variables. And yet, it is very
important to study those: in particular, the structure of the nucleons is the central area of experimental research. So, let us mention few works on that related with instanton effects. 

In 1990's Diakonov and collaborators have developed 
a version of the chiral bag model based on  the  ILM, and  
calculated certain leading and next-to-leading twist nucleons DA's, for a review see
\cite{Polyakov:1997ea}. Unfortunately, it was done in the
large $N_c$ limit, which missed important 
elements, such as  the  instanton-induced diquarks \cite{Rapp:1997zu}.
 The wave functions 
for  the $\Delta$, the nucleon and even their  5-light-quark component
 were derived  recently in~\cite{Shuryak:2019zhv}, with  the  t'Hooft residual interaction. It provides
 the first quantitative derivation of the $anti$quark PDFs inside the nucleon, explaining its flavor asymmetry.

\subsection{Twist and  chiral structures of the DA's of the pion}

We start this section from a generic discussion of
chiral symmetry and its breaking. Naively, in a theory with $massless$ quarks chiral symmetry is exact. If it remains unbroken, hadrons diagonal in chirality 
such as $\overline q_L q_L+\overline q_Rq_R$,  and non-diagonal  in chirality  such as $\overline  q_L q_R+\overline q_R q_L$
would simply be different species, with different masses and wave functions. Since chiral symmetry
is spontaneously broken in the QCD vacuum (and at low temperatures $T<T_c$), all such would be species are mixed together.  

Starting from the QCD sum rule days of 1970's, it is known
that hadrons can be excited by local operators with
different chiral structure. For example, positive pions can be
excited $both$ by (chiral non-diagonal) pseudoscalar operator
$(\bar di \gamma_5 u)$ and (diagonal) axial current
$(\bar d \gamma_\mu\gamma_5 u)$. Yet the pion role  in these two correlation functions 
is very different. While in the pseudoscalar  correlator  the pion practically dominates from very small distances on,  
in the axial-vector correlator the $A_1$ meson dominates, and only at rather large distances  a pion tail appears. The coupling to the axial-vector  current  $f_\pi$,
is relatively small, and vanishes  if chiral symmetry  is restored. 

In the case of the pion,
the   DA is
$\pi^+(p)\rightarrow q_{if\alpha}(k)\bar q_{jg\beta}(k-p)$ corresponds formally to the connected amplitude

\begin{eqnarray}
\label{WF1}
&&\int_{-\infty}^{+\infty}\frac{p^+dz^-}{2\pi}e^{ixp\cdot z}\left<0\left|\overline{d}_\beta(0)[0,z]u_\alpha(z)\right|\pi^+(p)\right>\nonumber\\
&&=
\bigg(+\frac{if_\pi}4 \gamma^5\bigg(\slashed{p}\,\varphi_{\pi^+}(x)
-\chi_\pi \varphi_{\pi}^P(x)+i \chi_\pi
\sigma_{\alpha\beta}\frac{p^\alpha p^{\prime\beta}}{p\cdot p^\prime}  {\varphi_{\pi}^{\prime\,T}(x)\over {6}}\bigg)\bigg)_{\alpha\beta}
\end{eqnarray}
and its conjugate

\begin{eqnarray}
\label{WF1X1}
&&\int_{-\infty}^{+\infty}\frac{p^{\prime -}dz^+}{2\pi}e^{-ixp^\prime\cdot z}\left<\pi^+(p^\prime)\left|\overline{u}_\beta(z)[z,0]d_\alpha(0)\right|0\right>\nonumber\\
&&=
\bigg(-\frac{if_\pi}4 \gamma^5\bigg(\slashed{p}^\prime\,\varphi_{\pi^+}(x)
+\chi_\pi \varphi_{\pi}^P(x)-i \chi_\pi
\sigma_{\alpha\beta}\frac{p^\alpha p^{\prime\beta}}{p\cdot p^\prime}  {\varphi_{\pi}^{\prime\,T}(x)\over {6}}\bigg)\bigg)_{\alpha\beta}
\end{eqnarray}
up to twist-3.  $[x,y]$  refers to the gauge link and  $\sigma_{\alpha\beta}=\frac i2[\gamma_\alpha, \gamma_\beta]$. (\ref{WF1}-\ref{WF1XX}) are explicitly odd under P parity.

Note that the 4-vector $p'_\mu$ appears  in the DA of a pion with 4-vector $p_\mu$, in reference to the conjugate
light-cone direction, with  generally  no relation to the second pion.
In the DA of a  pion with momentum $p'_\mu$, the exchange  $p\leftrightarrow p^\prime$ needs to be enforced,
effectively flipping the sign of the last term.
Also, note that  we dropped the contribution

\be
\frac{f_\pi\chi_\pi}4\sigma_{\mu\nu}p^\mu\bigg[\bigg(\frac{\partial}{\partial k_{\perp\nu}}\frac{\varphi_\pi^T(x)}6\bigg)+\frac{\varphi_\pi^T(x)}6\frac{\partial}{\partial k_{\perp\nu}}\bigg]
\ee
as it involves the dependence on $k_\perp$ which we have ignored in both the soft and hard blocks.

(\ref{WF1}-\ref{WF1X1})  can be  inverted, to recast the pion twist-2 and twist-3 light-cone wavefunctions in  explicit form

\begin{eqnarray}
&&\varphi_{\pi^+}(x)=
\frac  1{if_\pi}\int_{-\infty} ^{+\infty} \frac{dz^-}{2\pi}e^{ixp\cdot z}\left<0\left|\overline{d}(0)\gamma^+\gamma_5[0,z]u(z)\right|\pi^+(p)\right>\nonumber\\
&&\varphi^P_{\pi^+}(x)=
\frac  {p^+}{f_\pi\chi_\pi}\int_{-\infty} ^{+\infty}  \frac{dz^-}{2\pi}e^{ixp\cdot z}\left<0\left|\overline{d}(0)i\gamma_5[0,z]u(z)\right|\pi^+(p)\right>\nonumber\\
&&\frac{\varphi^{T\prime}_{\pi^+}(x)}6=
\frac  1{f_\pi\chi_{\pi}}\frac {p^\mu p^{\prime \nu}p^+}{p\cdot p^\prime}\int _{-\infty} ^{+\infty} \frac{dz^-}{2\pi}e^{ixp\cdot z}\left<0\left|\overline{d}(0)\sigma_{\mu\nu}\gamma_5[0,z]u(z)\right|\pi^+(p)\right>
\end{eqnarray}
with all DA  normalized to 1. 
The leading twist-2 DA  $\varphi_\pi(x)$ is chirally-diagonal.  Although it characterizes the axial-vector strength in the pion, 
it is traditionally referred to without the label $A$ or axial, a convention we will hold. Its normalization to 1 is fixed by 
 the weak  pion decay constant $f_\pi\approx 133\,{\rm MeV}$,    

\begin{equation}
\label{WF11}
\left<0\left|\overline{d}(0)\gamma^\mu{(1-\gamma^5)}u(0)\right|\pi^+(p)\right>=
-{\rm Tr}\bigg(\gamma^\mu{(1-\gamma^5)}\bigg(\frac{if_\pi}4\gamma^5\slashed{p}\bigg)\bigg)\,\int_0^1dx\,\varphi_{\pi^+}(x)\equiv 
if_\pi\,p^\mu
\end{equation}
Isospin symmetry and charge conjugation force
$\varphi_\pi(x)=\varphi_\pi(\overline x)$.
As pointed out initially in  \cite{Geshkenbein:1982zs}, there are two twist-3 and {\em chirally non-diagonal} independent  DA 
$\varphi^P_\pi(x)$ and $\varphi_\pi^T(x)$, characterizing the pseudoscalar and tensor strength in the pion respectively.  The
latters are tied by the current identity

\be
\partial^\nu\big(\overline{d}(0)\sigma_{\mu\nu}\gamma_5u(z)\big)
=-\partial_\mu\big(\overline{d}(0)i\gamma_5 u(z)\big)+m\,\overline{d}(0)\gamma_\mu\gamma_5u(z)
\ee
and share the same couplings.
The value of the  dimensionful coupling constant $\chi_\pi$ can be fixed by the divergence of the axial-vector  current
and the PCAC relation

\begin{eqnarray} \label{eqn_div_of_axial}
&&(m_u+m_d) \left<0\left|\overline{d}(0)i\gamma^5u(0)\right|\pi^+(p)\right>=\nonumber\\
&&-(m_u+m_d)\,{\rm Tr}\bigg(i\gamma^5\bigg(\frac{if_\pi}4\gamma^5 \chi_\pi\bigg)\bigg)\,\int_0^1dx\,\varphi_{\pi}^P(x)
=(m_u+m_d)\,f_\pi\chi_\pi
\end{eqnarray}
with $\varphi_{\pi}^P(x)$  normalized to 1.  Using the Gell-Mann-Oakes-Renner  relation
\begin{equation}
f_\pi^2m_\pi^2=-2(m_u+m_d)\left<\overline{q}q\right>
\end{equation} 
with $|\left<\overline q  q\right>|\approx (240\,{\rm MeV})^3$, which yields
\begin{equation}
\label{CHI}
\chi_\pi=
\frac{m_\pi^2}{(m_u+m_d)}
\end{equation}
Furthermore,  quark masses are not physical quantities
and their numerical values depend on the definition, in particular on chosen normalization point $\mu$. PDG tables use rather high value $\mu=2\, GeV$, appropriate e.g. to lattice simulations with fine lattices. For usage in DA's more appropriate are values at ``soft"
normalization, as used in hadronic spectroscopy, which are about twice larger than PDG values. Therefore we will use for $\chi_\pi\approx 1.2\, GeV$.

Note that while 
the coupling to the pseudoscalar current is numerically large compared to $f_\pi$, this term
flips the quark chirality. For the vector form factor, 
this term contributes typically subleading 
corrections $\sim \chi_\pi^2/Q^2$.
However, as we have shown in our summary results, its contribution
is far from being negligible in the kinematical region of interest.

Asymptotically,  twist-3 contributions combined  together give subleading contributions to the pion form factor at large $Q^2$ as can be seen in (\ref{eqn_Vapi}).
Indeed, the asymptotic  limits of these DA's are $\varphi_{\pi}^P(x)\rightarrow 1$ and $\varphi_T(x)\rightarrow 6x\bar x$ owing to their conformal spin, with 
$\varphi_T^\prime (x)\rightarrow 6( \bar x- x)$. When inserted in  (\ref{eqn_Vapi}) the twist-3 contribution simplifies at asymptotic $Q$ with the result

 \begin{eqnarray} \label{TWIST-2Q}
\frac{f_\pi^2\chi_\pi^2}{Q^4}  \int  dx_1 dx_2 \, {1\over \bar x_1\bar x_2}
\bigg[\bigg(\frac 1{\bar x_2}-1\bigg)+(\bar x_2-x_2)\bigg(\frac 1{\bar x_2}+1\bigg)=2\bar x_2\bigg]=2\frac{f_\pi^2\chi_\pi^2}{Q^4}  \int \frac {dx_1}{\bar x_1}\,\,
\end{eqnarray}
which is clearly subleading. 
 However, in the semi-hard domain of interest for this work, one does not
  expect  the twist expansion to converge.
Moreover,  as was also noted already in  \cite{Geshkenbein:1982zs},  while suppressed asymptotically, 
P and T contributions  are actually enhanced by a large prefactor,
and are in fact dominant over the leading axial term.

\subsection{The twist and chiral structure of DA's \\of the transversely polarized vector mesons}

The DA's of the vector mesons, with longitudinal and transverse polarizations, are extensively discussed
in the literature. We refer to~\cite{Ball:1996tb,Ball:1998fj} for a thorough discussion of the leading and subleading twist contributions,
and how their couplings are related. We will mostly discuss the transversely polarized  $\rho^+$, with the following  twist-2 and twist-3 DA

\begin{eqnarray}
\label{WF1XX}
\varphi_{\rho_\perp}(x)=&&g_\perp^{(v)}(x)=\frac{p^+}{f_\rho m_\rho}\int _{-\infty}^{+\infty}\frac {dz^-}{2\pi}e^{ixp\cdot z}\,
\left<0\left|\overline{d}(0)\slashed{\epsilon}_\perp u(z)\right|\rho^+(p)\right>\nonumber\\
\varphi^T_{\rho_\perp}(x)=&&\phi_\perp(x)=\frac{p^+}{2f^T_\rho p\cdot p^\prime}\int_{-\infty}^{+\infty}\frac{dz^-}{2\pi}e^{ix p\cdot z}
\left<0\left|\overline{d}(0)(\slashed{\epsilon}_\perp\slashed{p}^\prime-\epsilon_\perp\cdot p) u(z)\right|\rho^+(p)\right>\nonumber\\
\varphi^{\prime\,A}_{\rho_\perp}(x)=&&g_\perp^{\prime\,(a)}(x)=\frac{2ip^+}{3f_\rho m_\rho p\cdot p^\prime}\epsilon^{\mu\nu\rho\sigma}{\epsilon_{T\mu}}p^\prime_\rho p_\sigma
\int _{-\infty}^{+\infty}\frac {dz^-}{2\pi}e^{ixp\cdot z}\,
\left<0\left|\overline{d}(0)\gamma_\mu\gamma_5u(z)\right|\rho^+(p)\right>\nonumber\\
\end{eqnarray}
with non-vanishing $\epsilon^\mu_\perp \neq 0$ only for $\mu=1,2$. 
The first  labeling refers to our notations, and the second  labeling to  the notations used
in~\cite{Ball:1996tb}. 

Expressions (\ref{WF1XX})  can be  inverted, to  give  the transverse rho twist-2 and twist-3 light-cone wavefunctions in  explicit form

\begin{eqnarray}
\label{WF1}
&&\int_{-\infty}^{+\infty}\frac{p^+dz^-}{2\pi}e^{ixp\cdot z}\left<0\left|\overline{d}_\beta(0)[0,z]u_\alpha(z)\right|\rho^+(p)\right>\nonumber\\
&&=
\bigg(+\frac  i4 \slashed\epsilon_\perp\bigg(f_\rho m_\rho\varphi_\rho(x) -f_\rho^T\slashed{p}\varphi_\rho^T(x)\bigg)
+  \frac{f_\rho m_\rho}{4p\cdot p^\prime}\epsilon_{\mu\nu\rho\sigma}\gamma^\mu\gamma_5\epsilon^\nu p^\rho p^{\prime\sigma}\frac{\varphi_\rho^{A\prime}(x)}4\bigg)_{\alpha\beta}
\end{eqnarray}
and its conjugate

\begin{eqnarray}
\label{WF1X1}
&&\int_{-\infty}^{+\infty}\frac{p^{\prime -}dz^+}{2\pi}e^{-ixp^\prime\cdot z}\left<\rho^+(p^\prime)\left|\overline{u}_\beta(z)[z,0]d_\alpha(0)\right|0\right>\nonumber\\
&&=
\bigg(-\frac  i4 \slashed\epsilon_\perp\bigg(f_\rho m_\rho\varphi_\rho(x) +f_\rho^T\slashed{p}^\prime\varphi_\rho^T(x)\bigg)
-  \frac{f_\rho m_\rho}{4p\cdot p^\prime}\epsilon_{\mu\nu\rho\sigma}\gamma^\mu\gamma_5\epsilon^\nu p^\rho p^{\prime\sigma}\frac{\varphi_\rho^{A\prime}(x)}4\bigg)_{\alpha\beta}
\end{eqnarray}
with again all DA  normalized to 1.  The parameter
  $ f_\rho\approx 210\,{\rm MeV}$ is fixed from the electromagnetic decay $\rho\rightarrow e^+e^-$,
  
  \begin{equation}
\label{WF2XX}
\left<0\left|\overline{d}(0)\gamma^\mu u(0)\right|\rho_\perp^+(p)\right>=
-\int_0^1dx\,{\rm Tr}\bigg(\gamma^\mu\bigg(\frac i4\slashed{\epsilon}_T({f_\rho} m_\rho\, \varphi_{\rho_\perp^+}(x)-{f^T_\rho} \slashed{p}\, \varphi^T_{\rho}(x))\bigg)\bigg)
\equiv i f_\rho\, m_\rho\,\epsilon_T^\mu
\end{equation}
  Lattice evaluation of the tensor coupling  ~\cite{Braun:2016wnx} give $ f_\rho^T/f_\rho\approx 0.6 $.
The reason the first and third lines in the above equations have the same coupling are extensively discussed 
in~\cite{Ball:1996tb} (see Appendix). 
Note that the first vector and last axial components are  chirality diagonal, the tensor term is chirality flipping.

\subsection{Very special mesons  $\eta'(958)$  and  $a_0(1450)$.}

The light vector mesons $\rho,\omega$ (and their strange counterparts$K^*,\phi$) 
 considered in the previous subsection are ``the most normal" mesons, generally described  as
a pair of constituent quarks  rather weakly bound 
by a confining potential.  In this respect they are different
from the pions, deeply bound by 
 instanton-induced forces. 
 
 However, there is  one more family of unusual mesons that we already mentioned  in (\ref{eqn_hooft_structure}),
 which appears in the instanton-induced t'Hooft effective Lagrangian  with negative sign, 
 and which corresponds to $repulsive$ forces. In that discussion, done for simplicity in the old two-flavor notations, they were called
 pseudoscalar isoscalar $\eta$ and scalar isovector $\delta$, respectively. In the real world with three light flavors $u,d,s$ they correspond
 to  the  $\eta'(958)$ and its
chiral partner $a_0(1450)$, in current notations. (The index zero is for the spin $J$ and not the  charge which, like for  the pion and rho, would be $+1$).
Such instanton-induced repulsion  makes them significantly heavier than   $\rho,\omega$. (Recall that for
the  mesons one should think in terms of mass squared).  

Technically, the calculation of the $\eta'$ two- (and three-) point correlation functions 
is  difficult to calculate because of the ``disconnected"
two-loop quark diagrams. 
But, since $a_0$  is its chiral partner  with flavored
 states coupled to the charged $(\bar d u)$ operator,  it produces only one-quark-loop diagram,  and  may serve as a reasonable substitute.
 The corresponding correlator in Euclidean time has been calculated in the Interacting Instanton Liquid Model
 in \cite{Schafer:1996wv}, together with many other mesonic two-point functions.
The latters strongly show a strong decrease with the distance, a behavior consistent with a strong instanton-induced repulsion.
Unlike other channels, no fit for the particle mass and coupling constant was done in this channel.  The  study mentions that  apparently there was no state
below the ``continuum threshold" at about $1.5\, {\rm GeV}$. No identification with   $a_0(1450)$ was made at that old analysis.
There may  be lattice studies of this scalar isovector channel, but we are not aware of such.  

The first calculation of the 
$\eta'$ DA in \cite{Shuryak:2019zhv}
 has been already mentioned, with its shape shown in Fig.\ref{fig_pionwf}(b). Indeed, it  has a shape that is quite different from the other mesons. One may anticipate that it should be similar for the $a_0$, a chiral partner to the $\eta'$ .

 One can define  the DA of the scalar mesons in the same form as for
the pion, just omitting $\gamma_5$, namely

\begin{eqnarray}
\label{WF1X1}
&&\int_{-\infty}^{+\infty}\frac{p^+dz^-}{2\pi}e^{ixp\cdot z}\left<0\left|\overline{d}_\beta(0)[0,z]u_\alpha(z)\right|a_0^+(p)\right>\nonumber\\
&&=
\bigg(+\frac{if_{a_0}}4 \bigg(\slashed{p}\,\varphi^V_{a_0}(x)
-\chi_{a_0}^S \varphi_{a_0}^S(x)+i\chi_{a_0}^T
\sigma_{\alpha\beta}\frac{p^\alpha p^{\prime\beta}}{p\cdot p^\prime}  {\varphi_{a_0}^{\prime\,T}(x)\over {6}}\bigg)\bigg)_{\alpha\beta}
\end{eqnarray}
and its conjugate

\begin{eqnarray}
\label{WF1X2}
&&\int_{-\infty}^{+\infty}\frac{p^{\prime -}dz^+}{2\pi}e^{-ixp^\prime\cdot z}\left<a_0^+(p^\prime)\left|\overline{u}_\beta(z)[z,0]d_\alpha(0)\right|0\right>\nonumber\\
&&=
\bigg(-\frac{if_{a_0}}4 \bigg(\slashed{p}^\prime\,\varphi^V_{a_0}(x)
-\chi_{a_0}^S\varphi_{a_0}^S(x)+ i\chi_{a_0}^T
\sigma_{\alpha\beta}\frac{p^\alpha p^{\prime\beta}}{p\cdot p^\prime}  {\varphi_{a_0}^{\prime\,T}(x)\over {6}}\bigg)\bigg)_{\alpha\beta}
\end{eqnarray}
up to twist-3. 
The three functions introduced here, and their couplings, of course have nothing to do with those of the pion.
In particular, the short-distance repulsive quark interaction is expected to make those couplings
to be much smaller numerically, as they correspond to the wave function at the origin.

The notations in (\ref{WF1X1}-\ref{WF1X2}) parallel the pion ones,  but with big differences in the values of the
parameters. Indeed, it is readily seen that the divergence of the vector current relates to the scalar matrix element
with 

\begin{equation} \chi_{a_0}^S={m_{a_0}^2 \over m_d-m_u }
\end{equation}	
which is similar to  $\chi_\pi$ in 	(\ref{CHI}), but with the difference
of the quark masses instead of their sum. In the isospin symmetric limit $m_d=m_u$, 
$\chi_{a_0}^S\rightarrow \infty$. Also, the mass of the $a_0$  in (\ref{CHI}),
does not vanish in the  chiral limit $m_{u,d}\rightarrow 0$, in which case
we also have $\chi_{a_0}^S\rightarrow \infty$. 
In a way, our parallel use of the notations with the pion is a bit misleading,
as $f_{a_0}\rightarrow 0$ in the chiral and/or the isospin symmetric limit.
Hence, it is more appropriate to use the finite combinations

\begin{equation} 
f_{a_0} \chi_{a_0}^S \rightarrow  (f_{a_0}^S)^2 , \,\,\, \,\,\,  f_{a_0} \chi_{a_0}^T \rightarrow (f_{a_0}^T)^2 ,
\end{equation} 
for the scalar and tensor  in (\ref{WF1X1}-\ref{WF1X2}), and disregard the small vector contribution.

The reader may also notice that 
there is a relation  between the  scalar and tensor
couplings, stemming from the condition that, with the 
asymptotic wave functions, certain cancellation 
must take place, so that
 their common contribution should be $\sim 1/Q^4$
as they are generically twist-3 structures. For instance, in the results quoted in (\ref{eqn_Vadel}) 
the asymptotic cancellation suggests that $f^{S2}_{a_0}=f^{T2}_{a_0}$ for $\varphi_{a_0}^S(x)\rightarrow 1$ and $\varphi^T_{a_0}(x)\rightarrow 6\bar x x$,
which is the choice of parameters used in~\ref{FFa0}.

\section{Discussion and outlook} \label{sec_summary}
\subsection{Nonperturbative  quark-quark interactions at  the
	few-${\rm GeV}^2$ scale}

We start by emphasizing the  chief motivation and result of this work. 
In the momentum transfer range of interest  $Q^2\sim { few}\,{\rm GeV}^2$, the quark-quark interactions are 
{\em much more complex} than just the lowest-order one-gluon exchange. Clearly, some
 nonperturbative
effects (and  higher order gluon diagrams)
are needed to quantitatively explain many
observations, with the mesonic form factors addressed here  being just
the simplest examples.

We argued that, while in this kinematic domain 
the dynamics is complex, the
  {\em collinear factorization} framework
  should still hold. Indeed, it is
based on the purely  kinematic  separation of the hard probing scale $Q^2$ from the soft internal scale 
$\langle p_\perp^2\rangle \sim 0.1\, {\rm GeV}^2$.  The separation of those scales still allows
 to separate any exclusive process into two parts:  (i) the  (quasi) local {\em hard block
	operator}, and (ii)   the {\em light front distributions}.

In the introductory section \ref{sec_NJL} we 
indicated, that the NJL 4-fermion
operators, fitted to chiral phenomenology,
have magnitude comparable   to 
those from perturbative one-gluon exchange. We argued
therefore, that  
in order to understand the magnitude of  the  hadronic form factors (and other exclusive reactions) 
one has to  include them. 

In this spirit, we performed a 
relatively long calculation of only a part
of the nonperturbative effects we can evaluate at this point, namely the
 instanton-induced ones. Our results  confirmed that
their contributions are indeed {\em comparable or larger} than the perturbative ones in the region discussed,
 $Q^2\sim
2-10\, {\rm GeV}^2$.

We further emphasized the  importance of 
including  a complete density matrix of mesons,
with different chiral structures, rather than
relying on leading twist ones.
In particular, taking together the perturbative axial and pseudoscalar  density matrix, and the gluon and instanton contributions, 
we found a  reasonable magnitude and $Q$ dependence for the total 
vector pion form factors. In fact it matches smoothly with the  data (and the monopole fit) at the lower end of the domain. Note that this happened in  a rather
nontrivial way, and without any parameter specially fitted. 

In the course of this work our understanding of  the
instanton-induced effects has changed. 
In particular, the anticipated dominance of the zero mode
contribution to $V_d,S_d$ did not occur, and  was found even to vanish.

A direction we took in this work aims at 
 {\em as many form factors as possible}, 
 for  a {\em large variety of mesons},
all evaluated in the same framework. 
We separately identified
  the effect of instantons into the  hard block, for the 
 scalar, vector   and gravitational form factors.
We then convoluted those with the full form of  the density matrix,  
 for the pseudoscalar, (transversely polarized) vector and even scalar mesons
 
 For technical reasons, we restricted our analysis to hadrons made of light quarks. Indeed, only in this case  we have  analytic expressions for the quark  propagators in the instanton background. 
 However, extensions to strange (and perhaps even charmed) quarks
 can be done, with perturbative inclusion of their masses.

Since the results were already presented 
upfront in section \ref{sec_results},  we will not
repeat  our comments here.
Instead, we will now take a wider perspective and
speculate on how these results
can be combined with other theoretical and phenomenological 
inputs,   to attack a general problem of understanding
forces acting between quarks.

\begin{figure}[h!]
	\centering
	\includegraphics[width=0.3\linewidth]{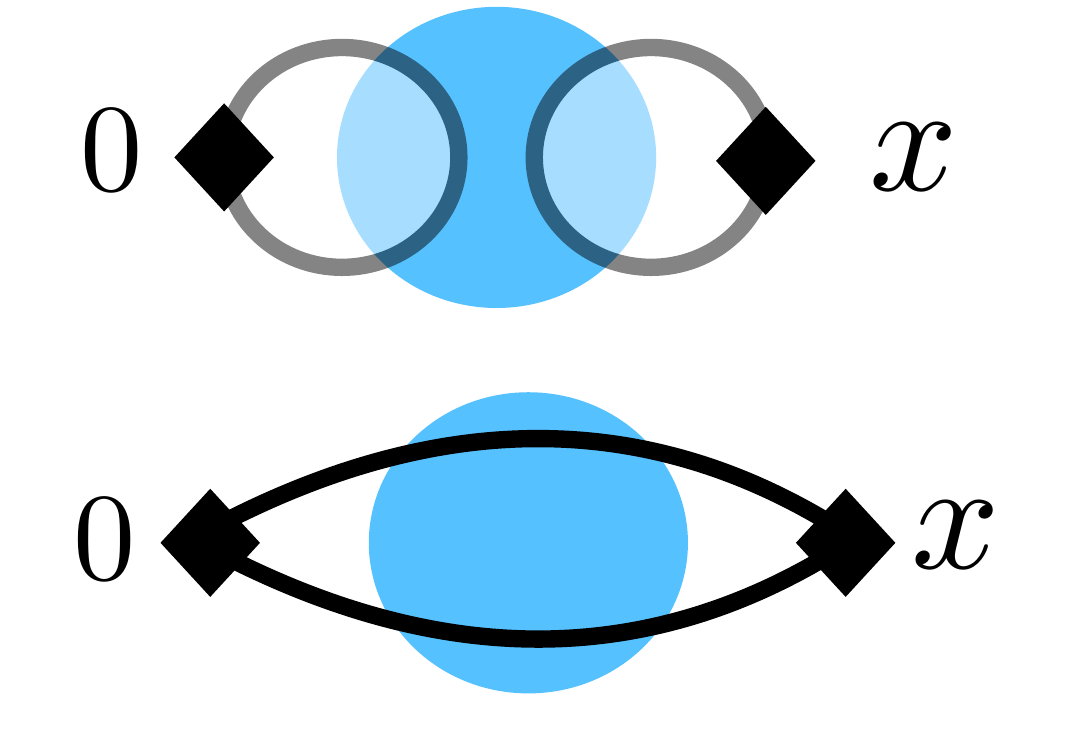}
	\caption{Two diagrams for in-vacuum correlation functions. The rhombuses correspond to two operators $\left<{\cal O}(0) {\cal O}(x)\right>$ inserted at Euclidean coordinates $0$ and $x\sim 1/Q$. The  blue circles refer to a nonperturbative background field.}
	\label{fig:corr}
\end{figure}

We recall  that  the important inputs  were provided by the point-to-point correlation functions at $intermediate$ distances $x\sim 1/Q$, 
with the scale of interest $Q^2\sim 	few-\,{\rm GeV}^2$.
The setting is schematically explained in Fig.\ref{fig:corr}. 
In a way, these studies revealed what can be called ``the form factors
of the QCD vacuum". 
 As discussed in detail e.g. in~\cite{Shuryak:1993kg,Schafer:1996wv}, at small distances they are described by pQCD diagrams, and at large distances by ``meson exchanges". 
At intermediate distances, of interest here, one 
finds a rather rich  channel-dependent set of correlators. This richness was first historically emphasized  in  the title of Ref~\cite{Novikov:1981xi}: `` Are all hadrons alike?", and quantitatively reproduced by instanton-based semiclassical theory.

The point-to-point correlation functions on the lattice
provided   {\it wave functions at the origin} in the form of constants like $f_\pi,\chi_\pi,f_\rho$, needed for
form factors. The strong splitting between the light $\pi-\sigma$ with $m_\pi^2 \approx 0$ and $m_\sigma^2\approx 0.2\, {\rm GeV}^2$ on the one hand,
and $\eta^\prime-\delta$ with $m_{\eta',\delta}^2\sim 1\, {\rm GeV}^2$ on the other, historically provided 
a  motivation for  the dominance of instanton-induced  forces described by 
the  effective 't Hooft  Lagrangian
 $$
{\cal O}_{\rm tHooft}\sim (\bar u_L u_R )(\bar d_L d_R) +(L \leftrightarrow R)$$

There are direct lattice evidences (e.g. \cite{Faccioli:2003qz}) that this operator is  indeed 
dominant in  the vacuum. One can certainly do now much
more systematic 
lattice  studies of multiple correlation functions at small/intermediate distances, and quantify
the strength of {\em all relevant} 4-fermion operators $${\cal O}_\Gamma=(\bar q \Gamma q)(\bar q \Gamma q) $$
That will put the NJL-type modeling of quark-quark forces on a more 
quantitative basis.

\begin{figure}[h!]
 	\centering
 	\includegraphics[width=0.3\linewidth]{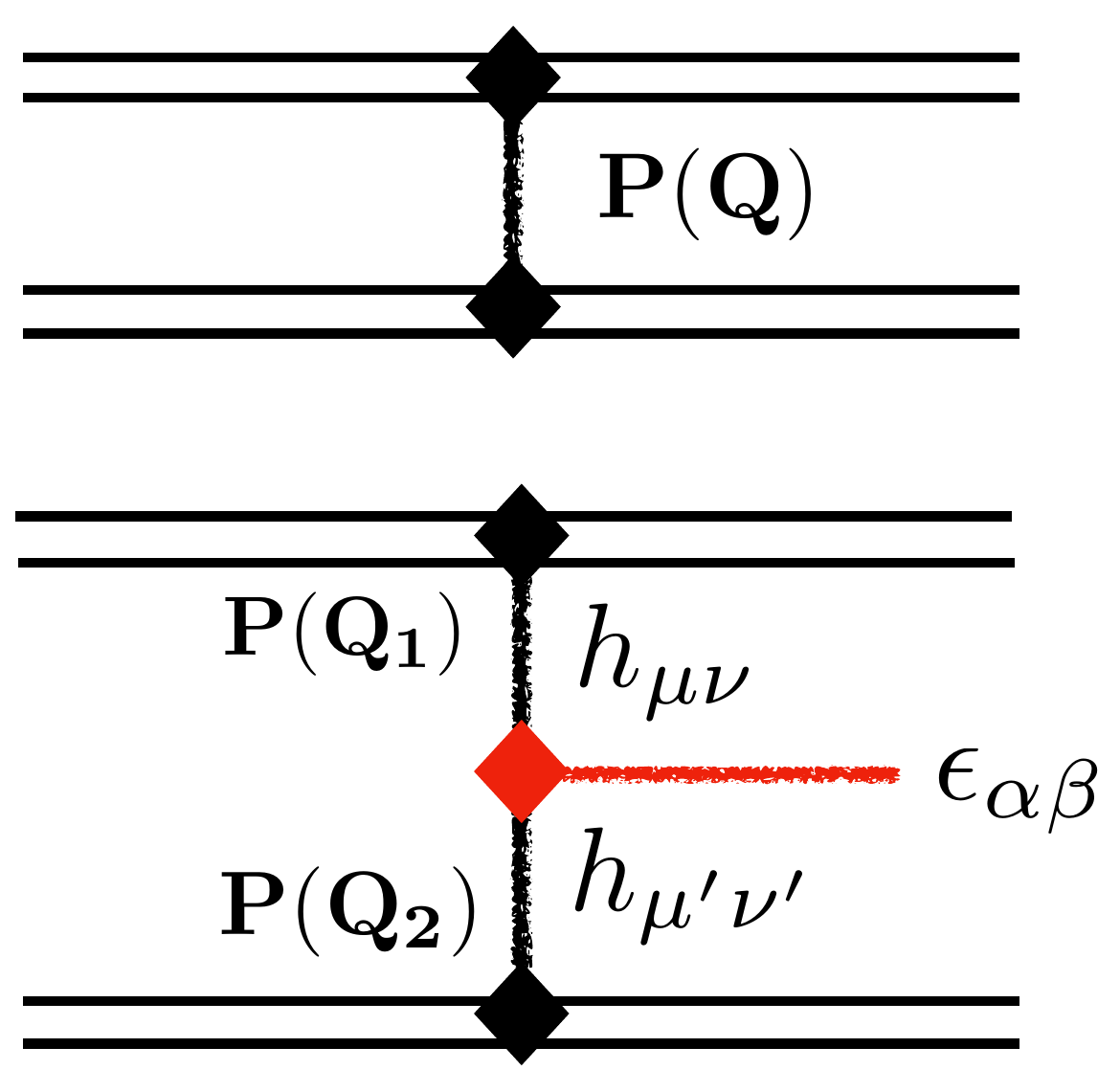}
 	\caption{ Elastic high energy collision of hadrons (upper) and double-diffractive production of a  hadron (lower plot).
 	Note that Pomeron-Pomeron-(tensor) Hadron may have many different index structures, but it is the ``triple-graviton" one which explains the data~\cite{Iatrakis:2016rvj}. }
 	\label{fig:pomerons}
 \end{figure}

Let us also mention here  another area in which 
an interesting phenomenology of quark interaction, in the same range of momentum transfer, has been developed: 
the physics of {\em Pomerons and diffractive processes}. 
In Fig.~\ref{fig:pomerons} we schematically show two
basic processes, the  high energy elastic scattering and the double-diffractive production
(sometimes called Pomeron-Pomeron collisions).
 
In the lowest perturbative order, the Pomeron is just a $two-gluon$ exchange. In higher orders it is given by ladder diagrams producing the so 
called BFKL Pomeron~\cite{Kuraev:1977fs,Balitsky:1978ic}. At intermediate $Q$ one also think of it as (Reggeized) 
exchange of glueballs. It is worth recalling, that
the first one on the Pomeron trajectory is not the lowest scalar glueball, but a tensor one $J^{PC}=2^{++}$. This correlates well with recent
demonstrations \cite{Iatrakis:2016rvj,Ewerz:2016onn} that the Pomeron is also an object possessing a symmetric polarization $tensor$ $h^{\mu\nu}$.
Finally, note that in the holographic models of QCD the 
Pomeron and tensor glueballs are just certain 
parts of Reggeized {\em graviton exchanges} that sum up to a close string exchange~\cite{Rho:1999jm,Brower:2006ea}.
Taking this into account, one may expect to find
among the quark-quark forces the operator
containing the product of two stress tensors
 $${\cal O}_{TT}=(\bar q \partial_\mu \gamma_\nu q)(\bar q \partial_\mu \gamma_\nu q) $$

The problem however remains: we do not entirely understand the mechanisms of the 
quark-antiquark scattering , at any level of precision.
The instantons are not the only nonperturbative 
objects in the QCD vacuum.  The (nearly 60 years old) NJL Lagrangian is very important, but still it is not the only quark interaction.

\subsection{Where should  further progress happen?}\label{sec-lat}

As we emphasized  in the Introduction, in spite
of significant efforts, 
experimental measurements of 
the pion and kaon form factors have hardly entered
the semi-hard domain of $Q^2$ discussed. Perhaps with
a new facility such as the  EIC in Brookhaven, there will be
further experimental progress.

In the near future, we anticipate a rapid progress in lattice calculations of the mesonic form factors. 
The simulations with physical light
masses are currently possible, and the subtleties of  chiral dynamics are  under control. 

\begin{figure}[t!]
\begin{center}
\includegraphics[width=7cm]{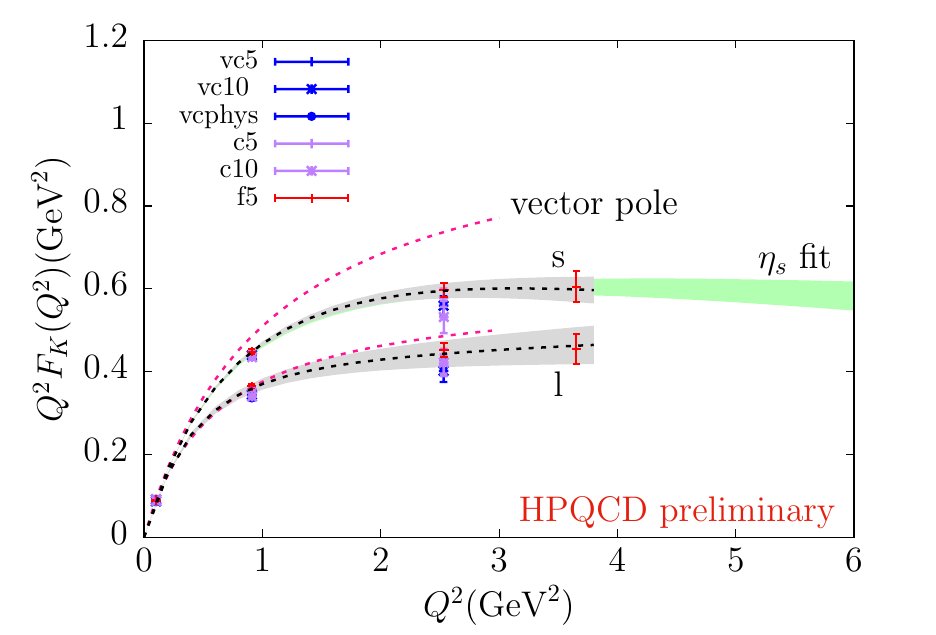}
\includegraphics[width=7cm]{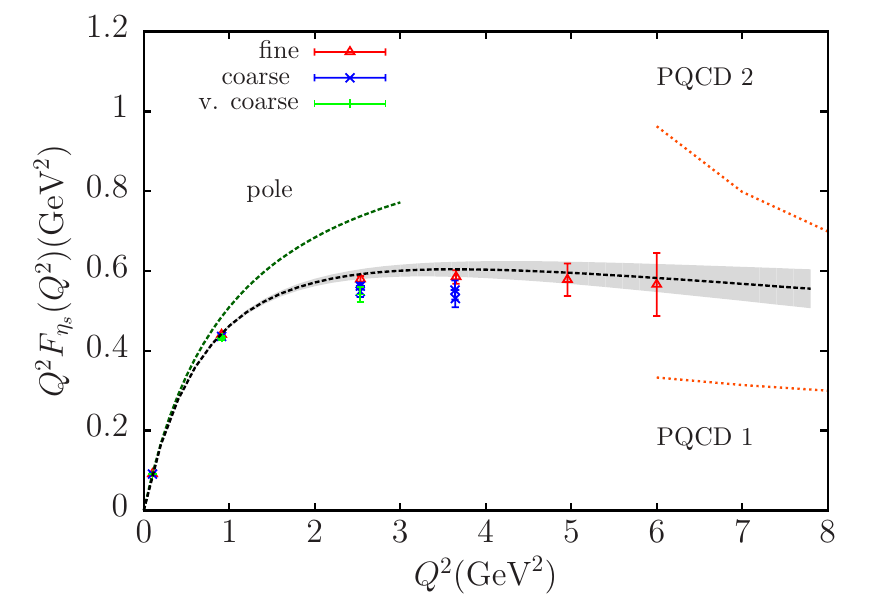}
\includegraphics[width=7cm]{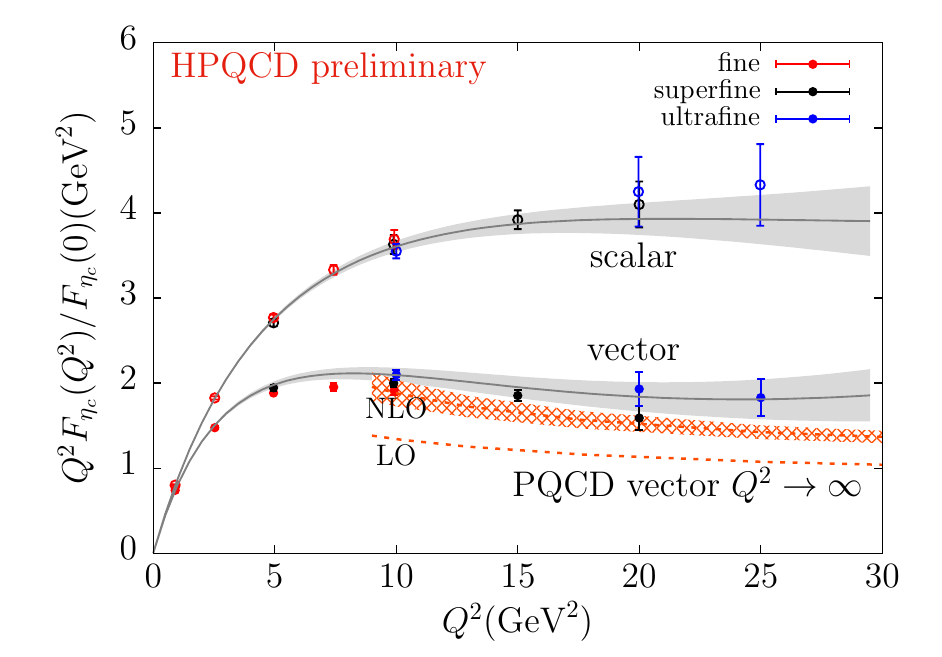}
\caption{Mesonic form factors calculated on the lattice by  the HPQCD collaboration:
 $K$-meson~\cite{Davies:2019nut}, $\eta_s$~\cite{Koponen:2017fvm} and $\eta_c$~\cite{Davies:2019nut}, top to bottom.
}
\label{fig_lat}
\end{center}
\end{figure}

Current 
 lattice studies (of the vector pion form factors  in~\cite{Brandt:2013ffb,Alexandrou:2017blh}
and scalar form factors~\cite{Gulpers:2013uca})
are restricted to
 momentum transfer range $Q^2< 1\, {\rm GeV}^2$. 
In order to get to larger momentum transfer, one needs
  lattices with smaller lattice spacing. With this in mind 
 we have picked up 
a sample of  mesonic form factors  carried by the HPQCD collaboration, see   Fig.~\ref{fig_lat}. 
The strategy of HPQCD  is to approach the problem
gradually, from heavier to lighter quarks. The natural expectation is  that  the physics of the heavy quark system is simpler, since their  nonrelativistic wave functions,
and some  other aspects are 
under better theoretical control. Heavier quark flavors
are expected to be less involved in nonperturbative interactions. Starting with the $b,c$ system, and going down to 
the strange quarks, we note that they become more relativistic, and more sensitive to the details
of chiral symmetry breaking and its  nonperturbative origins. 
On the lattice one can of course dial any value for the quark masses.

  In the vector channels,  mesons made of different flavors do not mix much, and one might think that  the 
  $\rho\rightarrow \phi \rightarrow J/\psi$ sequence can be smoothly connected via  a change in mass.
  However
  in
   the light pseudoscalar channels we know the  mixing is very strong. Furthermore, it is complicated by the broken chiral U$_A(1)$ symmetry,
   due to which  $\eta'$
   is not a partner to the octet states at all.  (This is  a strong indication of the 
   dominance of 't Hooft-like interaction, which is flavor-nondiagonal by construction.)

  In order not to deal with such issues, in~\cite{Koponen:2017fvm},  an artificial particle called $\eta_s$ was invented. It is different from the physical $\eta$ or $\eta'$.
 Its definition  can be explained as follows:
 imagine that there are {\em two more} species of strange quarks, $s'$ and $s''$, which have the same mass as 
 the standard $m_s$.  Their additional  properties are: (i)  they are not identical, so the pair $\bar s' s"$ cannot annihilate,
 and the  correlators  consist  only of the connected 
 one-loop diagram, while the (very costly) disconnected two-loop diagram does not exist. 
 This construction avoids the calculation of that diagram which is technically challenging; (ii) Also for simplicity,
 one of these presumed quarks  is given an electric charge $e_{s'}=1$, and the other one neutral, with  $e_{s''}=0$, eliminating  half of the diagrams.   
  It was numerically found that  in the range  $Q^2\approx 2-9\, {\rm GeV}^2$,  the $\eta_s$ form factor times $Q^2$
  \begin{equation}
  Q^2 F_{\eta_s}(Q^2)\approx 0.6 \, {\rm GeV}^2 
  \end{equation} 
  is approximatly constant, 
with no indication to reach the pQCD asymptotic value, which is significantly smaller. 
Almost  identical results were recently obtained for the $K$ meson form factor. 

For the charmed $\eta_c$ meson, the vector form factor was calculated to significantly higher momentum transfer $Q^2\sim 25\, {\rm GeV}^2$. 
Again,  $Q^2 F_{\eta_s}$ remains approximately constant, also about
twice larger than the asymptotic value.  Lattice data for the  scalar form factor 
indicate that it (times $Q^2$) is also about  constant, in contradiction 
with the chirality flip suppression rule predicted by the gluon exchange
mechanism. Furthermore, the numerical value of this constant is $larger$ than for vector form factor, by another factor 2! 

Multi-gluon diagrams lead to  a correction of order $(1+1.18\alpha_s)$~\cite{Melic:1998qr}, that 
can partly help to bridge the gap between the lattice results and the pQCD for the vector form factor.
However, this correction would not help for the scalar form factor, as  gluon exchange  cannot create the necessary chirality flip, except through
a penalty factor $m_f/Q$. However, we explain the 
presence of scalar form  factors by non-vanishing {\it cross-terms}, between
two chiral structures in the distributions.

Taken literally, our calculations are done for zero current quark masses, while the lattice data under consideration are for  massive  $s,c$ quarks.  So, they are not related directly.
  But, the strange quark mass is small enough $m_s^2 \ll Q^2$, and  it should not matter much, and in fact 
 it can perhaps be included perturbatively in the  propagators. 
What we found is that the  sum  of the  perturbative contribution with $Q^2 F(Q^2)$ weakly growing with $Q^2$, plus the  instanton contributions,
amounts to  a  near-constant  as observed on fine lattices by the  HPQCD collaboration.

The lattice configurations can be made ``more smooth" by
various procedures (e.g., the gradient flow) by which one can gradually remove high-momentum gluons, while keeping 
instantons (the action minima). Measuring formfactors
during such procedure, one is expected to see the results
changing, to their respective versions including only instanton-induced effects.
Our calculations can therefore be considered as predictions
for formfactors,  modified by such smoothening procedures. 

Finally,  instantons are only some fraction of the nonperturbative 
forces between quarks. We just calculated what we could. Instantons  are the only background for which full massless quark propagators  are available.  It may well be that, with better
measurements/calculations of these formfactors one may  need  some  non-perturbative
mechanisms of quark-antiquark interactions, other than the one-gluon exchange and instantons. Perhaps, for heavy quarks,  one can proceed along the line of Ref~\cite{Chernyshev:1995gj},
starting from the heavy quark limit.

\vskip 1cm
{\bf Acknowledgements} 

One of us (ES) thanks Vladimir Braun and the
other organizers of the Mainz 2019 workshop on light-front distribution amplitudes for the invitation,  to its participants, and in particular
 Christine  Davies, whose talk triggered 
  the present work. We thank Viktor Chernyak
for helpful comments on the first version of our paper.
This work is supported by the Office of Science, U.S. Department of Energy under Contract No. DE-FG-88ER40388.

\appendix

\section{Notations and kinematics} \label{sec_notations}

The definition of momenta are given in Fig.\ref{fig_diags}(a). Note that in the Breit frame
the initial meson momentum  $p_\mu=(0,0,-Q/2,Q/2)$,
and the final meson momentum is  $p'_\mu=(0,0,Q/2,Q/2)$, where
we ignored the meson mass $p_\mu^2\approx 0$.
Note that we put the energy as the 4-th
component rather than  the 0-th as we do in Euclidean
notations, and that we call the two first components $transverse$, for the momenta and polarization vectors.

Since we consider as an example mesons with charge
$+1$ or $\bar d u$ flavors, the upper line in Fig.\ref{fig_diags}(a) corresponds to  a $u$ quark,
and its direction is assumed to be left-to-right.
The other line ($\bar d$ with underlined momenta $\underline k_i$) has flow of baryon charge in the opposite direction, right-to-left, which is reflected
in the definition of its momenta with opposite sign. Therefore, in our notations 
$$p^\mu=k_1 ^\mu - \underline k_1 ^\mu,\,\,\,\,{p'}^\mu=k_2 ^\mu - \underline k_2 ^\mu$$
For completeness, let us mention that momentum conservation corresponds to $p^\mu+q^\mu={p'}^\mu$.
In Minkowskian kinematic we
use the standard Dirac ``slash" notations 
$$ \pslash \equiv p^\mu \gamma_\mu $$
Our set of gamma matrices are in the chiral basis,
meaning that $\gamma_5$ is diagonal.



\section{Instanton field  and its Fourier transform}\label{app_defs}

Throughout, we will use the conventions and notations developed in~\cite{Balitsky:1993ki,Moch:1996bs,Vandoren:2008xg} for the instanton calculus.
Superposition of instantons only makes sense if they are all in the so called singular gauge
\begin{equation}A^a_\mu(x)= {2\over g} \bar \eta^a_{\mu\nu} {\rho^2 (x_\nu-z_\nu) \over (x-z)^2((x-z)^2+\rho^2)} \end{equation}
$$= {2\over g} \bar \eta^a_{\mu\nu} (x_\nu-z_\nu)\big[ {1 \over (x-z)^2}- {1\over ((x-z)^2+\rho^2)}\big] $$

 All Fourier transforms will be carried using

\begin{equation}
\label{ZMF}
\mathbb{F} (\pm k)\equiv \int d^4x\,e^{\mp ik\cdot x}\,\mathbb{F} (x)
\end{equation}
The instanton   Fourier transform is

\begin{equation}A^a_\mu(k)={4\pi^2\over g} \bar \eta^a_{\mu\nu} {\partial \over \partial k_\nu}  \bigg( {1\over k^2}-{\rho \over k} K_1(k\rho)\bigg) 
\label{inst_field}
\end{equation}
where $K_1$ is the Bessel function. Quite characteristically, one finds two terms, one decaying as a power of $k$ as
$\langle A^*_\mu(k) A_\mu(k) \rangle\sim n/k^6$,  supplemented by a
 term which decreases exponentially at large k, $\sim e^{-k\rho}$.
 The former term comes from the point-like gauge topological singularity at the origin $\sim x_\nu/x^2$, and therefore it does not depend on instanton size $\rho$.
 It is spurious. The latter originates from the  regular bracket.

Following~\cite{Balitsky:1993ki,Moch:1996bs,Vandoren:2008xg}, 
we use the short hand matrix-valued notation
 $x\equiv \sigma_\mu x^\mu$ and $\overline{x}\equiv \overline\sigma_\mu x^\nu$, with 
the covariantized Pauli matrices in Euclidean and Minkowski space defined as 

\begin{eqnarray} \label{eqn_4dsigma}
{\rm Euclidean}:&&\qquad \sigma_\mu=(1,-i\vec\sigma)\qquad \overline\sigma_\mu=(1, +i\vec\sigma)\qquad 
\sigma_\mu\overline\sigma_\nu+\sigma_\nu\overline\sigma_\mu=2\eta_{\mu\nu}\nonumber\\
{\rm Minkowski}:&&\qquad \sigma_\mu=(1,-\vec\sigma)\qquad \overline\sigma_\mu=(1, +\vec\sigma)\qquad 
\sigma_\mu\overline\sigma_\nu+\sigma_\nu\overline\sigma_\mu=2g_{\mu\nu}\nonumber
\end{eqnarray}
with metric $g^{\mu\nu}=(+,-,-,-)$, $\eta^{\mu\nu}=\delta^{\mu\nu}$, and satisfying the identities

\begin{equation}
\sigma^\mu\overline{\sigma}^\nu-\sigma^\nu\overline{\sigma}^\mu=2i\overline{\eta}^{a\mu\nu}\tau^a\qquad\qquad
\overline\sigma^\mu{\sigma}^\nu-\overline\sigma^\nu{\sigma}^\mu=2i{\eta}^{a\mu\nu}\tau^a
\end{equation}
with the $\eta$-tHooft symbol.
The spinor indices are $\alpha,\beta=1,2$, and the color indices are  $i,j=1,2, ... N_c$. We will carry the  analytical continuation from 
Euclidean to Minkowski space using the prescription for space-like momenta $q_M^2\leq 0$

\begin{equation}
q_E^2\rightarrow -q_M^2+i0 \qquad \qquad \rho\sqrt{q_E^2}\rightarrow \rho\sqrt{-q_M^2}\
\end{equation}
For time-like momenta $q_M^2>0$,  it is more appropriate to use  the double prescription 

\begin{equation}
q_E^2\rightarrow -q_M^2+i0 \qquad \qquad \rho\sqrt{q_E^2}\rightarrow \rho\sqrt{q_M^2}\qquad \qquad {\rm arg}\rho=-{\rm arg}\sqrt{q^2}
\end{equation}
The analytical continuation  in the instanton size $\rho\rightarrow -i\rho$ compensates for the extra phase obtained when analytically continuing in momentum.
Since we are formally integrating over the instanton size distribution which is fixed by the saddle point, this continuation is absorbed by the $\rho$-integration measure
(\ref{dn_dist}).


\section{Fermionic zero modes }\label{app_zero}

Our  conventions for the $\gamma^5$ matrix in Euclidean and Minkowski space are respectively

\be
\gamma_E^5=
\begin{pmatrix}
-1& 0\\
0 &1
\end{pmatrix}\qquad\qquad 
\gamma_M^5=
\begin{pmatrix}
1& 0\\
0 &-1
\end{pmatrix}
\ee
In Weyl notations, the Euclidean Dirac spinor reads

\begin{eqnarray}
\label{NOTE1}
\Psi(x)&=&
\begin{pmatrix}
K^i_\alpha(x)\\
\phi^\alpha_i(x) \\
\end{pmatrix}
\qquad
\Psi^\dagger(x)=({K}_i^{\dagger\alpha}(x),\phi_\alpha^{\dagger i}(x))\equiv   (\overline{\phi}^\alpha_i(x), \overline{K}^i_\alpha(x))
\end{eqnarray}
The Euclidean fermionic action splits into left $K$ and right $\phi$ copies

\be
\overline K\sigma\cdot (\partial-igA)K+\overline \phi\,\overline\sigma\cdot(\partial -igA)\phi
\ee
with  $\overline K=\phi^\dagger$ and  $\overline\phi=K^\dagger$ using  

\be
\gamma_E^\mu=
\begin{pmatrix}
0& \bar\sigma_E^\mu\\
\sigma_E^\mu &0
\end{pmatrix}\qquad\qquad 
\gamma_M^\mu=
\begin{pmatrix}
0& \bar\sigma_M^\mu\\
\sigma_M^\mu &0
\end{pmatrix}
\ee

The instanton admits a left-handed zero mode $K^i_\alpha(x)$  satisfying $\sigma\cdot D\,K=0$,
and the anti-instanton a right-handed zero mode $\phi^\alpha_i(x)$ satisfying $\overline\sigma\cdot D\,\phi=0$, 
which are eigenstates of $(1\pm\gamma_E^5)/2$, and conjugate of each other. 
In terms of the Euclidean Weyl spinors,  the instanton zero mode  and its 
conjugate are


\begin{eqnarray}
\label{LR}
K^i_\alpha(x)&=&\frac{\rho^{\frac 32}}{\pi x^4}\frac{(\overline{x}\epsilon U)^i_\alpha}{\Pi_x^{\frac 32}}
=\frac{2\pi\rho^{\frac 32}}{\Pi_x^{\frac 32}}\,(\overline{S}_0(x)\epsilon U)^i_\alpha
\nonumber\\
K^{\dagger\alpha}_i(x)&=&\frac{\rho^{\frac 32}}{\pi x^4}\frac{(U^\dagger\epsilon x)^\alpha_i}{x^4\Pi_x^{\frac 32}}
=\frac{2\pi\rho^{\frac 32}}{\Pi_x^{\frac 32}}\,(U^\dagger\epsilon {S}_0(x))_i^\alpha\equiv {\overline\phi}^\alpha_i(x)
\end{eqnarray}
Here $\epsilon$ is the antisymmetric spin 2-tensor with the normalization 
$\epsilon_{\alpha\sigma}\epsilon^{\sigma\beta}=\delta_\alpha^\beta$, and

\begin{equation}
\Pi_x=1+\frac{\rho^2}{x^2}\qquad S_0(x)=\frac{x}{2\pi^2x^4}\qquad  \overline{S}_0(x)=\frac{\overline x}{2\pi^2x^4}
\end{equation}
with $S_0(x)$ the free massless quark propagator.
The zero modes  are normalized to $\rho$,

\bea
\int d^4x\, K^\dagger(x) K(x)=\rho\qquad \qquad \int d^4x \,\phi^\dagger(x)\phi(x)=\rho
\eea

For the free Dirac spinors we will use the notation $\chi(k)=\chi_R(k)+\chi_L(k)$  (with Minkowski labeling) as  the sum of  free Weyl spinors,  that  satisfy

\begin{eqnarray}
\label{FREEX}
\slashed k\chi(k) =\slashed  k
\begin{pmatrix}
\chi_R(k) \\
\chi_L(k) \\
\end{pmatrix}
=
\begin{pmatrix}
0&\overline{k} \\
k& 0\\
\end{pmatrix}
\begin{pmatrix}
\chi_R(k)\\
\chi_L(k) \\
\end{pmatrix}
=
\begin{pmatrix}
\overline{k}\chi_L(k) \\
k\chi_R(k)\\
\end{pmatrix}
=0
\end{eqnarray}
with the free-wave ortho-normalizations

\begin{eqnarray}
\chi_{L,R}(k)\chi_{L,R}^\dagger(k)=k,\overline k\qquad \qquad \chi^\dagger_{L,R}(k)\chi_{R,L}(k)=0
\end{eqnarray}

\section{Details of the averaging in (\ref{NZ1})}\label{app_51}


 The first bracket in (\ref{NZ1}),
 
 \begin{eqnarray}
 \label{BRAKET}
\left<k_2\left|{\partial}\,\overline{S}\,\overline{\epsilon}(q)\,\overline{S}\,{\partial}\,\frac{1+\gamma_5}2
+\overline{\partial}\, {S}\,{\epsilon}(q)\,{S}\,\overline{\partial}\,\frac{1-\gamma_5}2
\right|k_1\right>
\end{eqnarray}
  when converted to the configuration
representation, is dominated by the large $x,y$-asymptotics of the propagators on mass-shell. This translates formally to
$\overline S(x,z)\rightarrow  \overline{S}_0(x-z)/\sqrt{\Pi_z}$ to the left, and $\overline S(z,y)\rightarrow  \overline{S}_0(z-y)/\sqrt{\Pi_z}$ to the right, and similarly for ${S}(x,z)$ and
${S}(z,y)$. With this in mind, (\ref{BRAKET}) gives

\begin{eqnarray}
\label{NZ2}
\left(\overline\epsilon(q)\frac{1+\gamma_5}2+\epsilon(q)\frac{1-\gamma_5}2\right)\,\int d^4z\,\frac {e^{iq\cdot z}}{\Pi_z}
\end{eqnarray}
independently of the color orientations $U$ with

\begin{eqnarray}
\label{NZ2X}
\int d^4z\,\frac {e^{iq\cdot z}}{\Pi_z}=-{\partial_{q}^2}
\int d^4z\,{e^{iq\cdot z}}\int_0^\infty d\lambda\, e^{-\lambda (z^2+\rho^2)}=
-4\pi^2\rho^4\,\bigg(\frac{K_1(\xi)}{\xi}\bigg)^{\prime\prime}_{\xi=\rho q}
\end{eqnarray}

which at large $q$ asymptotes $\sim e^{-\rho q}/(\rho q)^{ 3/2}$,  plus an additional contribution

\begin{eqnarray}
\label{REDUCED}
&&\bigg(\frac 1{ik_2}\sigma_\mu \,[U\sigma_\mu\, (-i\overline\partial_{q})\,U^\dagger] \,\epsilon(q)\left(\frac{1-\gamma_5}2\right)\nonumber\\
&&-\,\overline\epsilon(q)\, \overline\sigma_\mu \,[U(-i\partial_{q})\,\overline\sigma_\mu \,U^\dagger] \, \frac 1{i\overline k_1}\left(\frac{1+\gamma_5}2\right)\,\bigg)\,
 \int d^4z \,{e^{iq\cdot z}}\,\frac{\rho^2}{z^4\Pi_z^2}\nonumber\\
\end{eqnarray}
which depends on the color orientations with

\begin{eqnarray}
\label{REDUCED1}
\int d^4z\,{e^{iq\cdot z}}\,\frac {\rho^2}{z^4\Pi^2_z}= -\rho^2{\partial \over \partial \rho^2} \int d^4z\,{e^{iq\cdot z}}\,\frac {1}{z^2+\rho^2}   =
-2\pi^2 \rho^2\,\bigg( \frac{(\xi K_1(\xi))^\prime}{\xi}\bigg)_{\xi=q\rho}
\end{eqnarray}
which  asymptotes  $\sim e^{-\rho q}/(\rho q)^{1/2}$. 

The contribution (\ref{NZ2X}) amounts  to the instanton contribution to the electric form factor
on a {\it single quark line}. To understand the electric or magnetic nature of the contribution (\ref{REDUCED}-\ref{REDUCED1}), we average it
over the instanton color moduli using the identity

\begin{equation}
\label{COLOR}
\int dU\,U^i_\alpha\,U^{\dagger  \beta}_j=\frac 1{N_c}\delta^i_j\delta^\beta_\alpha
\end{equation}
to have

\begin{eqnarray}
\label{REDUCED2}
\bigg(-\frac {\overline k_2}{k^2_2}\,q\,\epsilon(q)\left(\frac{1-\gamma_5}2\right)
+\,\overline\epsilon(q)\, \overline{q}\, \frac {k_1}{k^2_1}\left(\frac{1+\gamma_5}2\right)\,\bigg)\,
\bigg(-\frac{4\pi^2\rho^4}{N_c}\,\frac 1{\xi}\bigg(\frac{(\xi K_1)^\prime}{\xi}\bigg)^\prime\bigg)
\end{eqnarray}
On a {\it single quark line} with $q=k_2-k_1$ and both ends on mass-shell,  (\ref{REDUCED2}) yields the vector vertex

\begin{eqnarray}
\label{REDUCED3}
\frac{q^2}{M_Q^2}\bigg(1+{\cal O}\bigg(\frac{k_{1,2}}q\bigg)\bigg)\,\left(\overline\epsilon(q)\frac{1+\gamma_5}2+\epsilon(q)\frac{1-\gamma_5}2\right)\,
\bigg(-\frac{4\pi^2\rho^4}{N_c}\,\frac 1{\xi}\bigg(\frac{(\xi K_1)^\prime}{\xi}\bigg)^\prime\bigg)
\end{eqnarray}
We regulated the emerging poles using $k_{1,2}^2=0\rightarrow -M_Q^2$ in Euclidean signature.
 (\ref{REDUCED3}) is seen as an additional instanton contribution to the 
electric form factor of a {\it single quark line}, much like (\ref{NZ2X}).



The  second bracket in (\ref{NZ1})

\begin{eqnarray}
\label{SECOND}
\left<\underline{k}_1\left|{\partial}\,\overline{S}\,{\partial}\,\frac{1+\gamma_5}2
+\overline{\partial}\, {S}\,\overline{\partial}\,\frac{1-\gamma_5}2
\right|\underline{k}_2\right>\
\end{eqnarray}
can be LSZ reduced exactly. More specifically, applying the LSZ reduction to the right  of the
second term  in (\ref{SECOND})  and retaining only the leading  $\underline{k}_2^2\rightarrow 0$ contribution give

\begin{eqnarray}
\label{NZ3}
\bigg[\int d^4x\,{e^{-i({\underline{k}_2-\underline{k}_1)\cdot x}}}\,\frac{i\overline{\underline{k}}_1}{\sqrt{\Pi_x}}
\bigg(1+\frac{\rho^2}{2x^2}\frac{[Ux\underline{\overline{k}}_2U^\dagger]}{\underline k_2\cdot x}\bigg(1-e^{i\underline{k}_2\cdot x}\bigg)\bigg)+{\cal O}(\underline k_{2})\bigg]\,\frac{1-\gamma_5}2
\end{eqnarray}
Similarly,  applying the LSZ reduction to the left of the first term in (\ref{SECOND})  and retaining only the leading  $\underline{k}_1^2\rightarrow 0$ contribution give

\begin{eqnarray}
\label{NZ4}
\bigg[\int d^4x\,{e^{-i({\underline{k}_2-\underline{k}_1)\cdot x}}}\,
\bigg(1+\frac{\rho^2}{2x^2}\frac{[U\underline k_1\overline{x}U^\dagger]}{\underline k_1\cdot x}\bigg(1-e^{-i\underline{k}_1\cdot x}\bigg)\bigg)\,\frac{i{\underline{k}}_2}{\sqrt{\Pi_x}}
+{\cal O}(\underline k_{1})\bigg]\,\frac{1+\gamma_5}2
\end{eqnarray}
In the limit  $\underline{k}_{1,2}\rightarrow 0$, the contributions  in (\ref{NZ3}-\ref{NZ4}) do not vanish unless the integrals develop singularities which
can only arise from the large x-asymptotic of the integrands, as the contributions for $x\approx 0$ are all finite. For instance, for  a {\it single quark line} 
(\ref{NZ4}) after color averaging gives

\begin{eqnarray}
\label{MORE}
&&{i{\underline{k}}_2}\int d^4x\,{e^{-i(\underline{k}_2-\underline{k}_1)\cdot x}}\,
\bigg(1+\frac{\rho^2}{2N_cx^2}\bigg(1-e^{-i\underline{k}_1\cdot x}\bigg)\bigg)\frac 1{\sqrt{\Pi_x}}+{\cal O}(\underline k_{1})\nonumber\\
=&&{i{\underline{k}}_2}\,\bigg((2\pi)^4\,\delta^4(\underline{k}_2-\underline{k}_1)-\frac{2\pi^2\rho^2}{{(\underline{k}_2-\underline{k}_1)}^2}\bigg(1-\frac 1{N_c}\bigg)-\frac{2\pi^2\rho^2}{{\underline k}_2^2}\,\frac 1{N_c}\bigg)+{\cal O}(\underline k_{1,2})
\nonumber\\
\approx &&{i{\underline{k}}_2}\,\bigg((2\pi)^4\,\delta^4(\underline{k}_2)-\frac{2\pi^2\rho^2}{{\underline k}_2^2}\bigg(\bigg(1-\frac 1{N_c}\bigg) +\frac 1{N_c}\bigg)\bigg) +{\cal O}(\underline k_{1,2})
\end{eqnarray}
with the approximation $\underline{k}_2-\underline{k}_1\approx \underline k_2$ near the Euclidean mass shell. 
This approximation  supports
retaining only the 1-contribution in (\ref{NZ3}-\ref{NZ4}), with the final result in Euclidean signature
 \begin{equation}
\label{NZ8}
{2\pi^2\rho^2}\,\bigg(\bigg(1-\frac 1{N_c}\bigg) +\frac 1{N_c}\bigg)\,\bigg(\frac 1{i{\underline{k}}_1}\frac {1-\gamma_5}2+
\frac 1{i\overline{\underline{k}}_2}\frac {1+\gamma_5}2\bigg)\end{equation}


\section{Rules for the diagrams with zero modes insertion}\label{RULESX}

\begin{figure}[h!]
\begin{center}
\includegraphics[width=7cm]{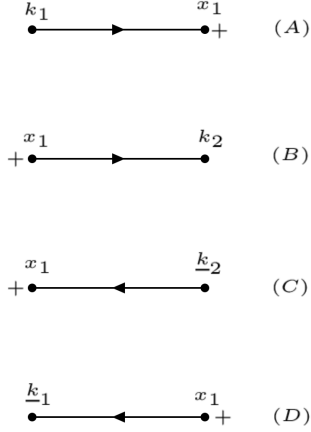}
\caption{Quark (A,B) and antiquark (C,D) zero modes entering and exiting an instanton labeled  by  $+$.}
\label{fig_fr1}
\end{center}
\end{figure}

In Fig.~\ref{fig_fr1} we illustrate the LSZ reduced Feynman diagram for incoming and outgoing waves stemming from an instanton zero mode.
We note that the LSZ reduction is not guaranteed unless the Euclidean field theory has a Hamiltonian interpretation, which is lacking in the 
instanton model of the QCD vacuum. Here it is understood as an algorithm.
More specifically, each LSZ reduced diagram corresponds to

\begin{eqnarray}
{\rm Fig}.~\ref{fig_fr1}A&=&\int d^4x_1\,\overline{\phi}(x_1)\overleftarrow{\overline\partial}_{x_1}\chi_L(k_1)e^{-ik_1\cdot x_1}
=\overline\phi(k_1)(i\overline{k})\chi_L(k_1)\nonumber\\ \nonumber\\
{\rm Fig}.~\ref{fig_fr1}B&=&\int d^4x_1\,\chi_R^\dagger(k_2)\overrightarrow{\partial}_{x_1}K(x_1)e^{+ik_2\cdot x_1}
=\chi_R^\dagger (k_2)(-ik_2)K(-k_2)\nonumber\\ \nonumber\\
{\rm Fig}.~\ref{fig_fr1}C&=&\int d^4x_1\,\overline{\phi}(x_1)\overleftarrow{\overline\partial}_{x_1}\chi_L(\underline{k}_2)e^{+i\underline{k}_2\cdot x_1}
=\overline\phi(-\underline{k}_2)(-i\overline{\underline{k}}_2)\chi_L(\underline{k}_2)\nonumber\\ \nonumber\\
{\rm Fig}.~\ref{fig_fr1}D&=&\int d^4x_1\,\chi_R^\dagger(\underline{k}_1)\overrightarrow{\partial}_{x_1}K(x_1)e^{-i\underline{k}_1\cdot x_1}
= \chi_R^\dagger (\underline{k}_1)(i\underline{k}_1)K(\underline{k}_1)\nonumber\\
\end{eqnarray}
with the instanton localized at $Z=0$. 
The integration over the collective $Z$-location of the instanton (anti-instanton)
in a given diagram gives rise to overall 4-momentum conservation

\begin{equation}
(2\pi)^4\delta(k_1+\underline{k}_1+q-k_2-\underline{k}_2)
\end{equation}
The LSZ reduction of the quark zero modes (\ref{LR})  amounts to the amputation of the free quark propagator in the large-x limit with
$\Pi_x\rightarrow1$. In momentum space this amounts to 

\begin{eqnarray}
\lim_{k^2\rightarrow 0}\chi_R^\dagger (k)\,ik\,K^i(-k)=-2\pi\rho^{\frac 32}\chi_{R\alpha}^\dagger (k)\epsilon^{\alpha\beta}U^i_\beta\nonumber\\
\lim_{k^2\rightarrow 0}\overline{\phi}_j(-k)(-i\overline{k})\chi_L(k)=
+2\pi\rho^{\frac 32}U^{\dagger\beta}_j\epsilon_{\beta\alpha}\chi_L^\alpha(k)
\end{eqnarray}

\begin{figure}[h!]
\begin{center}
\includegraphics[width=7cm]{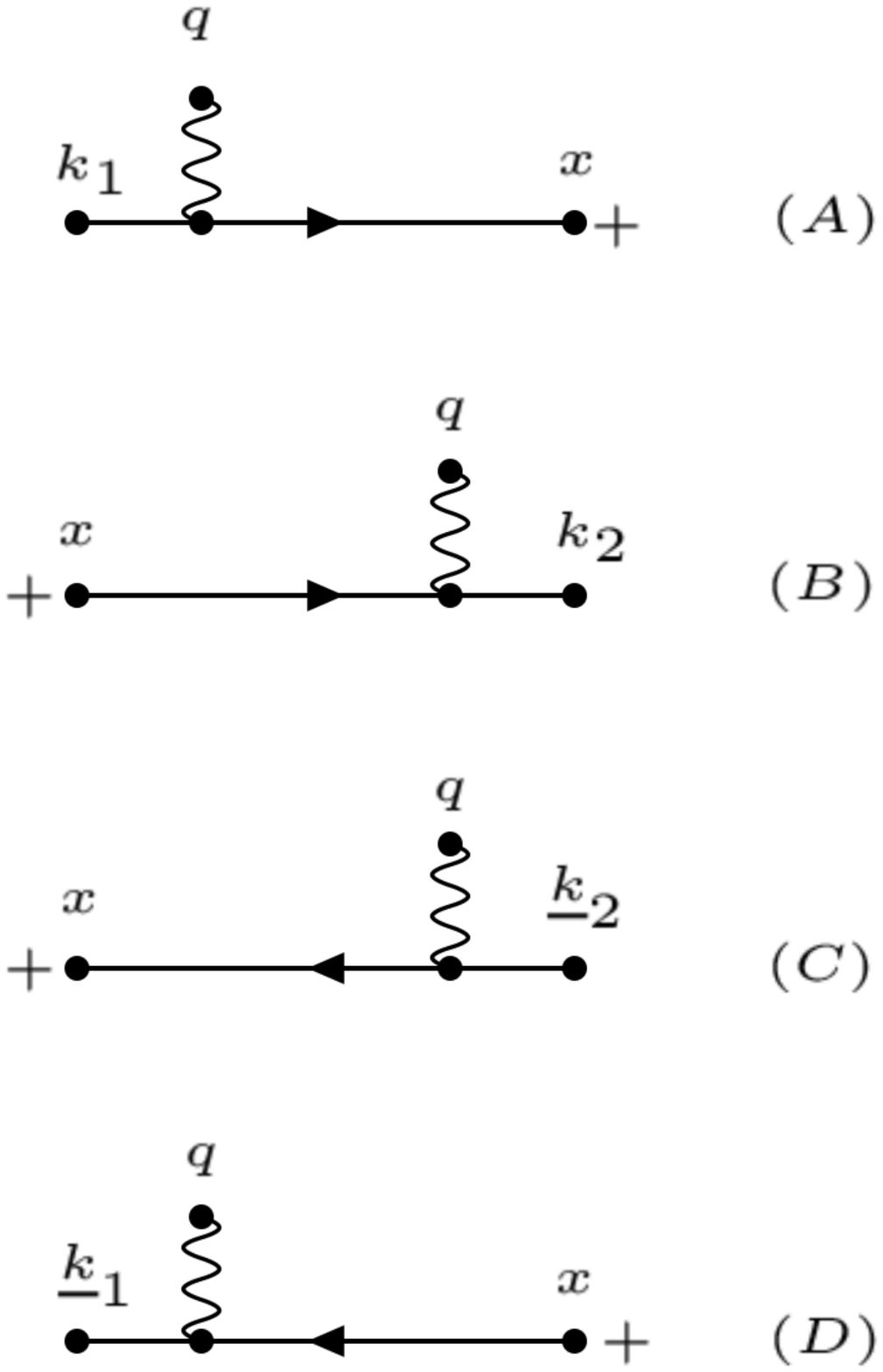}
\caption{Quark (A,B) and antiquark (C,D) absorbing or emitting a vector photon in an instanton background  labeled by $+$.}
\label{fig_fr2}
\end{center}
\end{figure}

In Fig.~\ref{fig_fr2} we show the Feyman graphs whereby a zero mode absorbs a virtual vector  with polarization 
$\epsilon_\mu(q)$ (absorption) 
and flips to a non-zero mode. The same rules
apply for a scalar or a pseudoscalar vertex  with the polarization set to $\pm 1$.  The extention to the energy-momentum 
tensor vertex were given above. 
Following the definitions established earlier, each of the vertex gives

\begin{eqnarray}
{\rm Fig}.~\ref{fig_fr2}A&=&\epsilon(q)\cdot {V}(q, k_1)\equiv \int d^4x\,e^{-iq\cdot x}\overline\phi(x)\,\overline\epsilon(q)\,S(x,k_1)\nonumber\\\nonumber\\
{\rm Fig}.~\ref{fig_fr2}B&=&\epsilon(q)\cdot \overline{V}(q, -{k}_2)\equiv\int d^4x\,e^{-iq\cdot x}\overline{S}(-k_2,x)\,\epsilon(q)\,K(x)\nonumber\\\nonumber\\
{\rm Fig}.~\ref{fig_fr2}C&=&\epsilon(q)\cdot {V}(q, -\underline{k}_2)\equiv\int d^4x\,e^{-iq\cdot x}
\overline\phi(x)\,\overline{\epsilon}(q)\,S(x, -\underline{k}_2)\nonumber\\\nonumber\\
{\rm Fig}.~\ref{fig_fr2}D&=&\epsilon(q)\cdot \overline{V}(q, \underline{k}_1)\equiv\int d^4x\,e^{-iq\cdot x}\overline{S}(\underline{k}_1,x)\,\epsilon(q)\,K(x)
\end{eqnarray}
The mixed Fourier transform of the non-zero mode propagators in Weyl form,  follows from (\ref{ZMF}) using

\begin{eqnarray}
S(x,y)=\bigg(S_0(x-y)\bigg(1+\rho^2\frac {Ux\bar yU^\dagger}{x^2y^2}\bigg)+\frac{\rho^2\sigma_\mu}{4\pi^2}
\frac{Ux(\bar x-\bar y)\sigma_\mu\bar yU^\dagger}{x^2(x-y)^2y^4\Pi_y}\bigg)\,\frac 1{(\Pi_x\Pi_y)^{\frac 12}}\nonumber\\
\overline{S}(x,y)=\bigg(\overline{S}_0(x-y)\bigg(1+\rho^2\frac {Ux\bar yU^\dagger}{x^2y^2}\bigg)+\frac{\rho^2\overline{\sigma}_\mu}{4\pi^2}
\frac{Ux\overline{\sigma}_\mu( x- y)\bar yU^\dagger}{\Pi_xx^4(x-y)^2y^2}\bigg)\,\frac 1{(\Pi_x\Pi_y)^{\frac 12}}
\end{eqnarray}
More specifically, the LSZ amputated propagators $S\bar k$ and $k\overline S$  are found to be

\begin{eqnarray}
\label{ZM-ZM}
ik\overline{S}(k,y)&\approx& \frac{e^{+ik\cdot  y}}{\Pi^{1/2}_y}
\bigg(1+\frac {\rho^2}{2y^2}\frac{Uk\bar{y}U^\dagger}{k\cdot  y}(1-e^{-ik\cdot y})\bigg)
\nonumber\\
{S}(x,k)i\bar k&\approx& \frac{e^{-ik\cdot x}}{\Pi^{1/2}_x}
\bigg(1+\frac {\rho^2}{2x^2}\frac{Ux\bar{k}U^\dagger}{k\cdot x}(1-e^{+ik\cdot x})\bigg)
\end{eqnarray}
in agreement with those originally found in~\cite{Moch:1996bs}.
The new and more involved LSZ amputations   $\overline S k$ and $\bar k S$  are

\begin{eqnarray}
\label{ZM-ZM-X}
\overline{S}(x,k)ik&\approx&
 \frac{e^{-ik\cdot x}}{\Pi^{1/2}_x}
 \bigg(1-\frac{\rho^2}{2x^2}\frac{U k\bar xU^\dagger}{k\cdot x}\bigg(1-\frac {i}{k\cdot x}(1-e^{+ik\cdot x})\bigg)
- \frac {\rho^2}{M_Q^2}\frac {ik}{x^4\Pi_x}\bar\sigma_\mu U {x} \bar\sigma_\mu U^\dagger\bigg)
\nonumber\\
i\bar k{S}(k,y)&\approx&
\frac{e^{+ik\cdot y}}{\Pi^{1/2}_y}
 \bigg(1-\frac{\rho^2}{2y^2}\frac{Uy \bar k U^\dagger}{k\cdot y}\bigg(1+\frac i{k\cdot y}(1-e^{-ik\cdot y})\bigg)
- \frac {\rho^2}{M_Q^2}\frac {i\bar k}{y^4\Pi_+}\sigma_\mu U \sigma_\mu  \bar{y}U^\dagger\bigg)\nonumber\\
\end{eqnarray}
Note that we made use of the Euclidean regulator $-k^2\approx 0\rightarrow M_Q^2$ in the last contributions
appearing in (\ref{ZM-ZM-X}).
The LSZ amputated parts of the vertices  in Figs.~\ref{fig_fr2}A,B  read

\begin{eqnarray}
&&\lim_{k_1^2\rightarrow 0}\epsilon(q)\cdot {V}(q, {k}_1)(i\overline{k}_1)=\nonumber\\&&+(i2\pi\rho^{\frac 32})
U^\dagger\bigg(\frac{\epsilon{k}_1\overline{\epsilon}(q)}{2k_1\cdot q}F(\rho \sqrt{q^2})+
\bigg(\frac{\epsilon(q+k_1)\overline{\epsilon}(q)}{(q+k_1)^2}-\frac{\epsilon k_1\overline{\epsilon}(q)}{2k_1\cdot q}\bigg)F(\rho\sqrt{(q+k_1)^2})\bigg)\nonumber\\
&&\lim_{k_2^2\rightarrow 0}(-ik_2)\epsilon(q)\cdot \overline{V}(q, -{k}_2)=\nonumber\\&&-(i2\pi\rho^{\frac 32})
\bigg(\frac{\epsilon(q)\overline{k}_2{\epsilon}}{2k_2\cdot q}F(\rho \sqrt{q^2})+
\bigg(\frac{\epsilon(q)(\overline q-\overline{k}_2){\epsilon}}{(q-k_2)^2}
-\frac{\epsilon(q)\overline{k}_2{\epsilon}}{2k_2\cdot q}\bigg)F(\rho\sqrt{(q-k_2)^2})\bigg)U\nonumber\\
\end{eqnarray}
respectively, and similarly for Figs.~\ref{fig_fr2}C,D with $F(x)=xK_1(x)$ in terms of the  Mac-Donald function.

\section{Scalar vertex redux}\label{redux}

If we ignore the transverse momentum dependence in the vertices, a simpler derivation of the vector, scalar and gravitational 
vertices in the instanton background can be obtained. Here, we illustrate it for the case of the scalar insertion. Typically, we have

\be
\label{S1X}
\tilde{\cal S}(q, k_1)=\int d^4x e^{-iq\cdot x}\overline\phi (x)S(x, k_1)
=\int d^4x e^{-iq\cdot x}\overline\phi (x)\,\overline\sigma_\mu{\overrightarrow D}_{\mu}\Delta(x,k_1)
\ee
with $\Delta(x,k_1)$ the half-mass shell  Fourier transform of the scalar propagator in (\ref{PROSCALAR}). 
Here since $\sigma_\mu D_\mu K(x)=0$ with $K^\dagger=\overline\phi$ (\ref{NOTE1}),  it is easier to proceed through the  integration by parts  and obtain

\be
\label{S2X}
\tilde{\cal S}(q, k_1)
=\int d^4x e^{-iq\cdot x}\overline\phi (x)\,i\bar q\,\Delta(x,k_1)\approx \int d^4x e^{-iq\cdot x}\overline\phi (x)\,\frac{i\bar q}{\sqrt{\Pi_x}}\,\Delta_0(x,k_1)
\ee
with the free scalar propagator $\Delta_0(x, k_1)=e^{ik_1\cdot x}/k_1^2$ and $k_1^2$ on mass shell. 
Inserting the instanton zero mode in (\ref{S2X}) and carrying the integration give
($\xi=\rho\sqrt{q^2}$)

\be
\label{S3X}
\tilde{\cal S}(q, k_1)\approx \frac{2\pi \rho^{7/2} q^2\,K_1(\xi)}{k_1^2\xi}\,(U^\dagger \epsilon ik_1)
\ee
In terms of (\ref{S3X}) the scalar vertex reads

\begin{eqnarray}
\label{S4X}
&&\bigg[\bigg(\frac{8\kappa q^2K_1(\xi)}{-k_1^2\xi}\bigg)
(\chi^\dagger_R(k_2)\epsilon U)\bigg(U^\dagger \epsilon \frac{k_1}{2M_Q}\chi_R(k_1)\bigg)
+(\chi^\dagger_L(k_2)\epsilon U)\bigg(U^\dagger \epsilon \frac{\bar k_2}{2M_Q}\chi_L(k_1)\bigg)\bigg]\nonumber\\
&&\times\bigg[\frac{(2\pi\rho)^2}{M_Q}(\chi_R^\dagger(\underline k_1)\epsilon U)(U^\dagger \epsilon \chi_L(\underline k_2))\bigg]
\end{eqnarray}
(\ref{S4X}) is in agreement with (\ref{FIRST2})  after analytical continuation with $k_1^2\approx 0\rightarrow -M_Q^2$ and
color averaging.

\section{Fierzing vector and scalar}

Further rearrangements at the expense of length can be done using Fierzing to 
recombine the color contractions., through the identity

\begin{equation}
\label{FIERZ}
(\overline{\psi}{\bf M}\phi)\,(\overline{\omega}{\bf N}\lambda)=-\frac 14\sum_{\cal O}\,
(\overline{\psi}{\cal O}\lambda)\,(\overline{\omega}{\bf N}{\cal O}{\bf M}\phi)
\end{equation}
with ${\cal O}={\bf 1}, \gamma_5, \gamma_\mu, i\gamma_5\gamma_\mu, i\gamma_\mu\gamma_\nu$.
More specifically, the subleading contributions in $1/N_c$ in  (\ref{VFULL}) can be cast in the form

\begin{eqnarray}
&&e_u\times \bigg(\frac {-1}{N_c(N_c^2-1)}\bigg)\bigg(\frac{-1}{4}\bigg)\nonumber\\
&&\times\bigg[\,\,2\,\bigg(\overline{u}_R(k_2)d_L(\underline{k}_2)\bigg)\bigg(\overline{d}_R(\underline{k}_1)(\mathbb F_V(q,k_1)+\mathbb F_V(q,-k_2))u_L(k_1)\bigg)\nonumber\\
&&\,\,\,\,\,\,-\,\bigg(\overline{u}_R(k_2)\gamma_\mu\gamma_\nu d_L(\underline{k}_2)\bigg)\bigg(\overline{d}_R(\underline{k}_1)\gamma^\mu\gamma^\nu(\mathbb F_V(q,k_1)+\mathbb F_V(q,-k_2))u_L(k_1)\bigg)\nonumber\\
&&\,\,\,\,\,\,+2\,\bigg(\overline{u}_R(k_2)u_L(k_1)\bigg)\bigg(\overline{d}_R(\underline{k}_1)(\mathbb F_V(q,k_1)+\mathbb F_V(q,-k_2))d_L(\underline{k}_2)\bigg)
\nonumber\\
&&\,\,\,\,\,\,-\,\bigg(\overline{u}_R(k_2)\gamma_\mu\gamma_\nu u_L(k_1)\bigg)\bigg(\overline{d}_R(\underline{k}_1)\gamma^\mu\gamma^\nu (\mathbb F_V(q,k_1)+\mathbb F_V(q,-k_2))d_L(\underline{k}_2)\bigg)
\bigg]
\end{eqnarray}
and

\begin{eqnarray}
&&e_{\overline d}\times\bigg(\frac {-1}{N_c(N_c^2-1)}\bigg)\bigg(\frac{-1}{4}\bigg)\nonumber\\
&&\times\bigg[
\,\,2  \bigg( \overline{u}_{R}(k_2)d_L(\underline{k}_2)\bigg)\bigg(\overline{d}_{R}(\underline{k}_1)
(\mathbb F_V(q,\underline{k}_1)+\mathbb F_V(q,-\underline{k}_2)) u_L(k_1)\bigg)\nonumber\\
&&- \bigg( \overline{u}_{R}(k_2)\gamma_\mu\gamma_\nu d_L(\underline{k}_2)\bigg)\bigg(\overline{d}_{R}(\underline{k}_1)
(\mathbb F_V(q,\underline{k}_1)+\mathbb F_V(q,-\underline{k}_2)) \gamma^\mu\gamma^\nu u_L(k_1)\nonumber\\
&&\,\,+2\bigg( \overline{u}_{R}(k_2)u_L(k_1)\bigg)\bigg(\overline{d}_{R} (\underline{k}_1)\mathbb (\mathbb F_V(q,\underline{k}_1)+\mathbb F_V(q,-\underline{k}_2))d_L(\underline{k}_2)\bigg)\nonumber\\
&&-\bigg( \overline{u}_{R}(k_2)\gamma_\mu\gamma_\nu u_L(k_1)\bigg)\bigg(\overline{d}_{R} (\underline{k}_1)
(\mathbb F_V(q,\underline{k}_1)+\mathbb F_V(q,-\underline{k}_2))\gamma^\mu\gamma^\nu d_L(\underline{k}_2)\bigg)
\bigg]
\end{eqnarray}

The subleading contributions in $1/N_c$ in  (\ref{SFULL}) can be Fierzed using (\ref{FIERZ}) with the result

\begin{eqnarray}
&&e_u\times \bigg(-\frac 1{N_c(N_c^2-1)} \bigg)\bigg(-\frac 14\bigg)\nonumber\\
&&\bigg[
\bigg( \overline{u}_{L}(k_2)\gamma_\mu u_L(k_1) \bigg)\bigg(\overline{d}_{R}(\underline{k}_1)\gamma^\mu \mathbb F_S(q,-k_2)d_L(\underline{k}_2)\bigg)\nonumber\\
&&-\bigg( \overline{u}_{L}(k_2)\gamma_\mu\gamma^5 u_L(k_1) \bigg)
\bigg(\overline{d}_{R}(\underline{k}_1)\gamma^\mu\gamma^5 \mathbb F_S(q,-k_2)d_L(\underline{k}_2)\bigg)\nonumber\\
&&+2\bigg( \overline{d}_{R}(\underline{k}_1) d_L(\underline{k}_2)\bigg) 
\bigg(\overline{u}_{R}({k}_2)  \overline{\mathbb F}_S(q,k_1)u_R({k}_1)\bigg)  \nonumber\\
&&-\bigg( \overline{d}_{R}(\underline{k}_1)\gamma_\mu\gamma_\nu d_L(\underline{k}_2)\bigg) 
\bigg(\overline{u}_{R}({k}_2) \gamma^\mu\gamma^\nu \overline{\mathbb F}_S(q,k_1)u_R({k}_1)\bigg)  \nonumber\\
&&+\bigg( \overline{u}_{L}(k_2) \gamma_\mu d_L(\underline{k}_2)  \bigg)\bigg(\overline{d}_{R}(\underline{k}_1)\gamma^\mu    \overline{\mathbb {F}}_S(q,-k_2)u_L(k_1)\bigg)\nonumber\\
&&-\bigg( \overline{u}_{L}(k_2) \gamma_\mu\gamma^5 d_L(\underline{k}_2)  \bigg)\bigg(\overline{d}_{R}(\underline{k}_1)\gamma^\mu \gamma^5   
\overline{\mathbb {F}}_S(q,-k_2)u_L(k_1)\bigg)\nonumber\\
&&+2\bigg(\overline{u}_{R}(k_2)   d^i_L(\underline{k}_2)\bigg)
\bigg(\overline{d}_{R}(\underline{k}_1)  \overline{\mathbb F}_S(q,k_1)u_R(k_1)\bigg)\nonumber\\
&&-\bigg(\overline{u}_{R}(k_2)\gamma_\mu\gamma_\nu   d_L(\underline{k}_2)\bigg)
\bigg(\overline{d}_{R}(\underline{k}_1)\gamma^\mu\gamma^\nu  \overline{\mathbb F}_S(q,k_1)u_R(k_1)\bigg)
\bigg]\nonumber\\
\end{eqnarray}
and

\begin{eqnarray}
&&e_{\overline d}\times \bigg(-\frac 1{N_c(N_c^2-1)} \bigg)\bigg(-\frac 14\bigg)\nonumber\\
&&\bigg[
2\bigg( \overline{u}_{R}(k_2)  u_L(k_1)      \bigg)
\bigg(\overline{d}_{R}(\underline{k}_1) \mathbb F_S(q, -\underline{k}_2)d_{R}(\underline{k}_2)    \bigg)\nonumber\\
&&-\bigg( \overline{u}_{R}(k_2) \gamma_\mu \gamma_\nu u_L(k_1)      \bigg)
\bigg(\overline{d}_{R}(\underline{k}_1) \gamma^\mu \gamma^\nu \mathbb F_S(q, -\underline{k}_2)d_{R}(\underline{k}_2)    \bigg)\nonumber\\
&&+\bigg(\overline{d}_{L}(\underline{k}_1) \gamma_\mu  d_{L}(\underline{k}_2)\bigg)
\bigg(\overline{u}_{R}(k_2) \gamma^\mu \overline{\mathbb F}_S(q, \underline{k}_1)u_{L}(k_1)\bigg)        \nonumber\\
&&-\bigg(\overline{d}_{L}(\underline{k}_1) \gamma_\mu\gamma^5  d_{L}(\underline{k}_2)\bigg)
\bigg(\overline{u}_{R}(k_2) \gamma^\mu\gamma^5 \overline{\mathbb F}_S(q, \underline{k}_1)u_{L}(k_1)\bigg)        \nonumber\\
&&+ \bigg( \overline{u}_{R}(k_2) \gamma_\mu d_{R}(\underline{k}_2)\bigg)   
\bigg(\overline{d}_{R}(\underline{k}_1)   \mathbb F_S(q, -\underline{k}_2) \gamma^\mu     u_L(k_1)\bigg)   \nonumber\\
&&- \bigg( \overline{u}_{R}(k_2) \gamma_\mu\gamma^5 d_{R}(\underline{k}_2)\bigg)   
\bigg(\overline{d}_{R}(\underline{k}_1)   \mathbb F_S(q, -\underline{k}_2) \gamma^\mu  \gamma^5   u_L(k_1)\bigg)   \nonumber\\
&&+2\bigg( \overline{u}_{R}(k_2)  d_{L}(\underline{k}_2)   \bigg)    \bigg(\overline{d}_{L}(\underline{k}_1)\overline{\mathbb F}_S(q, \underline{k}_1)   u_L(k_1)\bigg) \nonumber\\
&&-\bigg( \overline{u}_{R}(k_2)\gamma_\mu\gamma_\nu  d_{L}(\underline{k}_2)   \bigg)    
\bigg(\overline{d}_{L}(\underline{k}_1)\overline{\mathbb F}_S(q, \underline{k}_1)  \gamma^\mu\gamma^\nu  u_L(k_1)\bigg)   
\bigg]
\end{eqnarray}

\bibliography{forces-v2}

\end{document}